\documentclass[useAMS,usenatbib]{mn2e}

\usepackage{amsmath}
\usepackage{graphicx}
\usepackage{amssymb}

\newcommand{\nl}{\mathrm{NL}}
\newcommand{\nside}{\textsc{\small{NSIDE}}}

\title[Analysis of NG CMB maps based on the N-pdf]
{Analysis of non-Gaussian CMB maps based on the N-pdf. Application
to WMAP data}
\author[P. Vielva and J. L. Sanz]
{P. Vielva$^1$, J. L. Sanz$^{1,2}$ \\
$^{1}$Instituto de F{\'\i}sica de Cantabria (CSIC - Univ. de Cantabria), Avda. Los Castros s/n, 39005 - Santander, Spain\\
$^{2}$CNR Istituto de Scienza e Tecnologie dell'Informazione, 56124, Pisa, Italy \\
\hspace{0.1cm}E-mails : vielva@ifca.unican.es, sanz@ifca.unican.es
\\}
\date{Accepted ???. Received ???; in original form \today}
\pagerange{\pageref{firstpage}--\pageref{lastpage}}
\pubyear{1998}

\begin{document}
\maketitle
\label{firstpage}

\begin{abstract}
We present a new method based on the N-point probability
distribution (pdf) to study non-Gaussianity in cosmic microwave background (CMB)
maps. Likelihood and Bayesian estimation are applied to a local
non-linear perturbed model up to third order, characterized by a
linear term which is described by a Gaussian N-pdf, and a second and
third order terms which are proportional to the square and the cube of the
linear one. We also explore a set of model selection techniques
(the Akaike and the Bayesian Information Criteria, the minimum description length,
the Bayesian Evidence and the Generalized Likelihood Ratio Test)
and their application to decide whether a given data set is better described by the
proposed local non-Gaussian model, rather than by the standard
Gaussian temperature distribution. As an application, we consider the
analysis of the WMAP 5-year data at a resolution of $\approx
2^\circ$. At this angular scale (the Sachs-Wolfe regime), the
non-Gaussian description proposed in this work defaults (under certain
conditions) to an
approximative local form of the weak non-linear
coupling inflationary model~\citep[e.g.][]{komatsu01} previously addressed
in the literature. For this particular case, we obtain an estimation
for the non-linear coupling parameter of $ -94 < {\rm f}_{\nl} < 154$ at 95\%
CL. Equally, model selection criteria also indicate that the Gaussian
hypothesis is favored against the particular local non-Gaussian
model proposed in this work. This result is in agreement with
previous findings obtained for equivalent non-Gaussian models and
with different non-Gaussian estimators. However, our estimator based
on the N-pdf is more efficient than previous estimators and, therefore,
provides tighter constraints on the coupling parameter at degree angular
resolution.
\end{abstract}

\begin{keywords}
cosmology: observations -- cosmology: cosmic microwave background --
methods: data analysis -- methods: statistical
\end{keywords}

\section{Introduction}
\label{intro}

The Cosmic Microwave Background (CMB) fluctuations, a relic
radiation originated around 400,000 years after the Big-Bang, is one
of the most outstanding sources for understanding the evolution and
energy/matter content of the Universe. The large amount of high
quality data provided by recent CMB experiments and other
complementary astronomical observations, have provided with a
consistent picture for a flat Universe filled with cold dark matter
(CDM) and dark energy in the form of a cosmological constant
($\Lambda$), plus the standard baryonic and electromagnetic
components: the \emph{concordance model} \citep[e.g.][]{komatsu08}.
However, beyond the strength of the CMB measurements to put
constraints on the cosmological parameters, like the ones already
provided by the NASA \emph{Wilkinson Microwave Anisotropy Probe}
\citep[WMAP,][]{hinshaw08} and the ones expected from the incoming
ESA \emph{Planck satellite}, the CMB is a unique tool to probe
fundamental principles and assumptions of the so-called
\emph{standard model}. In particular, the application of
sophisticated statistical analysis to CMB data might help us to
understand whether the temperature fluctuations of the primordial
radiation are compatible with the fundamental isotropic and Gaussian
predictions from the \emph{inflationary phase}. The basic
inflationary scenario relates the CMB fluctuations, as well as the
large-scale structure of the Universe, to the Gaussian quantum
energy density perturbations present during the early
Universe~\citep[see for instance][]{liddle00}. The
present homogeneity and isotropy of the Universe is compatible with
an inflationary era in the early universe and this idea is the only
way, nowadays, to explain efficiently current observations
ranging from galaxies to the CMB. In particular, this fundamental
hypothesis predicts that the CMB fluctuations follow an isotropic
and Gaussian random. In fact, the estimation of
the cosmological parameters defining the concordance
model is done by assuming these statistical properties.

The quality of current CMB data (in particular the ones provided by
the WMAP satellite) has allowed for a systematic probe of the
statistical properties of the relic radiation. Indeed, the
interest of the scientific community in this field has experimented a
significant growth, since the analysis of the WMAP data has reported
several hints for departure from isotropy and Gaussianity of the CMB
temperature distribution. Some of these works are the following.
\cite{park04} detected a Gaussianity
deviation with a genus-based statistic; \cite{vielva04} found a
significant non-Gaussian signature on the 1-year WMAP data on the
kurtosis of the Spherical Mexican Hat (SMHW) wavelet coefficients at
scales of around 10 degrees, pointing out a very large \emph{cold
spot} (CS) on the southern hemisphere as a possible source for this
non-Gaussianity. A posterior analysis \citep{cruz05} confirmed the
anomalous nature of the CS by performing an area-based statistical
analysis of the wavelet coefficients. This detection, as well as new
ones, were confirmed by analyzing WMAP with various wavelet bases
and several statistical estimators
\citep{mukherjee04,cayon05,mcewen05,cruz06,martinez06,cruz07a,mcewen06,pietrobon08,wiaux08}.
Isotropy deviations were also reported in different manners: an
anomalous alignment of the low multipoles of the CMB
\citep{copi04,deOliveira04,katz04,schwarz04,bielewicz05,land05b,abramo06,freeman06,land07};
north-south asymmetries of the CMB fluctuations
\citep{eriksen04a,eriksen04b,hansen04a,hansen04b,donoghue05,eriksen05,land05a,bernui06,bernui07,eriksen07,rath07,gordon07};
an anomalous variance value \citep{monteserin08}; unexpected correlation among the CMB phases
\citep{chiang03,coles04,chiang06}; unbalanced distribution of the
temperature extrema \citep{tojeiro06}; and anomalous
alignment of CMB structures \citep{wiaux06,vielva06,vielva07}.

All the previous analyses can be considered as \emph{blind
approaches}, since null tests were performed to probe the CMB
compatibility with the isotropic and Gaussian hypotheses.
Complementary to these works, the reader can also find in the
literature several studies where \emph{targeted} departures from
Gaussianity are explored, based on non-standard physical models. In
particular, several analyses have studied the WMAP data
compatibility with anisotropic universes: the Bianchi VII$_h$ model
\citep{jaffe06a,jaffe06b,jaffe06c,bridges07a,bridges08}. Recently,
some works have proposed non-rotational invariant models like
\cite{bohmer08} and \cite{ackerman07} \citep[exprored by][in the
context of WMAP data]{groeneboom08}. Although these non-rotational
invariant models are promising and provide us with anisotropic
templates that could help to fix some anomalies in WMAP data, more
work is still needed to connect these anisotropic patterns of the
CMB fluctuations to a satisfactory physical model describing the
evolution of the anisotropic field
\citep[][]{himmetoglu08a,himmetoglu08b}.

In addition to the previous analyses, several works have studied
different hypotheses to explore the anomalous nature of the CS; for
instance \cite{cruz08} explored the CS compatibility with different
non-standard models, pointing out that an explanation in terms of a
cosmic texture \citep[as proposed by][]{cruz07b} is much more
favored than other alternatives already discussed in the literature,
like a very large void in the large scale structure of the Universe
or contamination, in the form of the Sunyaev-Zeldovich emission, due
to a large and nearby galaxy cluster. However, the study of
non-standard inflationary models is the problem that has attracted a
larger attention. For instance, and for the WMAP case, the
non-linear coupling parameter f$_{\nl}$ that describes the
non-linear evolution of the inflationary potential \citep[see
e.g.][and references therein]{bartolo04} has been constrained by
several groups: using the angular bispectrum
\citep{komatsu03,creminelli06,spergel07,komatsu08}; applying the
Minkowski functionals
\citep{komatsu03,spergel07,gott07,hikage08,komatsu08}; and using
different statistics based on wavelets
\citep{mukherjee04,cabella05,curto08}. Besides a claim for f$_{\nl}
>$ 0 with a probability greater than 95\% \citep{yadav08}, there is a
general consensus on the WMAP compatibility with the predictions
made by the standard inflationary scenario at least at 95\%
confidence level. The current best limits \citep{curto08} are:
$-8 < {\rm f}_{\nl} < 111$ at 95\% CL.

The present paper is related to \emph{non-blind} or \emph{targeted}
probes for Gaussianity deviations, and, more
specifically, it addresses two common topics arose in these kind of
studies: the definition of an optimal estimator for the non-standard
model (and, therefore, the optimal estimation of the parameters that
define such a model) and the complementary issue of model selection.
The former aspect is, usually, quite complex, since, for many
physical models and realistic observational limitations (e.g.
incomplete sky coverage) there is not a
trivial solution and, therefore, non-optimal estimators are usually
proposed (which are posteriorly characterized by simulations). The
latter issue on model selection is equally complex, since,
generally, relies on heuristic principles. Some authors adopt a fully
Bayesian view, whereas others adopt asymptotic measurements for the
distance between two hypothesis, and others prefer to rely on the
information that can be obtained from the likelihood itself. In this
work we propose to study a non-Gaussian model for the CMB that is
a local non-linear expansion of the temperature fluctuations.
For this model ---that, at large scales, is considered
by some authors as an approximation to the non-linear coupling
parameter f$_{\nl}$--- we are able to build the exact likelihood on
pixel space. To work in pixel space allows one to include easily
non-ideal observational conditions, like incomplete sky coverage and
anisotropic noise. We show how, from this likelihood, it is straightforward to
obtain an analytical expression for the optimal estimator for the
non-Gaussian term. In addition, we also explore different model selection criteria, like the Akaike
information criterion \citep{akaike73}, the Bayesian information
criterion \citep{schwarz78}, the minimum description length
\citep{rissanen01}, the Bayesian
evidence, and the generalized likelihood ratio test.

The paper is organized as follows. In Section~\ref{sec:model} we
describe the physical model based on the local expansion of the CMB
fluctuations and derive the full posterior probability. In
Section~\ref{sec:statistics} we address the issue of the parameter
estimation (\ref{subsec:estimation}) and model selection
(\ref{subsec:selection}). The methodology is explored on WMAP-like
simulations in Section~\ref{sec:simulations} and it is applied to WMAP 5-year data
in Section~\ref{sec:wmap}. Finally, conclusions are given
in Section~\ref{sec:final}.

\section{The non-Gaussian model}
\label{sec:model}

Current observations indicate that CMB temperature fluctuations can
be well described by random Gaussian fluctuations, as predicted by
the standard inflationary scenario. However, as it was discussed in
the Introduction, these observations
still allow for a small departure from the Gaussian distribution,
which could reflect the role played by physical processes described
by non-standard models for the structure formation.

In this paper, we will focus on a parametric non-Gaussian model,
that accounts for a small and local (i.e. point-to-point)
perturbation of the CMB temperature fluctuations, around its
intrinsic Gaussian distribution:
\begin{equation}
\label{eq:physical_model} {\Delta T}_i = \left({\Delta T}_i\right)_G
+ a\left[\left({\Delta T}_i\right)_G^2 - \left\langle\left({\Delta
T}_i\right)_G^2\right\rangle \right] + b\left({\Delta
T}_i\right)_G^3.
\end{equation}
$\left({\Delta T}_i\right)_G$ (\emph{the linear term}) is the
Gaussian part, whose N-point probability density function (N-pdf) can
be easily described in terms of the standard inflationary model. The
second and third terms on the right-hand side are \emph{the
quadratic} and \emph{cubic perturbation terms}, respectively. Their
corresponding contribution to the observed CMB fluctuations
(${\Delta T}_i$) is governed by the $a$ and $b$ parameters. Subindex
$i$ refers to the direction corresponding to a certain pixelization
on the sphere, and the operator $\langle \cdot \rangle$ averages
over all the pixels defining the sky coverage.
Notice that the previous expression does not include instrumental
noise. We have adopted this simplification, since we aim to
focus the work on large-scale CMB data sets ($> 1^\circ$), where
(as it is the case for WMAP), the contribution from noise
is typically negligible.

Let us point out that the particular situation for $b\equiv0$ defaults
into a well known case that has been already addressed in the literature \citep[e.g.][]{komatsu01,cayon03,curto07}.
It describes a local approximation to the weak non-linear coupling inflationary model
\citep[e..g][]{komatsu01,liguori03} at scales larger than the
horizon scale at the recombination time (i.e. above the degree
scale). In this context, the $a$ factor is usally related to the non-linear
coupling parameter f$_{\nl}$ by: 
\begin{equation}
\label{eq:fnl} a \equiv \frac{3\mathrm{f}_{\nl}}{T_0},
\end{equation}
where $T_0 = 2,725$ mK is the CMB temperature, and we follow the sign convention
in \cite{liddle00} for the relation between the temperature fluctuations and the
gravitational potential at the Sachs-Wolfe regime.
Let us remark that, of course, this model do not pretend to incorporate all the
gravitational (like lensing) and non-gravitational effects, due to the evolution of the
initial quadratic potential model.
Indeed, the reason to select the specific model given by equation~\ref{eq:physical_model}
is twofold. One the one hand, it is
an useful parametrization for describing a small departure from Gaussianity, that allows
us to present a new methodology. On the other hand, it is a model previously addressed
by other authors (using other estimators), and, therefore, it is easier to make
a straightforward comparison among the results.

For simplicity, let us transform the Gaussian
part $\left({\Delta T}_i\right)_G$ into
a zero mean and unity variance random variable $\phi_i$, hence,
equation~\ref{eq:physical_model} can be rewritten as: 
\begin{equation}
\label{eq:model} x_i = \phi_i + \epsilon\left(\phi_i^2 - 1\right) +
\alpha\epsilon^2\phi_i^3,
\end{equation}
where:
\begin{equation}
\label{eq:equivalences} x \equiv \frac{1}{\sigma}{\Delta T}, \ \ \
\phi \equiv \frac{1}{\sigma}\left({\Delta T}\right)_G ,\ \ \
\epsilon \equiv {a \sigma},\ \ \ \alpha  \equiv
\frac{b\sigma^2}{\epsilon^2},
\end{equation}
and $\sigma^2 \equiv \left\langle\left({\Delta
T}_i\right)_G^2\right\rangle$ is the rms of the CMB fluctuations.
Let us remark that, since the proposed non-Gaussian model is a
perturbation of the standard Gaussian one, the non-linear parameters
have to satisfy: 
\begin{equation}
\label{eq:condition} \vert \epsilon \vert \ll 1,\ \ \ \vert
\alpha \vert \lesssim 1.
\end{equation}
It is straightforward to show that the normalized Gaussian field
$\phi$ satisfies: 
\begin{equation}
\label{eq:phiproperties} \langle \phi_i \rangle = \langle \phi_i^3
\rangle = 0,\ \ \ \langle \phi_i^2 \rangle = 1,\ \ \ \langle \phi_i
\phi_j \rangle = \xi_{ij},
\end{equation}
where $\xi_{ij}$ represents the normalized correlation between pixels $i$ and
$j$. The N-pdf of the $\bmath{\phi}=\{\phi_1, \phi_2, ...,
\phi_N\}$ random field (where $N$ refers to the number of pixels on
the sphere that are observed) is given by a multivariate Gaussian:
\begin{equation}
\label{eq:pdfphi} p(\bmath{\phi}) = \frac {1}{(2\pi
)^{N/2}(\det{\bmath{\xi}} )^{1/2}}e^{-\frac{1}{2}
\bmath{\phi}\bmath{\xi}^{-1}\bmath{\phi}^{t}},
\end{equation}
where $\bmath{\xi}$ denotes the correlation matrix and operator
$\cdot^{t}$ denotes standard matrix/vector transpose.

Our goal is to compute the N-pdf associated to the non-Gaussian
$\bmath{x}=\{x_1, x_2, ..., x_N\}$ field, as a function of the
non-linear coupling parameters based on the full N-pdf for the
underlaying Gaussian signal $\bmath{\phi}$. Hence, let us make first
the inversion of equation~\ref{eq:model} to find the expression of
$\phi_i$ as a function of $x_i$: 
\begin{equation}
\label{eq:inversmodel}
\phi_i = x_i - \epsilon (x_i^2 - 1)
+\epsilon^2[(2 - \alpha)x_i^3- 2x_i] + O(3).
\end{equation}

Obviously, since previous equation is a local transformation, the
Jacobian matrix is diagonal and, therefore, the Jacobian ($Z$) is
given by:
\begin{equation}
\label{eq:jacobian} Z = \det{\left[ \frac{\partial \phi_i}{\partial
x_j} \right]} = \prod_i \left( \frac{\partial \phi_i}{\partial
x_i}\right).
\end{equation}
It is more convenient to work with the log-Jacobian ($\log{Z}$),
which, taking into account equation \ref{eq:inversmodel} and
expanding the log function up to second order, is given by:
\begin{eqnarray}
\label{eq:log-jacobian1} \log{Z} & = & \sum_i \log{\left(
\frac{\partial \phi_i}{\partial x_i} \right)} \\
& =&  -2\epsilon\sum_i x_i + \epsilon^2\left[-2N + \left(4 - 3\alpha
\right)\sum_i x_i^2\right] + 0(3) \nonumber.
\end{eqnarray}
Now, taking into account equation~\ref{eq:model} and recalling that
$\bmath{\phi}$ is a Gaussian field, it is straightforward to prove
that the data $\bmath{x}$ satisfy: $\sum_i x_i = 0$ and
$\frac{1}{N}\sum_i x_i^2 = 1 + O(2)$. Therefore, the log-Jacobian is
given by: 
\begin{equation}
\label{eq:log-jacobian} \log{Z} = N\epsilon^2\left(2-3\alpha\right)
+ O(3),
\end{equation}
Finally, it is easy to calculate $p(\bmath{x} \vert \epsilon)$, the
N-point pdf of $\bmath{x}$ given the $\epsilon$ parameter: 
\begin{equation}
\label{eq:pdfx} p(\bmath{x}\vert \epsilon ) = p(\bmath{\phi} =
\bmath{\phi} (\bmath{x})) Z = p(\bmath{x}\vert 0)e^{l(\bmath{x}
\vert \epsilon)}
\end{equation}
where the probability $p(\bmath{x}\vert 0)$ is given by: 
\begin{equation}
\label{eq:pdfxH0} p(\bmath{x}\vert 0) = \frac {1}{(2\pi
)^{N/2}(\det{\left(\bmath{\xi}\right)} )^{1/2}}e^{-\frac{1}{2}
\bmath{x}\bmath{\xi}^{-1}\bmath{x}^{t}},
\end{equation}
i.e., it is the N-point Gaussian pdf (i.e. $p(\bmath{x}\vert 0)
\equiv p(\bmath{\phi} \equiv \bmath{x} ) )$, and the log-likelihood
$l( \bmath{x} \vert \epsilon)$ reads as: 
\begin{equation}
\label{eq:loglike1} l(\bmath{x}\vert \epsilon ) = \log{Z} -
\frac{1}{2}\left[\phi \bmath{\xi}^{-1} \phi^t - x \bmath{\xi}^{-1}
x^t \right].
\end{equation}
It is easy to show that  $l(\bmath{x} \vert \epsilon)$ is given by:
\begin{equation}
\label{eq:loglike} l(\bmath{x}\vert \epsilon ) \equiv N[\epsilon R -
\epsilon^2Q + O(3)],
\end{equation}
where the functions $R$ and $Q$ are given by:
\begin{equation}
\label{eq:R} R = \frac{1}{N}\bmath{x}\bmath{\xi}^{-1}
\left[\bmath{x}^2 - \mathbb{I}\right]^{t},
\end{equation}
and
\begin{equation}
\label{eq:Q} Q =  -2 + J + \alpha S,
\end{equation}
being $\mathbb{I}$ the unity vector of dimension $N$ and
\begin{equation}
\label{eq:J} J =
\frac{1}{N}\left\{2\bmath{x}\bmath{\xi}^{-1}\left[\bmath{x}\left(\bmath{x}^2-\mathbb{I}\right)\right]^{t}
+
\frac{1}{2}\left[\bmath{x}^2-\mathbb{I}\right]\bmath{\xi}^{-1}\left[\bmath{x}^2-\mathbb{I}\right]^{t}
\right\}
\end{equation}
and
\begin{equation}
\label{eq:S} S = \frac{3}{N}\left\{
\mathbb{I}\bmath{\xi}^{-1}\mathbb{I}^{t} -
\frac{1}{3}\bmath{x}\bmath{\xi}^{-1}{\bmath{x}^3}^{t} \right\}.
\end{equation}

Let us remark that $R$ and $Q$ could be seen as a kind of
generalized third-order and fourth-order moments, respectively (or,
equivalently, in terms of the spherical harmonic coefficients, to
the bispectrum and the trispectrum). As an example, for the
particular case of uncorrelated data (i.e., $\bmath{\xi} \equiv
\bmath{\delta}$), it is straightforward  to show that $R = k_3$ and
$Q = -5/2 + (3\alpha - 2)k_2 + (5/2 - \alpha)k_4$, with $k_n =
\frac{1}{N}\sum_i x_i^n$.

The N-pdf $p(\bmath{x}\vert \epsilon)$ given by
equation~\ref{eq:pdfx} contains all the required information, on the
one hand, to estimate the non-linear coupling parameter $\epsilon$
and, on the other hand, to perform a model selection (Gaussian vs.
non-Gaussian). These two aspects will be studied in
next Section.

Notice that the parameter $\alpha$ (or, equivalently, $b$) cannot be estimated in
this framework: it just appears as an arbitrary constant in the
definition of Q (equation~\ref{eq:Q}): it just controls the relevance
of $S$ in $Q$. If one were interested in obtaining a posterior probability of the data
$\bmath{x}$ given both non-linear parameters ($\epsilon$ and $\alpha$), then it
would be necessary to expand the local non-Gaussian model beyond $0(3)$.
However, we would always find an expression in which the non-linear
parameter controlling the highest order in the expansion, acts as an
arbitrary constant. For that reason, we keep terms up to $0(3)$.
There is a discussion within the field~\citep[e.g.,][]{okamoto02, kogo06, babich05, creminelli07}
about to what extend the cubic term in the description of
the weak non-linear coupling inflationary model (for which, we recall, the local non-Gaussian
model in equation~\ref{eq:physical_model} can be seen as an approximation
at large scales) is really negligible or not with respect to the quadratic contribution.
For the former scenario (negligibility of the cubic term), the particular value of $\alpha$ becomes an irrelevant
issue since, naturally, one will have that $S \ll -2 + J$ (of course, $\alpha$ should
always satisfy the condition given in equation~\ref{eq:condition}). However, for the latter
case different results could be obtained, depending on the specific value for $\alpha$.
In the following, we will discuss parameter estimation and model selection assuming $\alpha \equiv 0$
in equation~\ref{eq:Q}. However, we will explore (analyzing WMAP data and simulations) whether
the condition $S \ll -2 + J$ is naturally satisfied or not.

\section{Statistical analysis}
\label{sec:statistics}

In this Section we aim to address two aspects very much linked one
to the other: the estimation of the parameter defining the local
non-Gaussian model ($\epsilon$, Section~\ref{subsec:estimation}) and
the computation of some heuristic rules to decide whether a given
data set is better described by the local non-Gaussian model rather
than by the standard Gaussian one (Section~\ref{subsec:selection}).

\subsection{Parameter estimation}
\label{subsec:estimation}

The description of the full N-pdf for the non-Gaussian model
proposed in Section~\ref{sec:model} allows one to obtain an optimal
estimation of the non-linear coupling parameter $\epsilon$.

Let us recall that an \emph{optimal estimation} of the
$\epsilon$ parameter is possible, since it would be derived from the full pdf
for the non-Gaussian model, $p(\bmath{x}\vert \epsilon)$. In other
words, we could obtain an unbiased and minimum variance estimator.
This is possible, precisely, for the specific selection of the local
non-Gaussian model in equation~\ref{eq:model}: for other physical
non-Gaussian models it is not always trivial to obtain a full
description of the posterior probability of the data given the
parameters (at least, under realistic observational conditions like
incomplete sky coverage) and shortcuts have to
be taken by defining \emph{pseudo-optimal} estimators that are,
afterwards, validated with simulations.

\subsubsection{Parameter estimation from the log-likelihood}
\label{subsubsec:loglike}

We shall define the optimal estimator for the non-linear
parameter as the value, $\hat{\epsilon}$, that maximizes the
probability of $\bmath{x}$ given $\epsilon$. From
equation~\ref{eq:pdfx}, it is obvious that maximizing this
probability is equivalent to maximize $l(\bmath{x}\vert \epsilon)$
in equation~\ref{eq:loglike}. By derivation one obtains: 
\begin{equation}
\label{eq:est_loglike} \hat{\epsilon} = \frac{R}{2Q}.
\end{equation}
We can also estimate the error associated to $\hat{\epsilon}$ from
the Fisher matrix $F_{\hat{\epsilon}} \equiv -\frac{{\rm d}^2l}{{\rm
d}\epsilon^2} = 2NQ$: 
\begin{equation}
\label{eq:error_epsilon} \sigma_{\hat{\epsilon}} =
F_{\hat{\epsilon}}^{-1/2} = (2NQ)^{-1/2}.
\end{equation}
Notice that the error on the estimation of the parameter $\epsilon$ is
constant, up to the order considered in equation~\ref{eq:model}.

\subsubsection{Bayesian parameter estimation}
\label{subsubsec:bayes}

Within the Bayesian framework we can include any \emph{a priori}
information that we might have in relation to the $\epsilon$
parameter. In particular, following Bayes' theorem, the probability
of $\epsilon$ given the data $\bmath{x}$ read as:
\begin{equation}
\label{eq:bayes} p(\epsilon \vert \bmath{x})\propto p(\bmath{x}\vert
\epsilon )p(\epsilon ),
\end{equation}
\noindent where $p(\epsilon )$ is the \emph{prior} probability
function for the parameter $\epsilon$. Of course, for the case of
the local non-Gaussian model proposed in this work, there is not a
clear physical motivation to choose a particular prior.

Let us however explore, as an exercise, two simple scenarios, which
could be useful for more general purposes. We consider first a
\emph{uniform prior} given by:
\begin{equation}
\label{eq:uniform_prior} p(\epsilon) \left \{ \begin{array}{ll}
\frac{1}{\epsilon_{\rm M} - \epsilon_{\rm m}} & \mathrm{if} \ \ \epsilon \in \left[ \epsilon_{\rm m}, \epsilon_{\rm M} \right] \\
\\
0 & {\rm otherwise} \\
\end{array} \right. ,
\end{equation}
where, obviously, the range allowed to $\epsilon$ is such that
$\epsilon \ll 1$ for any $\epsilon \in \left[ \epsilon_{\rm m},
\epsilon_{\rm M} \right]$. For this particular case, it is trivial
to show that the Bayesian estimation for the non-linear coupling
parameter ($\bar{\epsilon}$) is equivalent to the one obtained
via the maximum-likelihood estimation  (i.e, $\bar{\epsilon} \equiv
\hat{\epsilon}$) if $\hat{\epsilon} \in \left[ \epsilon_{\rm m},
\epsilon_{\rm M} \right]$.

The second case we want to address corresponds to a Gaussian prior
$p(\epsilon)$, described by a most probable value $\epsilon_*$ and a
dispersion $\sigma_*$:
\begin{equation}
\label{eq:gaussian_prior} p(\epsilon) =
\frac{1}{\sqrt{2\pi}\sigma_*} e^{-\frac{\left(\epsilon -
\epsilon_*\right)^2}{2{\sigma_*}^2}}.
\end{equation}
By deriving the posterior probability, it is trivial to obtain the
Bayesian estimation for the non-linear coupling parameter
($\bar{\epsilon}$):
\begin{equation}
\label{eq:est_bayes} \bar{\epsilon} = \frac{N {\sigma_*}^2 R +
\epsilon_*}{2 N {\sigma_*}^2 Q + 1}.
\end{equation}
For the particular case of $\sigma_* \rightarrow 0$, i.e. a very
strong prior for $\epsilon$, peaked around $\epsilon_*$, one
trivially obtain $\bar{\epsilon}  \equiv \epsilon_*$, that is, the
prior dominates Bayesian estimation, leading to a most probable
value for $\epsilon$ equal to the maximum value for the prior. Also
trivially one finds that, for a non-informative scenario (i.e., $\sigma_*
\rightarrow \infty$), $\bar{\epsilon}  \equiv \hat{\epsilon} = R/2Q$.

\subsection{Model selection}
\label{subsec:selection}

In this subsection we aim to calculate under which conditions
(according to different model selection criteria) a given
observation $\bmath{x}$ is better described by a local non-Gaussian
model as the one described by equation~\ref{eq:model} with $\vert
\epsilon \vert
> 0$ (hereinafter $H_1$) rather than by a Gaussian random field
described just in term of the N-point correlation function
$\bmath{\xi}$ (hereinafter $H_0$).

Some of the model selection
approaches investigated in this paper have been
previously applied to different astronomical/cosmological problems
For instance,~\citep{szydlowski06a,szydlowski06b,borowiec06} applied
the Akaike and the Bayesian information criteria (AIC and BIC, recpectively)
to study whether astronomical data sets favored simplest models for the
accelerating universe against more complex ones. This issue was also
addressed by~\cite{davis07} by analyzing the ESSENCE supernova survey data,
and appliying Bayesian evidence (BE) in addition to the AIC and the BIC
approaches, to study. These three model selection criteria (AIC, BIC and
BE) were also applied to study the impact of non-standard physical models
on the Friedmann equations~\citep{szydlowski08}.
\cite{liddle04} used the AIC and the BIC techniques to study, on the
one hand, whether WMAP
1-year data preferred a spatially flat cosmology versus a closed one, and, on the
other hand the significance of the running spectral index detected on
this WMAP data release. In a posterior work~\citep{liddle07}, BE was added
to the AIC and the BIC approaches to study the suitability of different cosmological
models to the WMAP 3-year data.

Bayesian evidence (BE) is, probably, the model selection criterion
that has attracted a greater interest from cosmologists during the
past years. In addition to the works mentioned above (where it was
compared with other model selection criteria), it has been also
applied to several problems where competing
cosmological models were explored. Some of these applications are
the following:~\cite{mukherjee06} followed a BE approach to study
cosmological models with different matter power spectra and dark
energy evolution models;~\cite{liddle06} studied different
dark energy evolving scenarios;~\cite{bridges06,bridges07b}
performed a model selection on the matter power spectrum from the BE
analysis of the WMAP;~\cite{bridges07a,bridges08} followed a similar
approach for analyzing the WMAP compatibility with anisotropic
Bianchi VII$_h$ models;~\cite{cruz07b,cruz08} used it
to decide whether the WMAP cold spot was compatible
or not with predictions from non-standard models like the
cosmological defects;~\cite{mukherjee08} studied the Planck ability
to discriminate between several re-ionization
models; the BE criterion was also
applied~\citep{carvalho08,feroz08b} to the problem of compact source
detection on microwave data; and more recently,~\cite{feroz08a}
investigated different properties of the constrained minimal
supersymmetric model (mSUGRA), using WMAP data.

\subsubsection{The Akaike information criterion (AIC)}
\label{subsubsec:aic}

The Akaike information criterion \citep[AIC,][]{akaike73} provides with
a selection index to decide among competing hypothesis, being the
model associated to the lowest index the most favored one. The
Akaike index corresponding to a given model or hypotheses $H_i$
defined by $p$ parameters and with a maximum value for the
log-likelihood of $\hat{l}$ is given by: 
\begin{equation}
\mathrm{AIC}(H_i) = 2\left( p - \hat{l} \right).
\end{equation}

From equation~\ref{eq:loglike}, one can find that $\mathrm{AIC}(H_1)
= 2\left( 1 - N\frac{R^2}{4Q}\right)$, whereas, trivially,
$\mathrm{AIC}(H_0) = 0$. Therefore, according to the AIC, the
decision rule reads as: 
\begin{equation}
\label{eq:aic_rule} \mathrm{AIC}:  \left \{ \begin{array}{ll}
H_0 & \mathrm{if} \ \ \frac{R^2}{Q}\leq \frac{4}{N} \\
\\
H_1 & \mathrm{if} \ \ \frac{R^2}{Q} > \frac{4}{N} \\
\end{array} \right.
\end{equation}

\subsubsection{Bayesian information criterium (BIC)}

\label{subsubsec:bic} This asymptotic bayesian criterion introduced by
\cite{schwarz78} is prior independent. It is based on the BIC
function, that provides a measurement of the goodness-of-fit of the
model to the data, taking into account the number of parameters
defining the model as well as the amount of data ($N$): 
\begin{equation}
\label{eq:bic} \mathrm{BIC}(H_i) = \left(-2\hat{l} + p\ln{N}
\right),
\end{equation}

\noindent where $p$ is the number of parameters defining the data
and $\hat{l}$ is the maximum value for the log-likelihood. As for
AIC, BIC provides a ranging index for competing hypothesis, where
the one with the lower BIC value is the most favored one. From
equation~\ref{eq:loglike}, one can find that (for our specific
problem) $\mathrm{BIC}(H_1) = -N\frac{R^2}{2Q} +
\ln{\left(N\right)}$ and $\mathrm{BIC}(H_0) = 0$. Therefore,
according to the BIC, the decision rule reads as: 
\begin{equation}
\label{eq:bic_rule} \mathrm{BIC}:  \left \{ \begin{array}{ll}
H_0 & \mathrm{if} \ \ \frac{R^2}{Q}\leq \frac{2}{N}\ln{N} \\
\\
H_1 & \mathrm{if} \ \ \frac{R^2}{Q} > \frac{2}{N}\ln{N} \\
\end{array} \right.
\end{equation}

\subsubsection{Minimum description length (MDL)}

\label{subsubsec:mdl} MDL \citep[e.g.][]{rissanen01} is an inference
approach mostly developed during the 80s and 90s, based on the key
idea that the more \emph{regular} a given data set is, the higher is
the \emph{compression} degree to which we can code the data, and,
therefore, the more we can \emph{learn} on the properties of the
data. Among many statistical applications, MDL is used to select
between competing models describing the data, selecting the one
allowing for a higher compression degree (which can be seen as an
alternative formulation of the Occam's Razor).

For our particular case described by equation \ref{eq:model},
the MDL measurement of compression is given by: 
\begin{equation}
\label{eq:mdl} \mathrm{MDL}(\epsilon ) = -\hat{l} + \frac{1}{2} \ln{\left(\frac{N}{2\pi
}\right)} + \ln{\int_{\epsilon \in \Omega}{\rm d}\epsilon
[\det{F_{\hat{\epsilon}}}]^{1/2}},
\end{equation}
where $F_{\hat{\epsilon}}$ is the Fisher matrix of $\hat{\epsilon}$.
Therefore, taking into account equations~\ref{eq:pdfx} and~\ref{eq:loglike},
one can easily compute $\mathrm{MDL}(H_1)$ and $\mathrm{MDL}(H_0)$
providing a decision rule that reads as: 
\begin{equation}
\label{eq:mdl_rule} \mathrm{MDL}:  \left \{ \begin{array}{ll}
H_0 & \mathrm{if} \ \ \frac{R^2}{Q}\leq \frac{4}{N} \ln{\left( {N\Omega} \sqrt{\frac{Q}{\pi}} \right)} \\
\\
H_1 & \mathrm{if} \ \ \frac{R^2}{Q} > \frac{4}{N} \ln{\left( {N\Omega} \sqrt{\frac{Q}{\pi}} \right)} \\
\end{array} \right.,
\end{equation}
where $\Omega$ is the interval where $\epsilon$ is defined.

\subsubsection{Generalized likelihood ratio test (GLRT)}
\label{subsubsec:glrt}

Generalized likelihood ratio test (GLRT) is one of the most common
approaches in model selection and its particular application to
solve astronomical/cosmological problems has been very extensive.

The criterion established by the GLRT to accept the alternative
hypothesis $H_1$ against $H_0$ is given by $p(\bmath{x} \vert
\hat{\epsilon}, H_1) > e^\nu p(\bmath{x} \vert 0, H_0)$ or, equivalently
by: 
\begin{equation}
\label{eq:maxloglike} \hat{l} \equiv l(\hat{\epsilon})=
N\frac{R^2}{4Q}
> \nu,
\end{equation}
where $\nu$ is an arbitrary value indicating the \emph{strength} in
choosing $H_1$ instead of $H_0$. Therefore, according to the GLRT,
the decision rule reads as:
\begin{equation}
\label{eq:glrt_rule} \mathrm{GLRT}:  \left \{ \begin{array}{ll}
H_0 & \mathrm{if} \ \ \frac{R^2}{Q} \leq \frac{4}{N}\nu \\
\\
H_1 & \mathrm{if} \ \ \frac{R^2}{Q} > \frac{4}{N}\nu \\
\end{array} \right.
\end{equation}
Notice that the case $\nu \equiv 1$ provides the same decision rule
as the AIC (equation~\ref{eq:aic_rule}), and that $\nu \equiv \ln{\sqrt{N}}$ corresponds
to the BIC case (equation~\ref{eq:bic_rule}).

\subsubsection{Bayesian evidence (BE)}

BE is defined as the average likelihood of the model $H_i$ in the
prior $p(\epsilon)$:
\begin{equation}
\label{eq:evidence} E_{H_i}(\bmath{x}) =  \int {\rm d}\epsilon\,
p(\epsilon, H_i )p(\bmath{x}\vert \epsilon, H_i ),
\end{equation}
where $p(\bmath{x}\vert \epsilon, H_i )$ is given by
equation~\ref{eq:pdfx}. Model selection in terms of the BE grounds
on the \emph{Bayes' factor}, $B_{10}$: 
\begin{equation}
B_{10}(\bmath{x}) = \frac{E_{H_1}(\bmath{x})}{E_{H_0}(\bmath{x})}.
\end{equation}
BE framework provides a rule to quantify how
strong the decision is. In the literature it is commonly accepted
the \emph{Jeffreys' scale} \citep{jeffreys61}, that provides a recipe
in terms of the logarithmic Bayes' factor. Roughly speaking, it is
commonly said that the
evidence for $H_1$ against $H_0$ is not significant if $0 \leq
\ln{B_{10}(\bmath{x})} <1$, mild if $1 \leq \ln{B_{10}(\bmath{x})}
\leq 3$ and strong if $\ln{B_{10}(\bmath{x})}  > 3$.

As it was already discussed, the alternative hypothesis $H_1$
representing the local non-Gaussian model (equation~\ref{eq:model})
does not offer any physical motivation for a particular prior. Even
thus, we study here the two particular cases already mentioned in
subsection~\ref{subsubsec:bayes}: the Gaussian
(equation~\ref{eq:gaussian_prior}) and the uniform
(equation~\ref{eq:uniform_prior}) priors.

For the former, it is straightforward to prove that:
\begin{equation}
\label{eq:b10_gaussian} B_{10} =
\frac{\sigma_{\hat{\epsilon}}}{\sqrt{\sigma_{\hat{\epsilon}}^2 +
\sigma_*^2}} e^{ -\frac{\epsilon_*^2}{2\sigma_*^2} +
\frac{\left(\epsilon_* \sigma_{\hat{\epsilon}}^2 + \hat{\epsilon}
\sigma_*^2\right)^2}{2 \sigma_*^2 \sigma_{\hat{\epsilon}}^2
\left(\sigma_{\hat{\epsilon}}^2 + \sigma_*^2\right)}},
\end{equation}
whereas for the latter, one can obtain:
\begin{equation}
\label{eq:b10_uniform} B_{10} = \sqrt{\frac{\pi}{2}}
\frac{\sigma_{\hat{\epsilon}}}{\epsilon_M - \epsilon_m}
e^\frac{\hat{\epsilon}^2}{2\sigma_{\hat{\epsilon}}^2}
\left[\mathrm{erf}\left(\frac{\epsilon_M -
\hat{\epsilon}}{\sqrt{2}\sigma_{\hat{\epsilon}}}\right) +
\mathrm{erf}\left(\frac{\hat{\epsilon} -
\epsilon_m}{\sqrt{2}\sigma_{\hat{\epsilon}}}\right)\right],
\end{equation}
where $\hat{\epsilon}$ is the maximum-likelihood estimation for the
non-linear parameter $\epsilon$ (equation~\ref{eq:est_loglike}) and
$\sigma_{\hat{\epsilon}}$ is the error on this estimation
(equation~\ref{eq:error_epsilon}). Finally, notice that for the
particular cases of $\sigma_* \gg \sigma_{\hat{\epsilon}}$ or
$\epsilon_M - \epsilon_m \gg \sigma_{\hat{\epsilon}}$ in equations~\ref{eq:b10_gaussian}
and~\ref{eq:b10_uniform}, respectively (i.e., a broad prior),
one obtains:
\begin{equation}
B_{10} \simeq
\frac{\sigma_{\hat{\epsilon}}}{\gamma}
e^\frac{\hat{\epsilon}^2}{2\sigma_{\hat{\epsilon}}^2},
\end{equation}
where $\gamma=\sigma_*$ (for the Gaussian prior)
or $\gamma=\left(\epsilon_M - \epsilon_m\right)/\sqrt{2\pi}$ (for the uniform prior).

\section{Application to WMAP simulations}
\label{sec:simulations}

In this Section we aim to explore the performance of the parameter
estimators and the model selection criteria described in the
previous Section. We apply them to CMB simulations of the WMAP
5-year data at \nside=32 HEALPix~\citep{gorski05} resolution ($\approx 2^\circ$).

\begin{figure}
\includegraphics[angle=270,width=8cm,keepaspectratio]{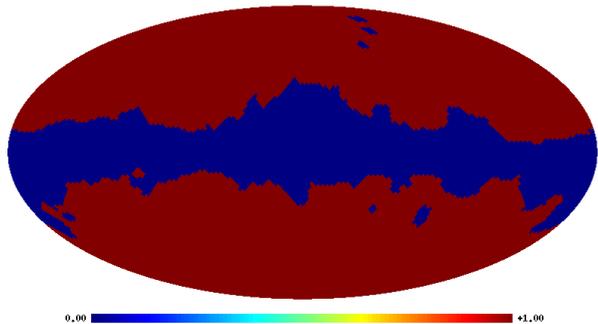}
\caption{\label{fig:mask}Mask at \nside=32 HEALPix resolution used
in this work. It corresponds to the WMAP KQ75 mask, although the
point source masking has not been considered, since the point
like-emission due to extragalactic sources is negligible at the
considered resolution. At this pixel resolution, the mask keeps around
69\% of the sky.}
\end{figure}

The procedure to generate a CMB Gaussian simulation ---$\left(\Delta
T\right)_G$ in equation~\ref{eq:physical_model}--- is as follows.
First, using the $C_\ell$ obtained with the cosmological parameters
provided by the best-fit to WMAP data alone~\citep[Table 6
in][]{hinshaw08}, we simulate WMAP observations (taking into account the corresponding
beam window functions) for the Q1, Q2, V1,
V2, W1, W2, W3, W4 difference assemblies at \nside=512 HEALPix
resolution. We obtain a single co-added CMB map through a
noise-weighted linear combination of the eight maps (from Q1 to W4).
Weights are proportional to the inverse mean noise variance. They are
independent on the position (i.e., they are uniform
across the sky for a given difference assembly) and they are
normalized to unity. Notice that we do not add a random noise
realization to each map, since we have checked that noise plays a
negligible role at the angular resolution in which we are interested in
($\approx 2^\circ)$. However, we perform the linear combination of
the difference assembly maps following the procedure described
above, since it will be the same process that we will follow with
the WMAP data.\footnote{Co-added WMAP 5-year data is made in this
way to produce a final map with a noise level smaller than, for instance,
the one that could be achieved just by averaging the 8 difference assembly maps,
assuring better a negligible noise contribution to the final map
at resolution of $\approx 2^\circ$.} Afterwards, the co-added map at
\nside=512 is degraded down to the final resolution of \nside=32.
Finally, a mask representing a sky coverage like the one allowed by
the WMAP KQ75 mask~\citep{gold08} is adopted. At \nside=32 the
mask keeps around
69\% of the sky (notice that we do not consider the masking due to
point sources, since at this resolution the contribution from individual
extragalactic point sources is negligible, see
figure~\ref{fig:mask}). Let us remark that observational constraints
like incomplete sky coverage can be easily taken into account by the
local non-Gaussian model proposed in this work, since it is
naturally defined in pixel space.

We have used 500,000 simulations of $\left(\Delta T\right)_G$, generated as
described above, to estimate the correlation matrix $\bmath{\xi}$
accounting for the Gaussian CMB cross-correlations. We have
computed this large number of simulations to assure an accurate description
of the CMB Gaussian temperature fluctuations. Additional 1,000 simulations were
also generated to carry out a statistical analysis on the performance
of the different parameters estimators and model selection criteria.

Each of these 1,000 $\left(\Delta T\right)_G$ simulations are
transformed into $\bmath{x}$ (following
equations~\ref{eq:physical_model} and~\ref{eq:equivalences}) to
study the response of the statistical tools as a function of the
non-linear $\epsilon$ parameter defining the local non-Gaussian
model proposed in equation~\ref{eq:model}.

\subsection{Parameter estimation}
\label{subsec:sims_estimation} Let us first consider the
estimation made via maximum-likelihood estimation
(Subsection~\ref{subsubsec:loglike}). As it was mentioned above,
we have generated 1,000 non-Gaussian WMAP-like observations
according to equation~\ref{eq:model} for a range of values of
$\epsilon$, in particular, we have considered $\epsilon \in \left[0,
0.035\right]$, or equivalently, in terms of the most common
coupling f$_\nl$ parameter (equation~\ref{eq:fnl}), we explore
f$_\nl \in \left[0,500\right]$. We only explore positive values of the
non-linear parameter, since the response of the proposed methodology
does not depend on the sign of $\epsilon$.

\begin{figure*}
\includegraphics[width=5cm,keepaspectratio]{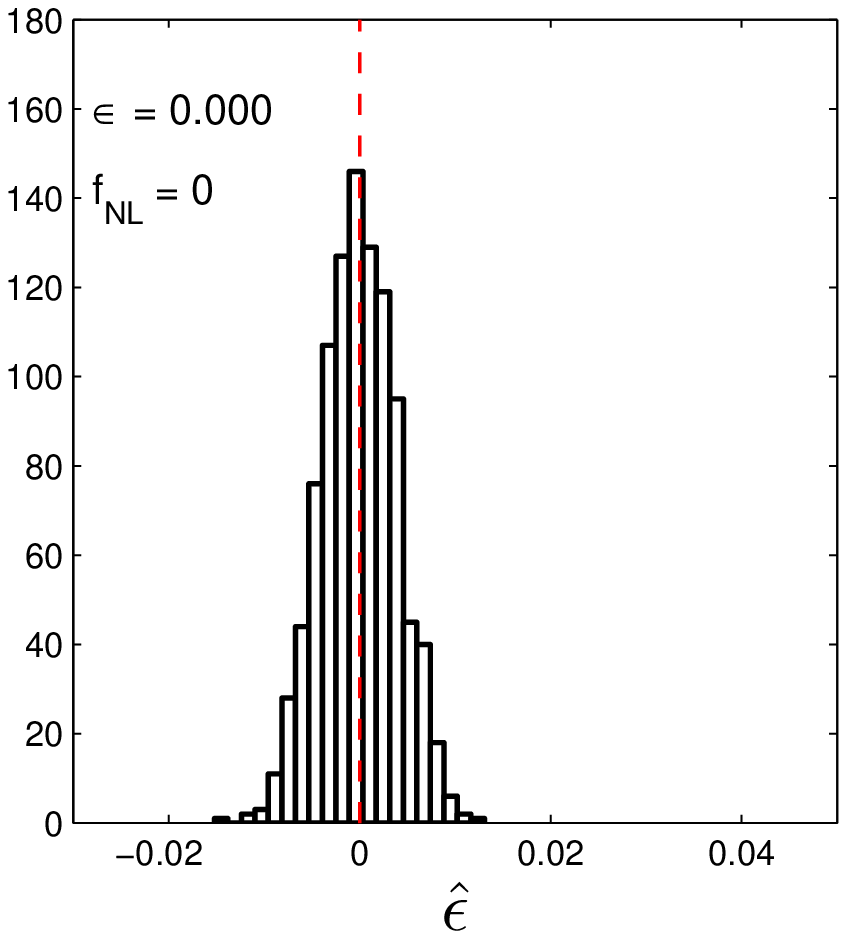}
\includegraphics[width=5cm,keepaspectratio]{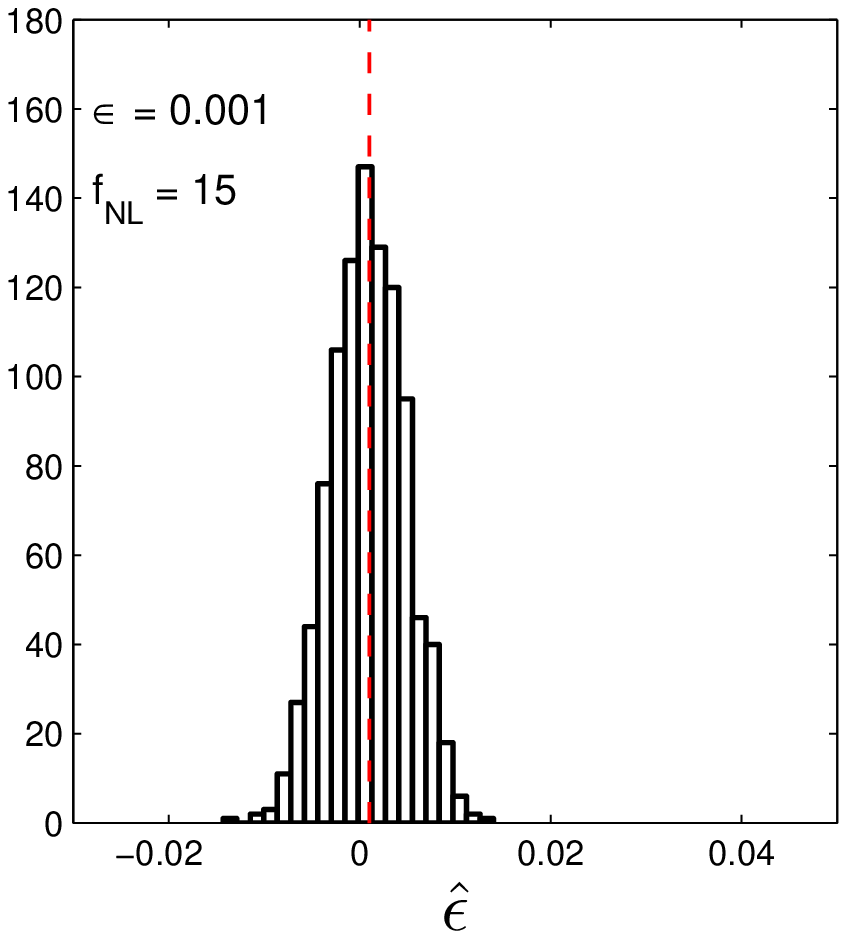}
\includegraphics[width=5cm,keepaspectratio]{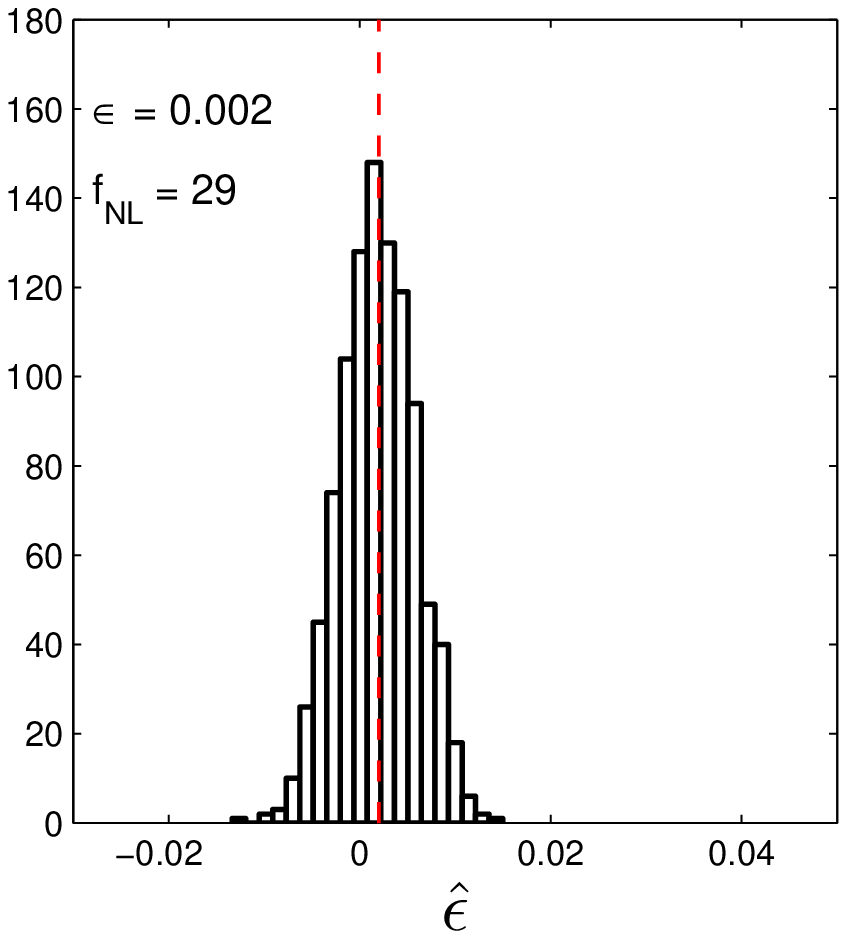}
\includegraphics[width=5cm,keepaspectratio]{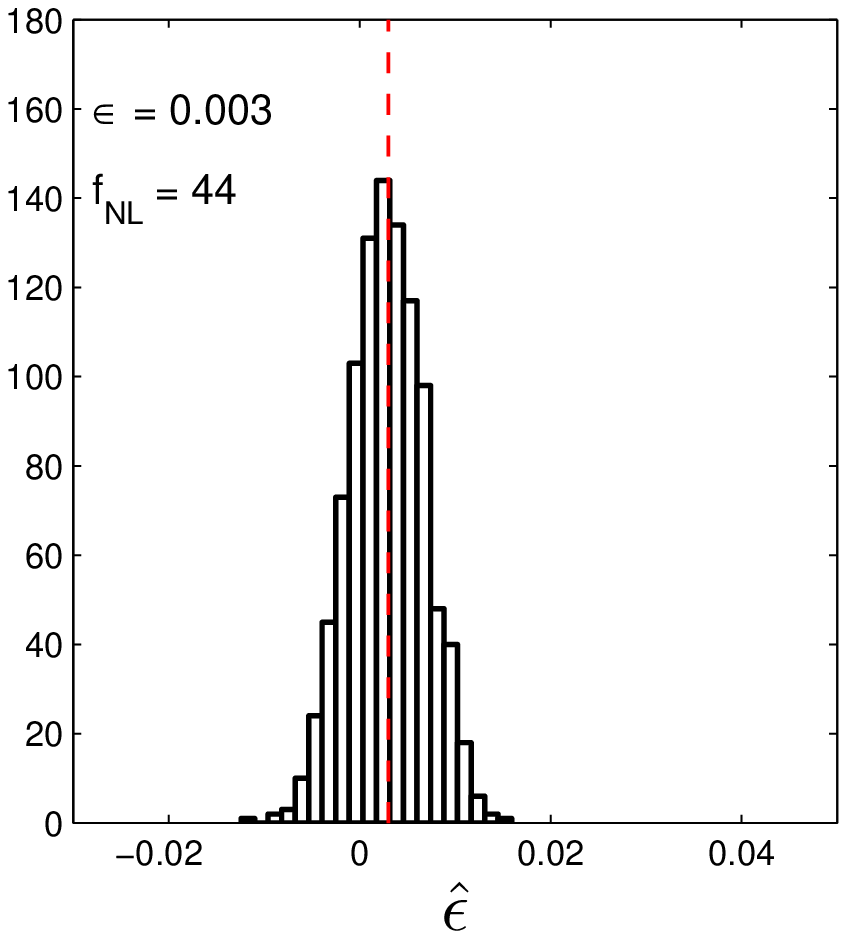}
\includegraphics[width=5cm,keepaspectratio]{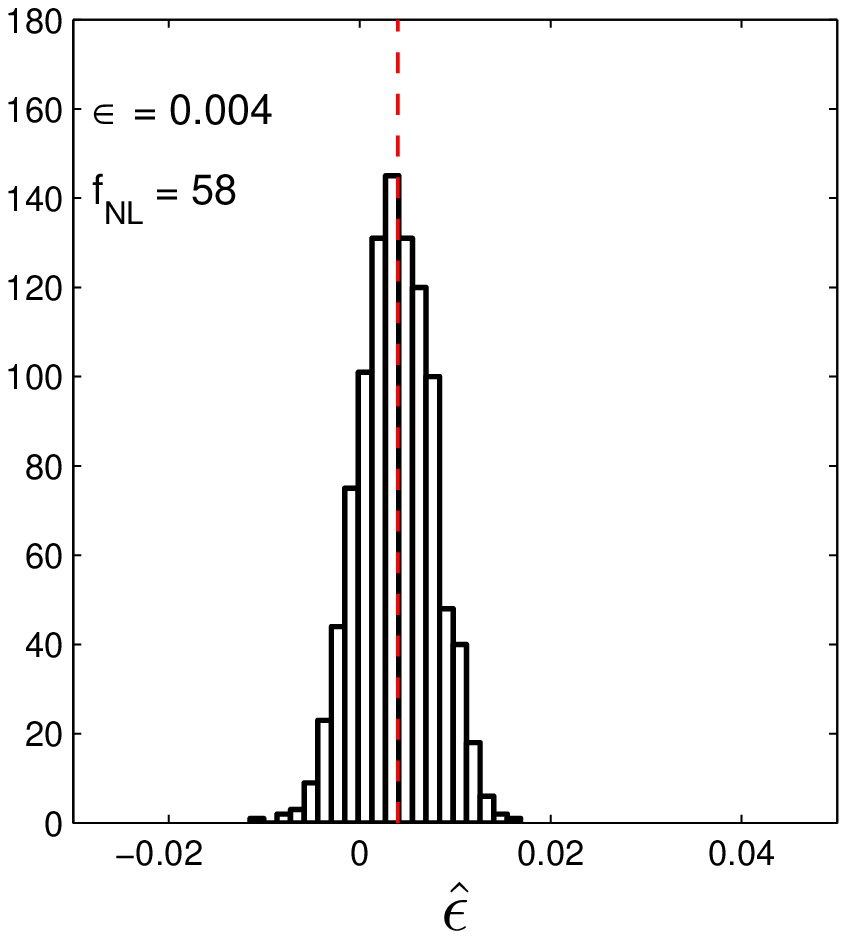}
\includegraphics[width=5cm,keepaspectratio]{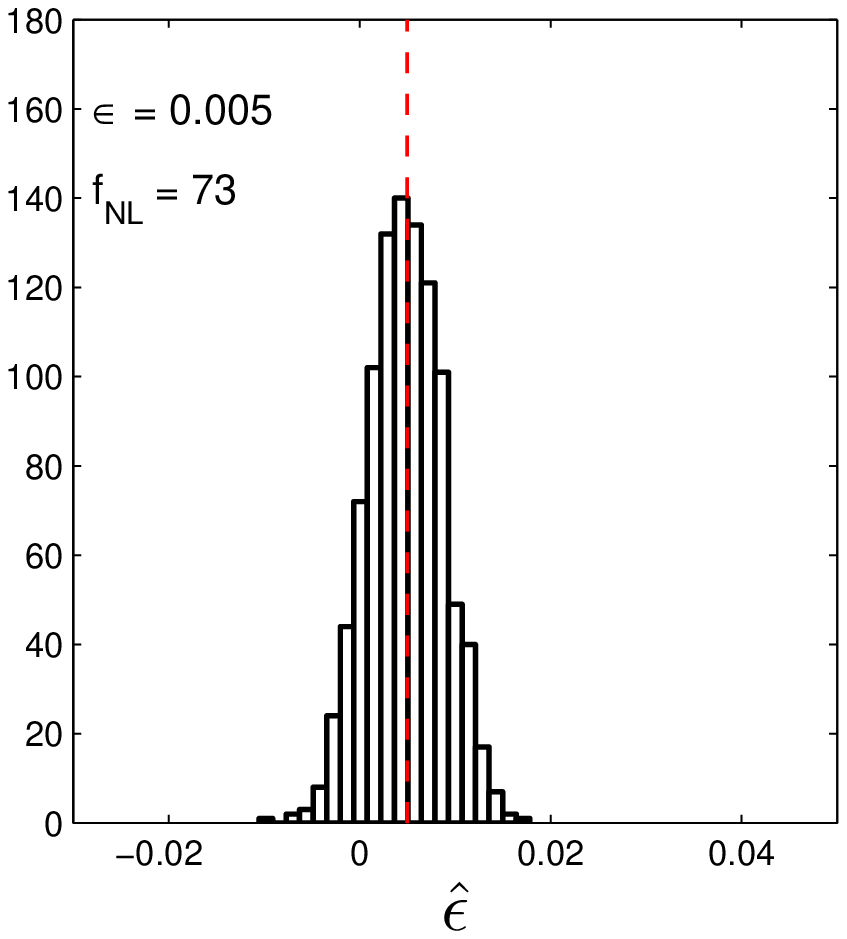}
\includegraphics[width=5cm,keepaspectratio]{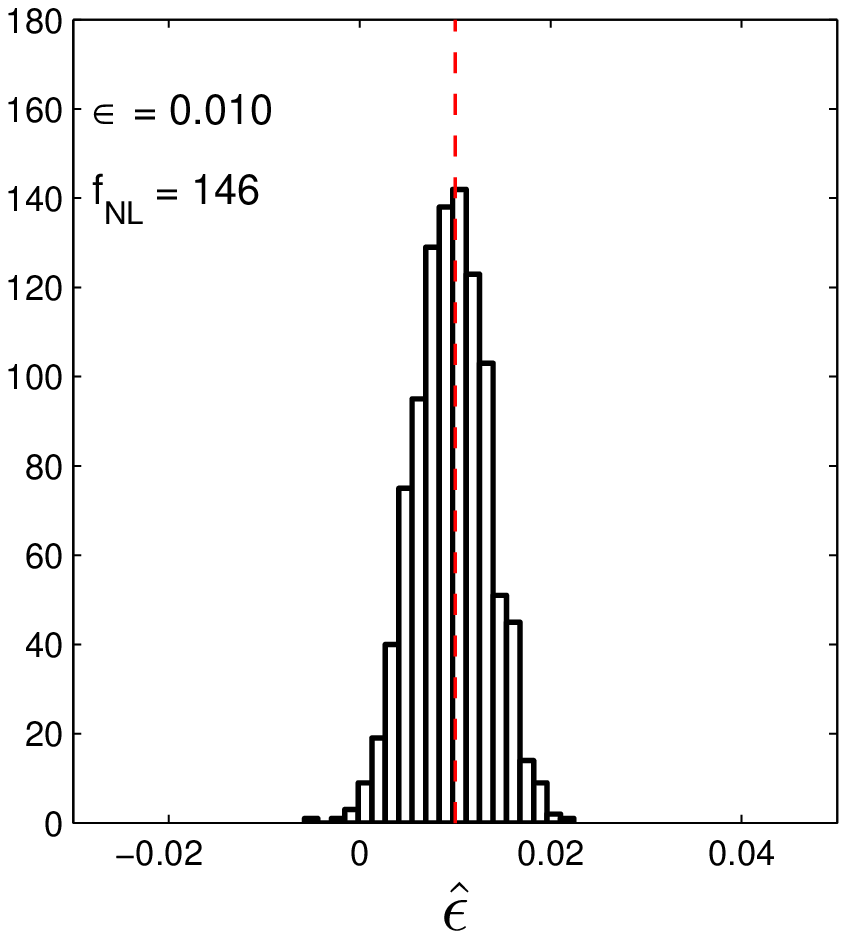}
\includegraphics[width=5cm,keepaspectratio]{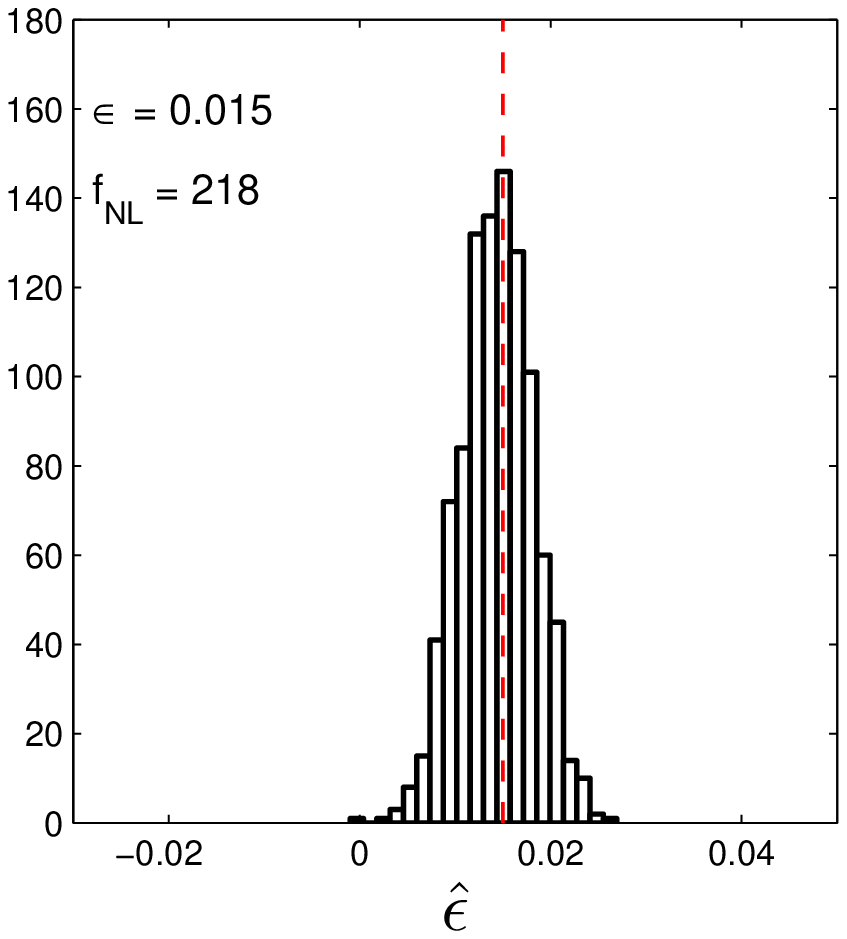}
\includegraphics[width=5cm,keepaspectratio]{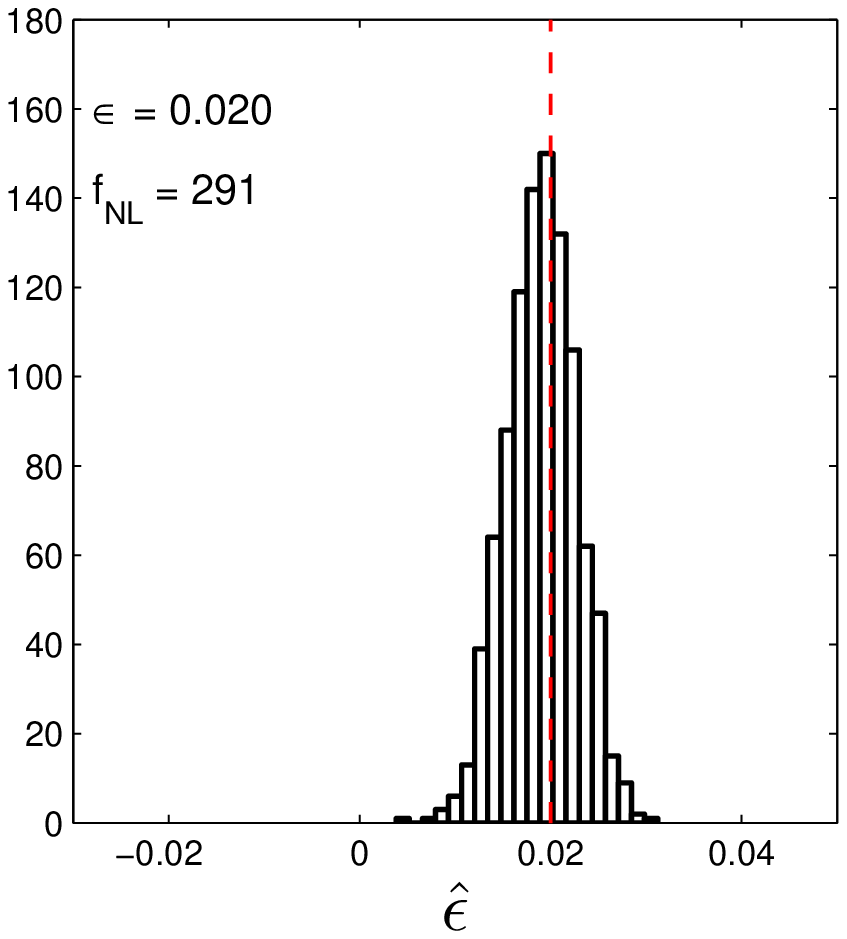}
\includegraphics[width=5cm,keepaspectratio]{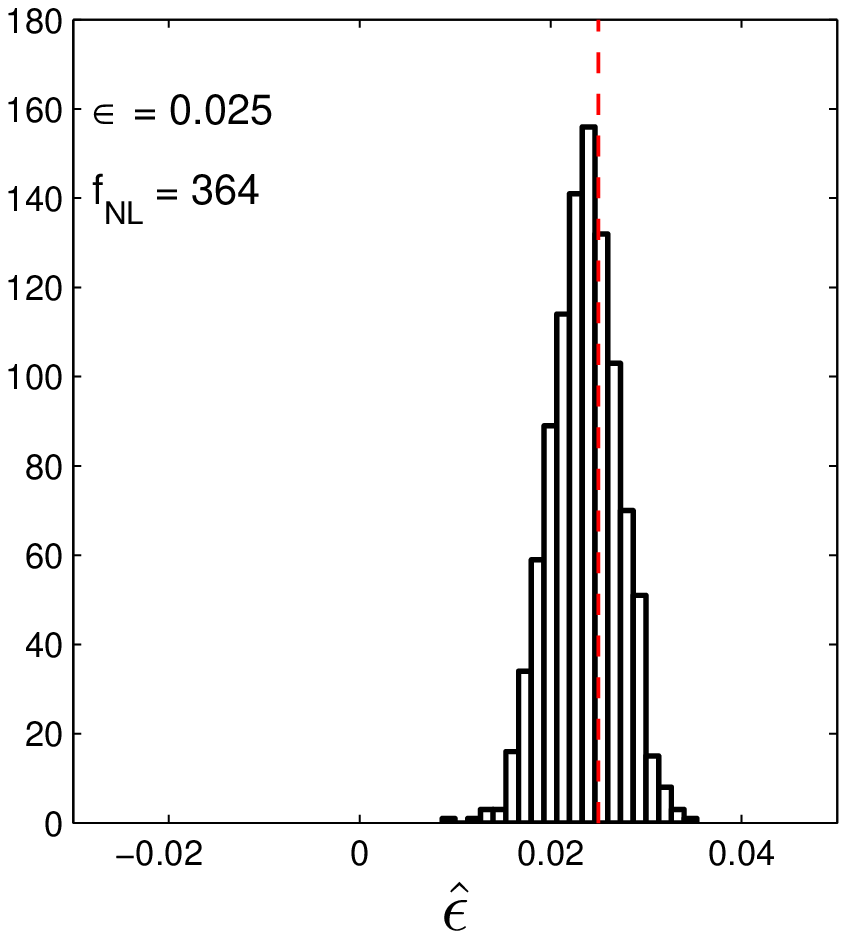}
\includegraphics[width=5cm,keepaspectratio]{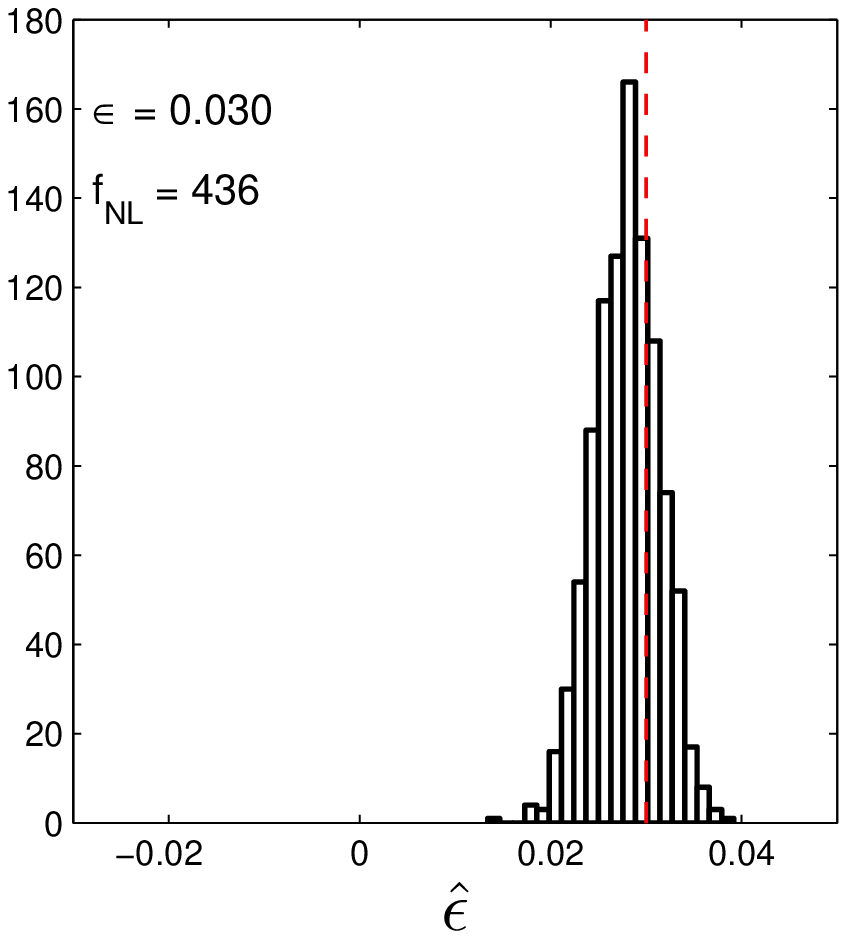}
\includegraphics[width=5cm,keepaspectratio]{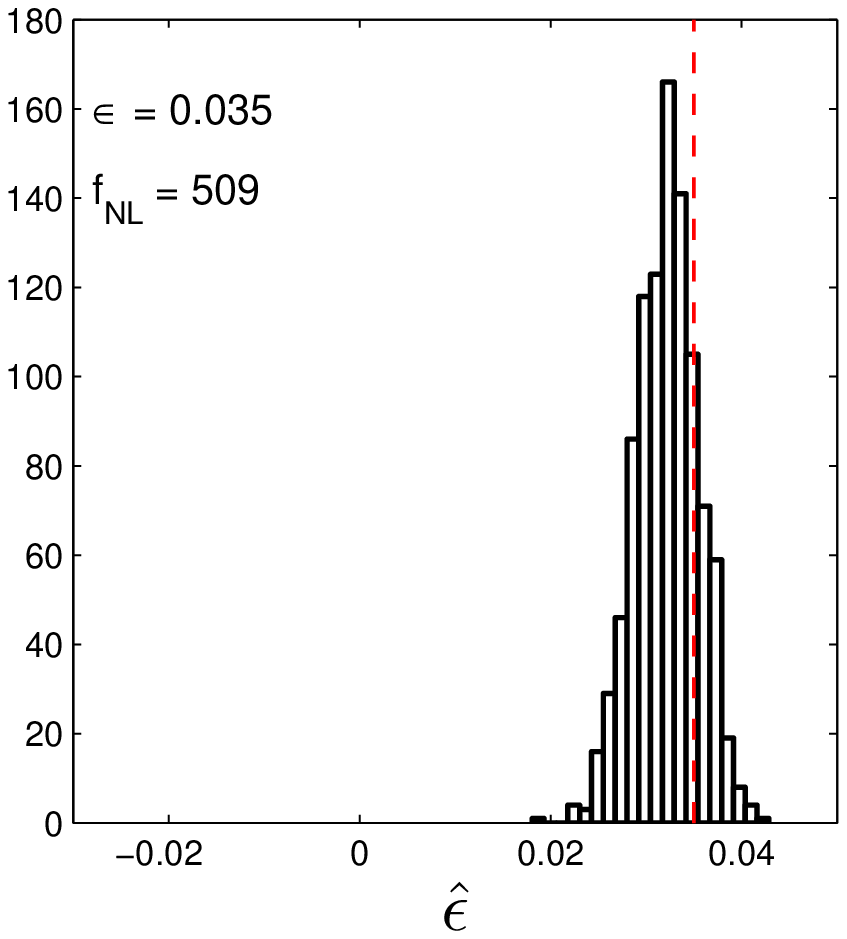}
\caption{\label{fig:fnl_sims}Distributions for the maximum-likelihood estimations
of the non-linear parameter $\hat{\epsilon}$ obtained by analyzing 1,000
simulations, according to the local non-Gaussian model given in equation~\ref{eq:model}.
Several values of $\epsilon$, or, equivalently of the coupling f$_\nl$ parameters
(both numbers are written in each panel) are explored. Vertical dashed lines indicate
the value of the $\epsilon$ used to generate each set of local non-Gaussian simulations.}
\end{figure*}

In figure~\ref{fig:fnl_sims} we present the $\hat{\epsilon}$
distributions obtained from the analysis of local non-Gaussian
simulations. From left to right and from top to bottom, the
panels show the cases: $\epsilon$ = 0, 0.001, 0.002, 0.003, 0.004, 0.005, 0.010, 0.015,
0.020, 0.025, 0.030 and 0.035.
Notice that, as expected, the parameter estimation is unbiased for \emph{reasonable} values
of the $\epsilon$ parameter: all the distributions are peaked around the value of
$\epsilon$ used to generate the simulations. Only for $\epsilon > 0.025$ (or,
equivalently, f$_\nl \gtrapprox 350$) a bias starts to appear. This effect comes from
the fact that these values of $\epsilon$ are not small enough to assure a local non-Gaussian
model as the one described by equation~\ref{eq:model}, i.e., a non-Gaussian model that is a
local perturbation of the underlaying CMB Gaussian signal. Also as expected (see equation~\ref{eq:error_epsilon}),
the width of these distributions does not depend on the particular value of $\epsilon$, if, once more,
$\epsilon$ is small enough to assure a proper expansion for the local non-Gaussian model. In
this regime, we obtain, on average, $\sigma_{\hat{\epsilon}} \approx 0.004$, or,
equivalently, $\sigma_{\hat{{\rm f}_\nl}} \approx 60$. Again, for $\epsilon > 0.025$
the width of the distributions starts to be slightly smaller, indicating an
inadequate value of the non-linear parameter. These two effects (bias of the maximum-likelihood
parameter estimation and dependence on the error on the parameter estimation) provide
a natural range (at least for a pixel resolution of $\approx 2^\circ$) where the non-linear parameter is allowed to take
values: $\epsilon \in \left[-0.025, 0.025\right]$.

This allowed range for $\epsilon$ can be seen as a \emph{natural prior}
$p\left(\epsilon\right)$, that could be used for performing a parameter estimation within
a Bayesian framework. Obviously (as discussed in Subsection~\ref{subsubsec:bayes}), the Bayesian
estimation made with this uniform prior produces the same estimations for $\epsilon$ already
reported from the maximum-likelihood.

Finally, we have also investigated whether the value of $S$ in equation~\ref{eq:Q}
is negligible as compared to $-2 + J$, which, as discussed in Section~\ref{sec:model},
would lead to a situation where the choice of a particular value for the non-linear
parameter $\alpha$ becomes an irrelevant problem. We found that, actually,  this is not the case:
on average, $\vert S/ (-2+J) \vert \approx 0.7$. This implies that, first,
different values of $\alpha$ could provide different results, not only to the ones
presented in this work, but also for other works in the
literature (where, we recall, it is assumed $\alpha \equiv 0$).
In particular, it is trivial to show that, values of $\alpha \lesssim 0.15$ would
affect the determination of the coupling parameter $\epsilon$ in $\approx 10\%$.
Second, as it has been discussed
by some authors~\citep{okamoto02,kogo06}, it would indicate that, for the
weak non-linear coupling inflationary model, the role played by cubic terms could be
non negligible as compared to quadratic contributions, since, to some extent, they
will contribute to the full trispectrum (as one can notice from equation~\ref{eq:Q}, where
$\alpha$ governs the role played by $S$). This results could be important when describing
more complete non-local non-Guassian model, since it would indicate the need of including
physical effects up to third order.

\subsection{Model selection}

\label{subsec:sims_selection}
We discuss which is the performance of the different model
selection criteria presented in Subsection~\ref{subsec:selection}.
For the particular case of BE, we assume the natural prior $p\left(\epsilon\right)$
found in the previous Subsection: $\epsilon \in \left[-0.025, 0.025\right]$.
As for the case of the parameter estimation previously discussed, we
have considered a set of local non-Gaussian models given by different
values of the non-linear parameter.

\begin{figure*}
\includegraphics[width=3.1cm,keepaspectratio]{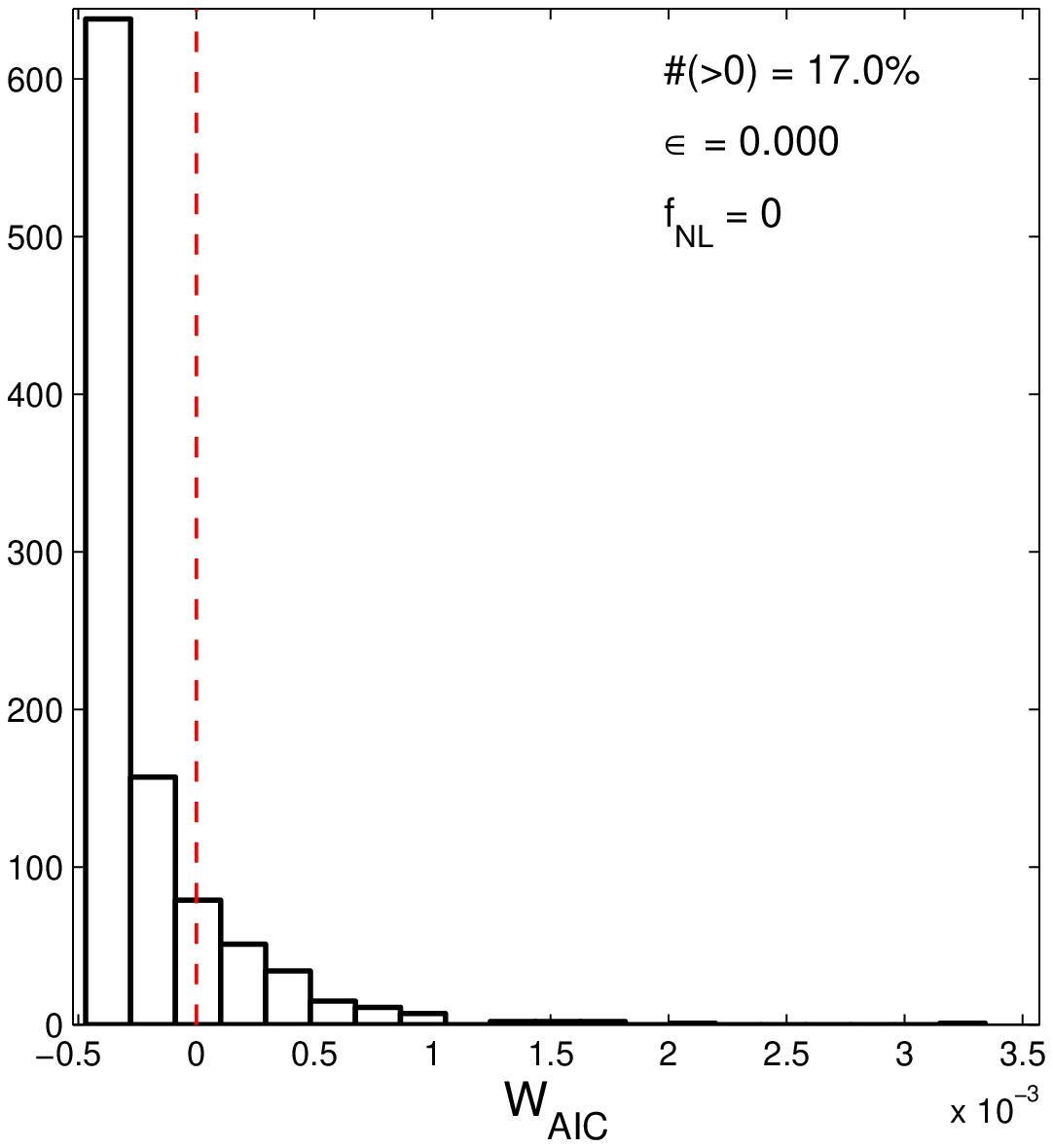}
\includegraphics[width=3.1cm,keepaspectratio]{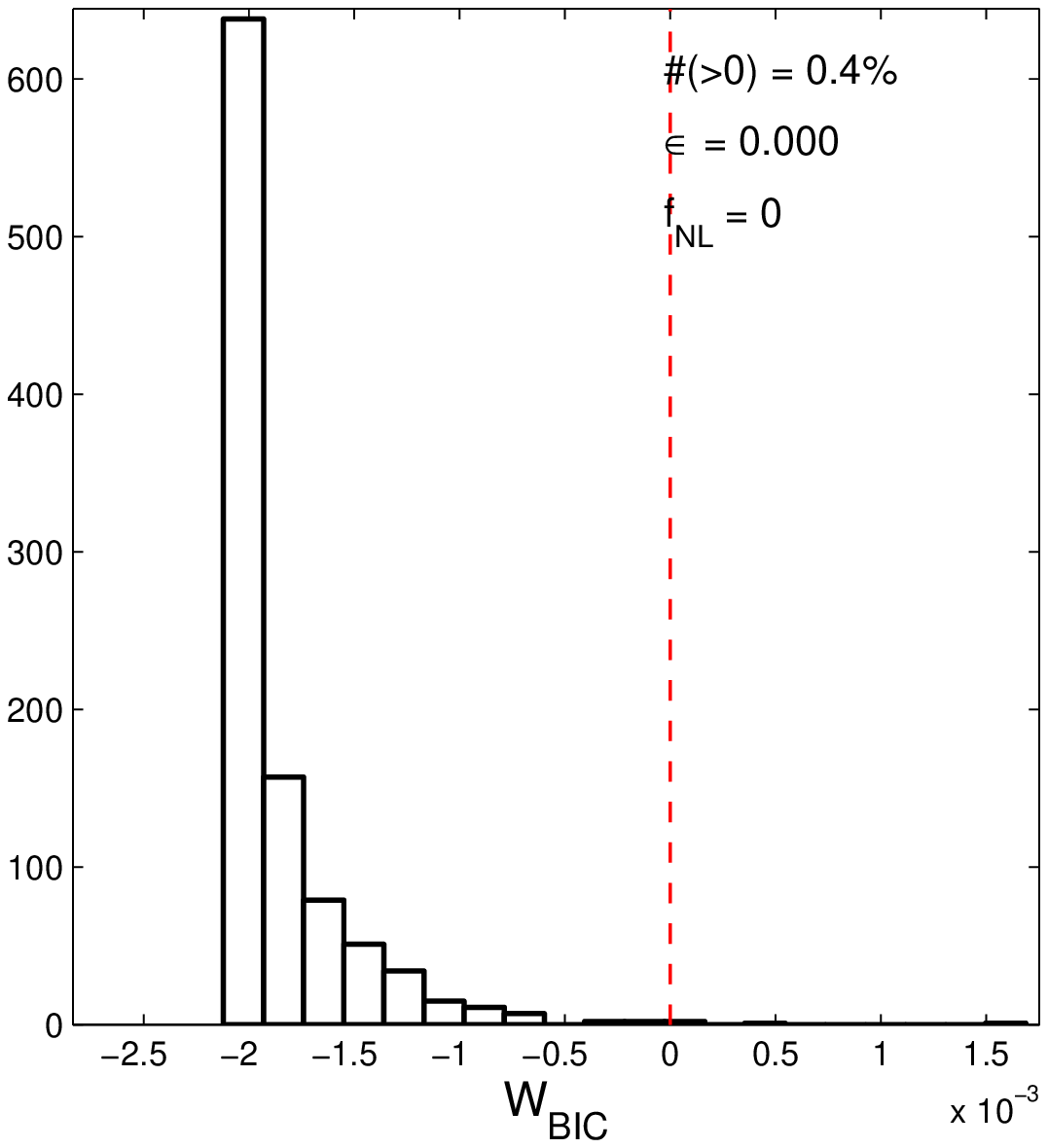}
\includegraphics[width=3.1cm,keepaspectratio]{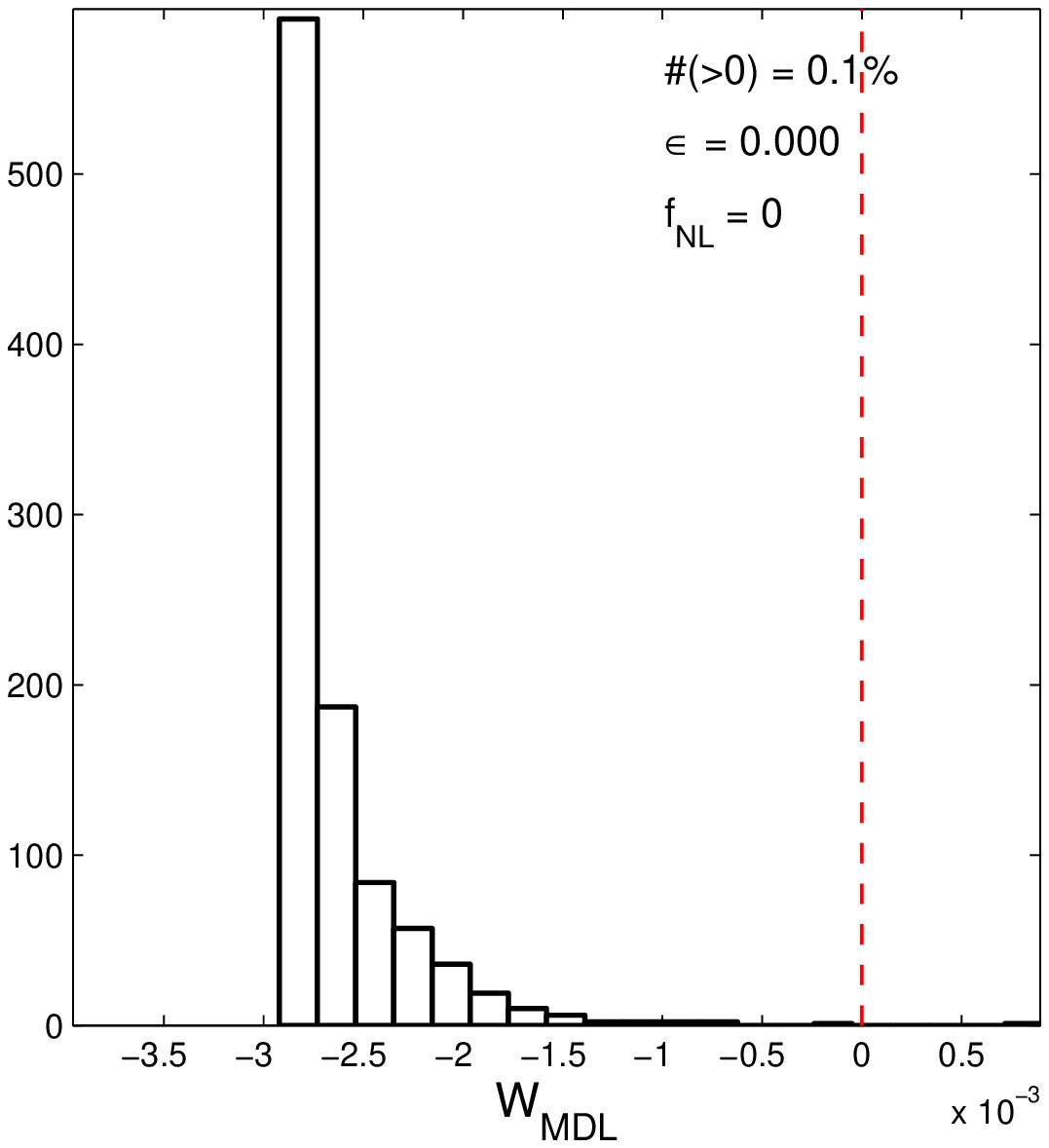}
\includegraphics[width=3.1cm,keepaspectratio]{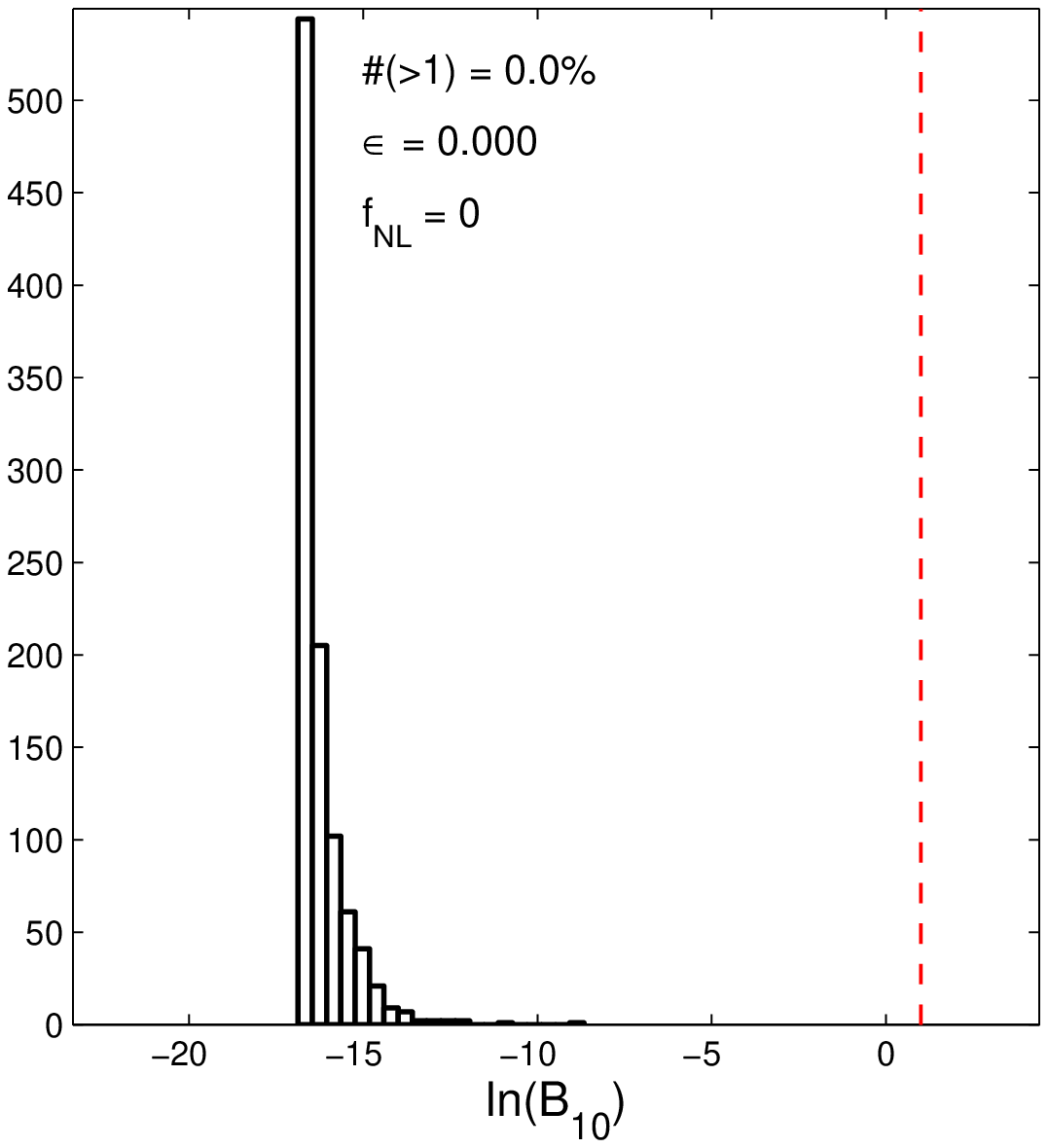}
\includegraphics[width=3.1cm,keepaspectratio]{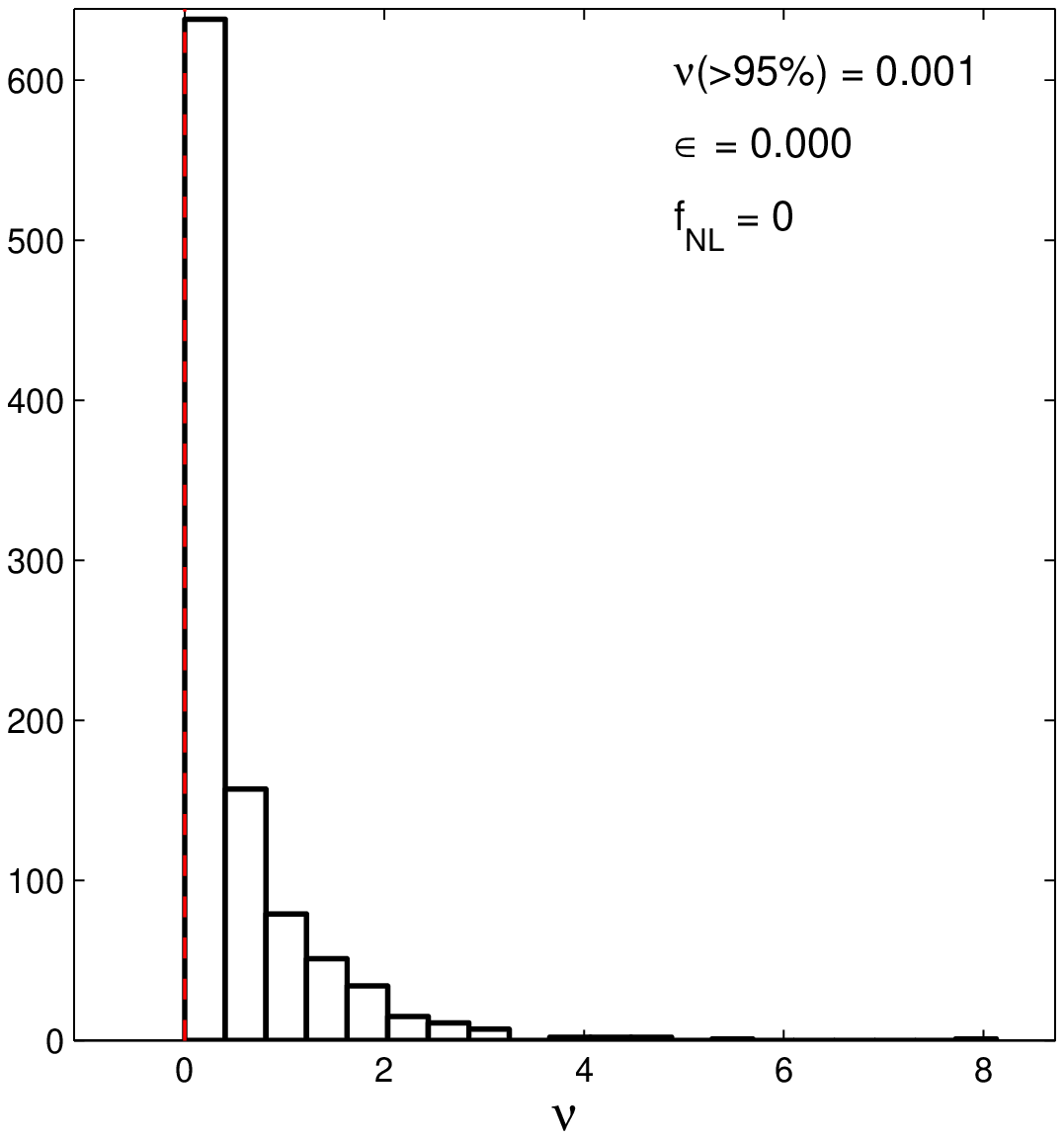}
\includegraphics[width=3.1cm,keepaspectratio]{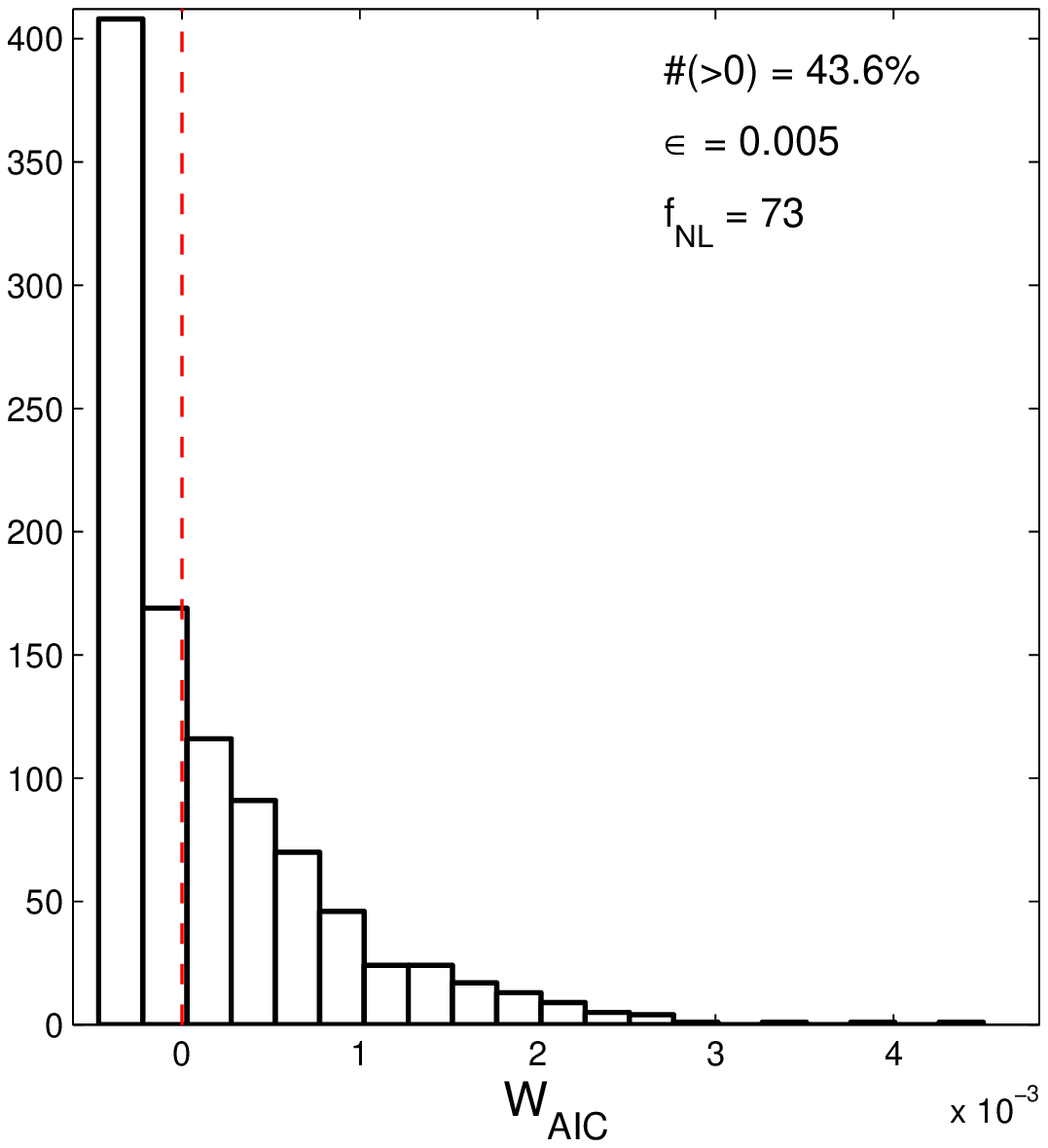}
\includegraphics[width=3.1cm,keepaspectratio]{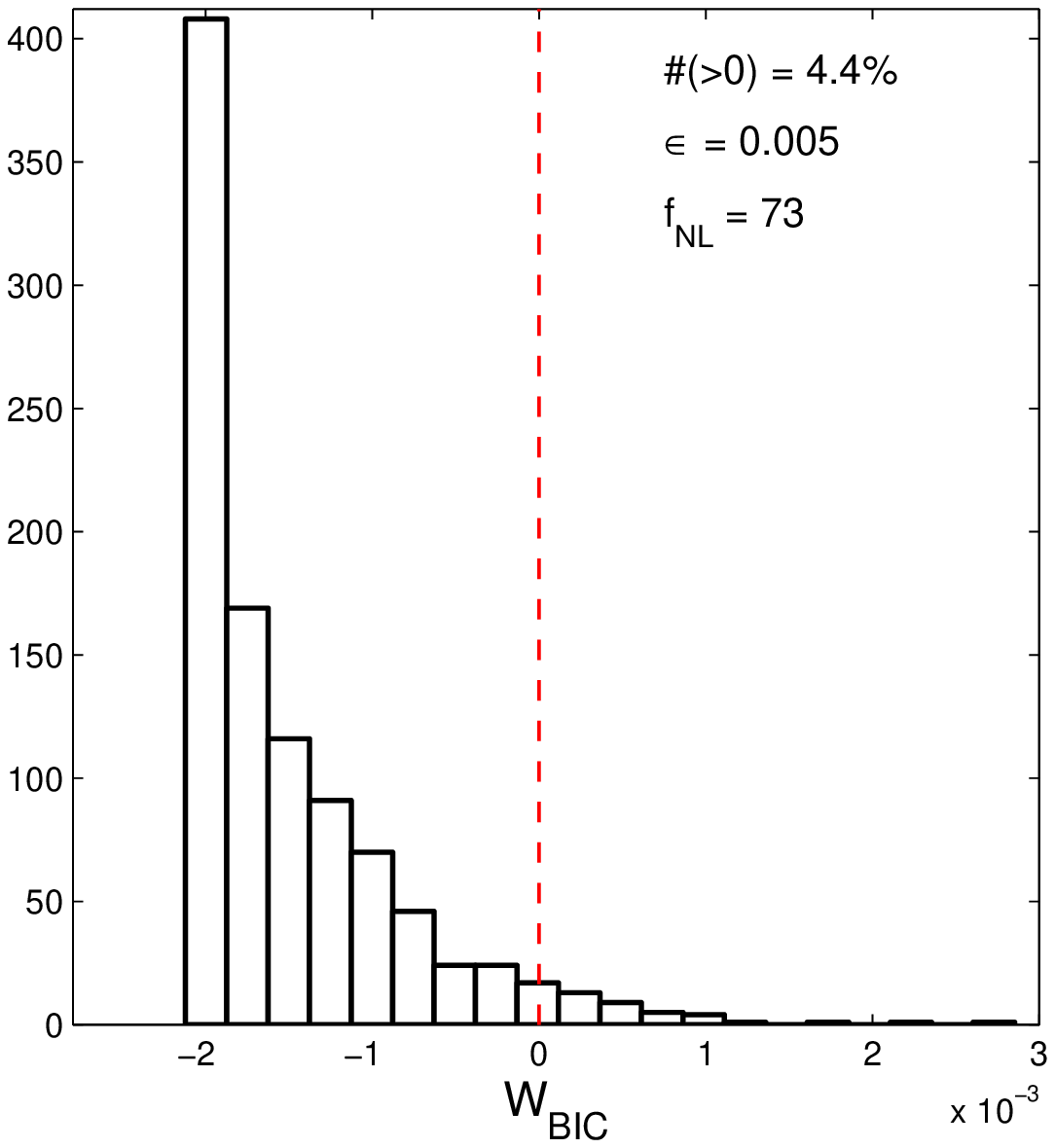}
\includegraphics[width=3.1cm,keepaspectratio]{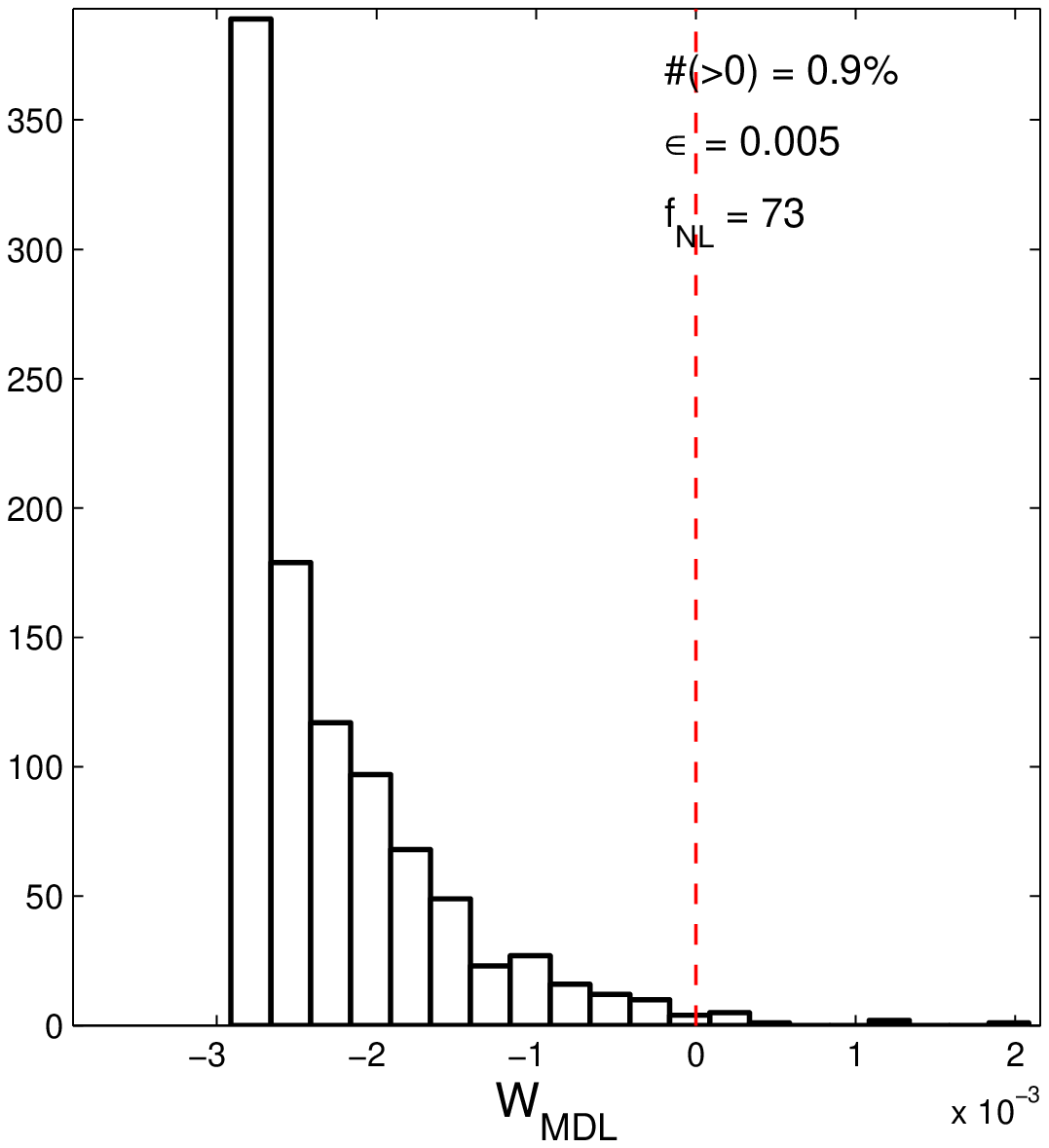}
\includegraphics[width=3.1cm,keepaspectratio]{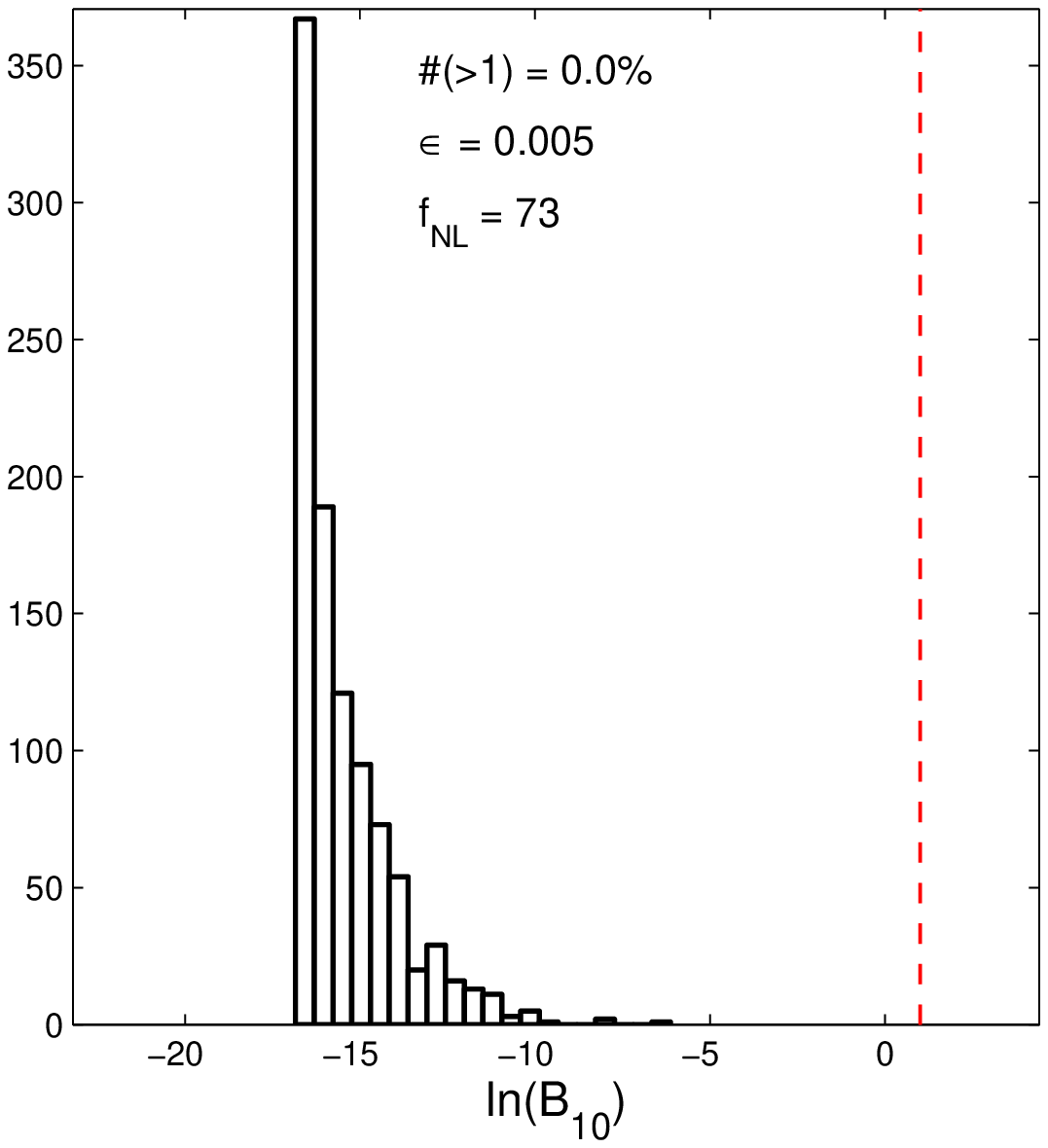}
\includegraphics[width=3.1cm,keepaspectratio]{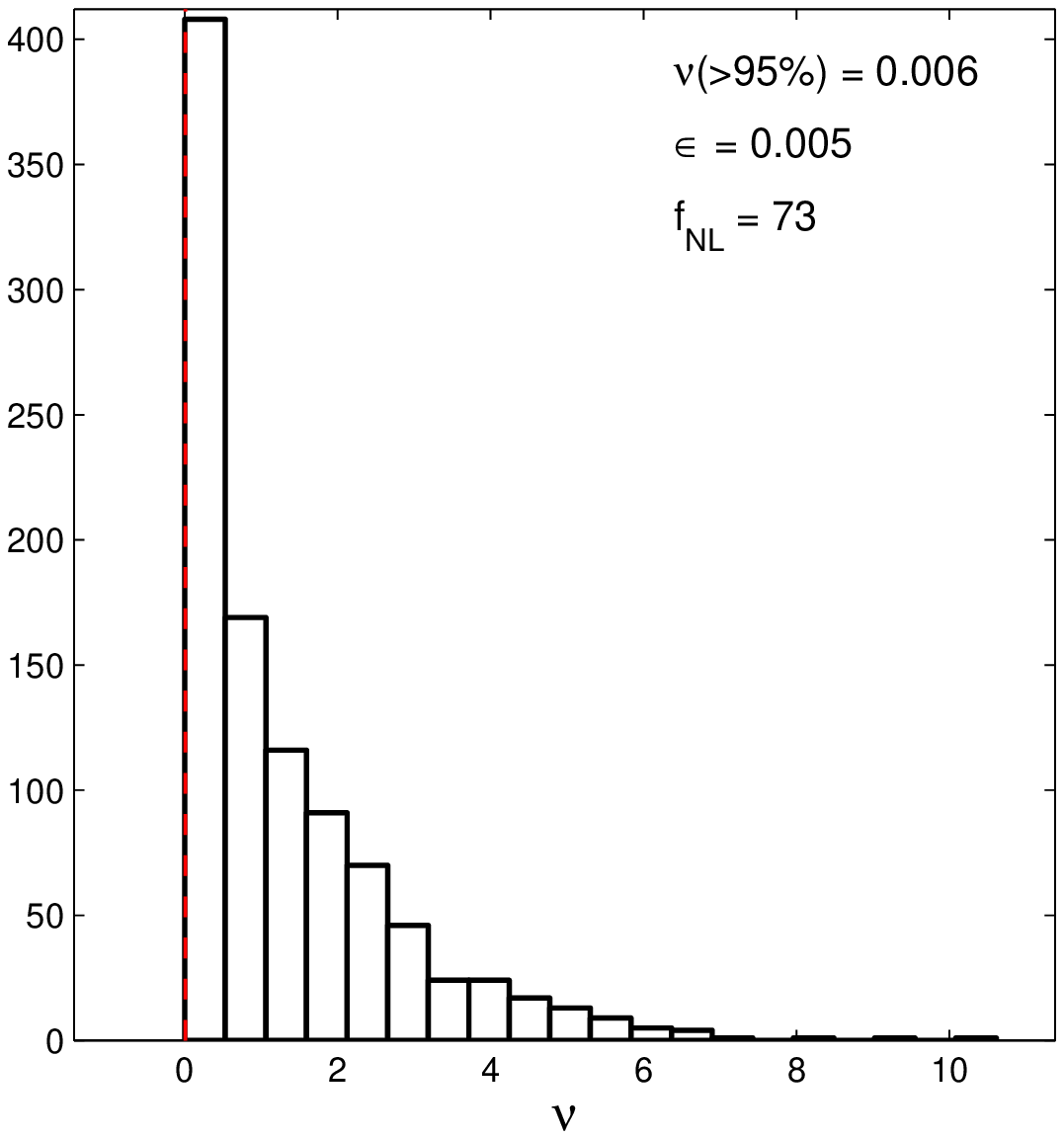}
\includegraphics[width=3.1cm,keepaspectratio]{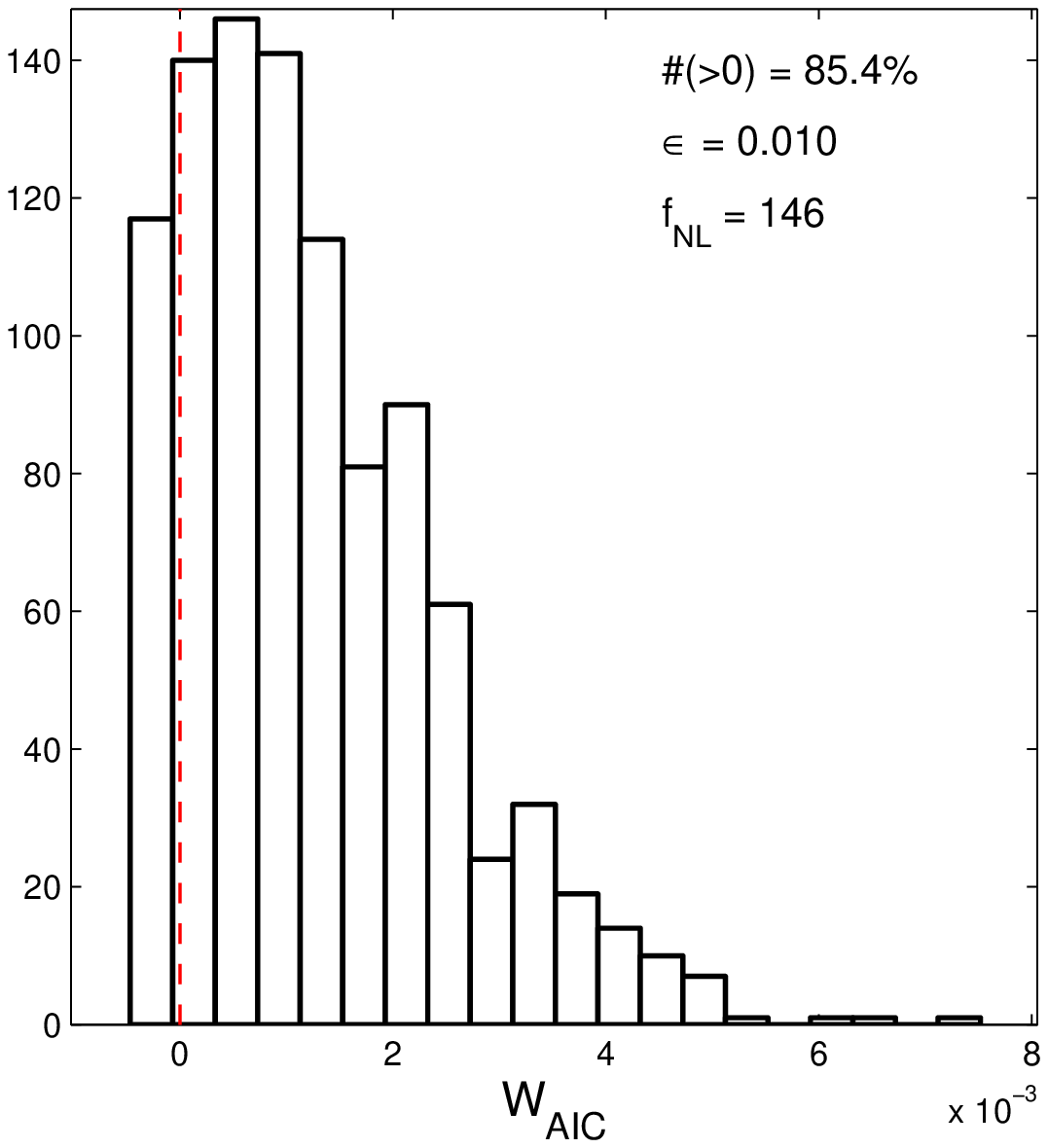}
\includegraphics[width=3.1cm,keepaspectratio]{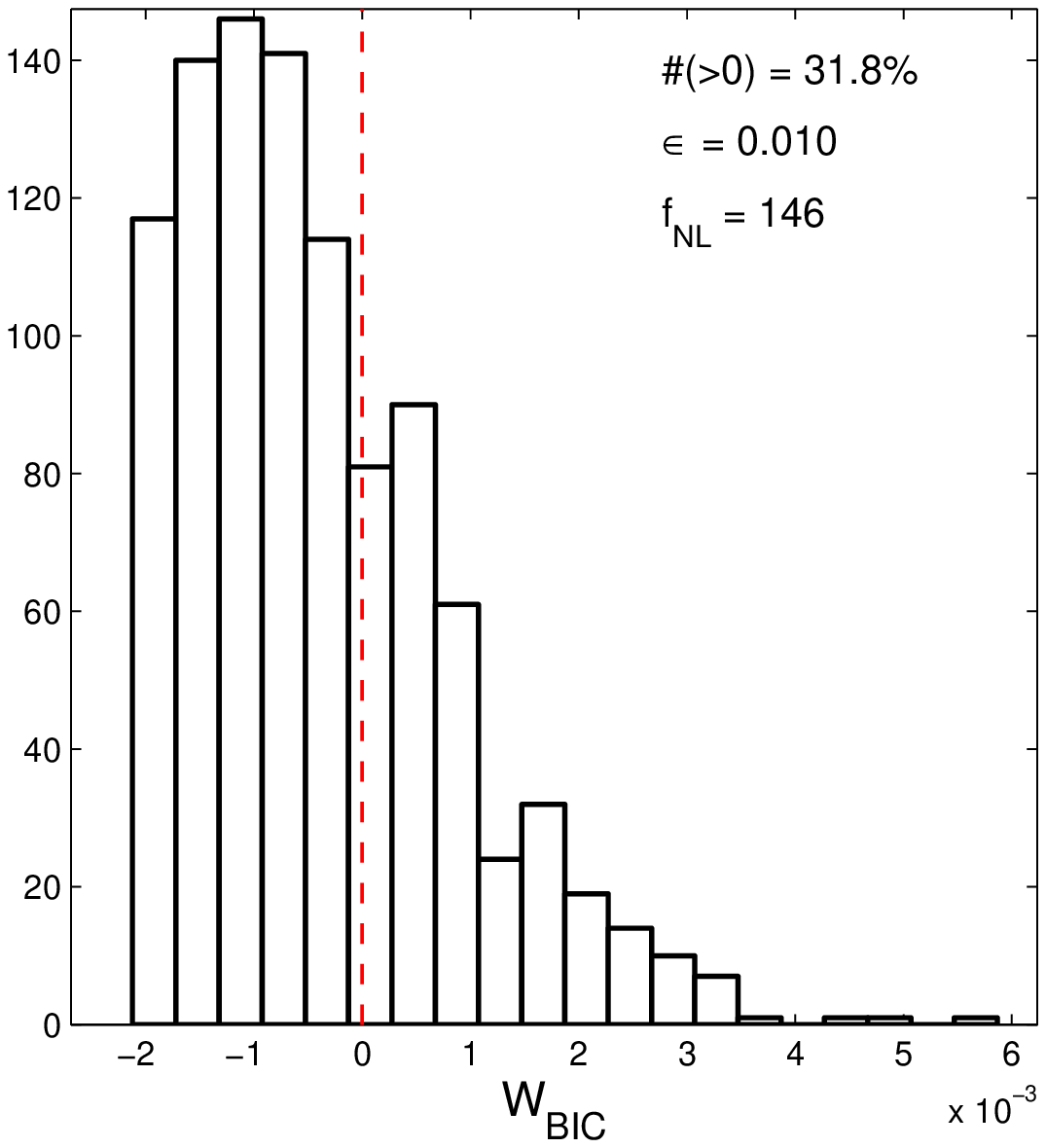}
\includegraphics[width=3.1cm,keepaspectratio]{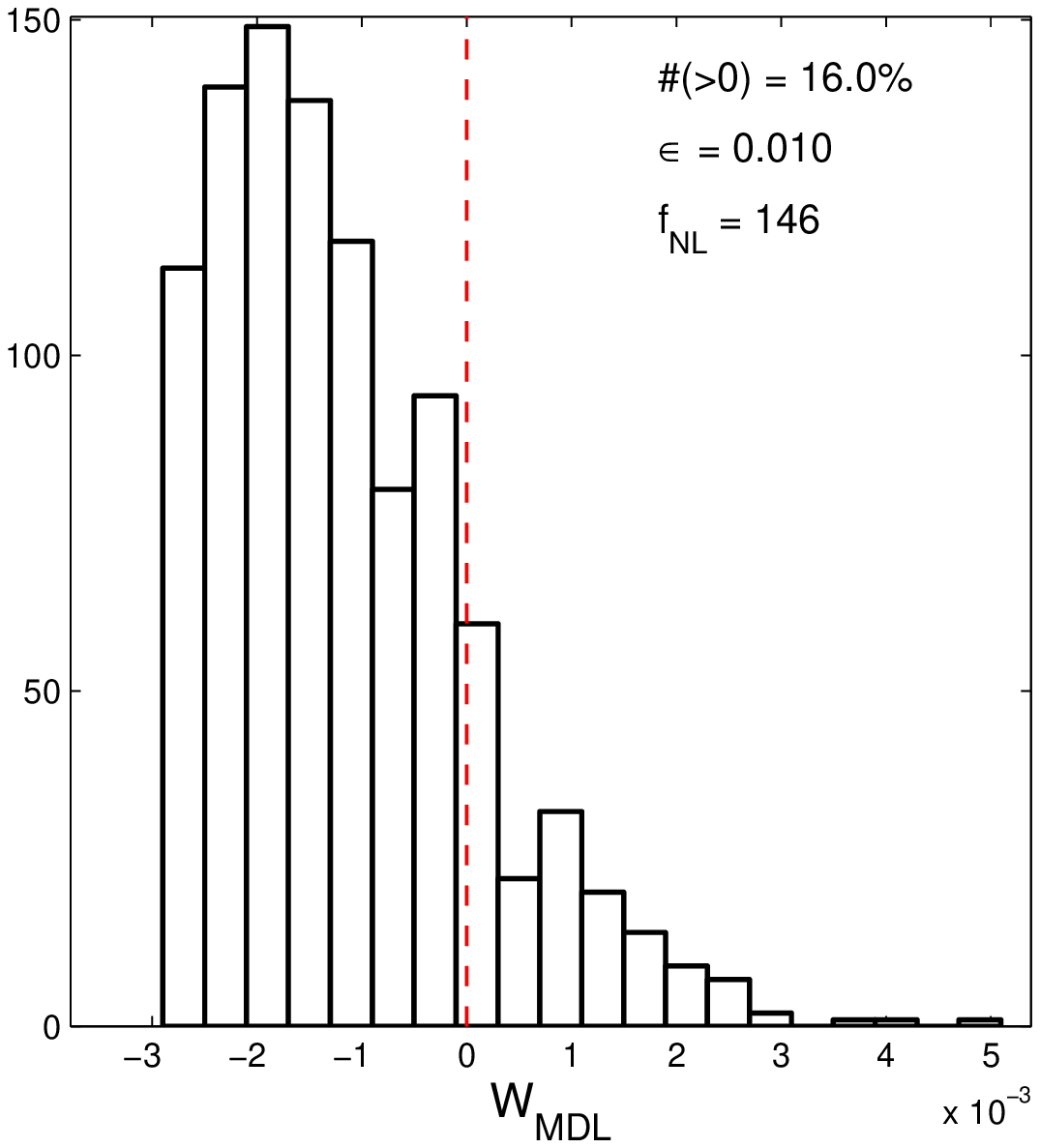}
\includegraphics[width=3.1cm,keepaspectratio]{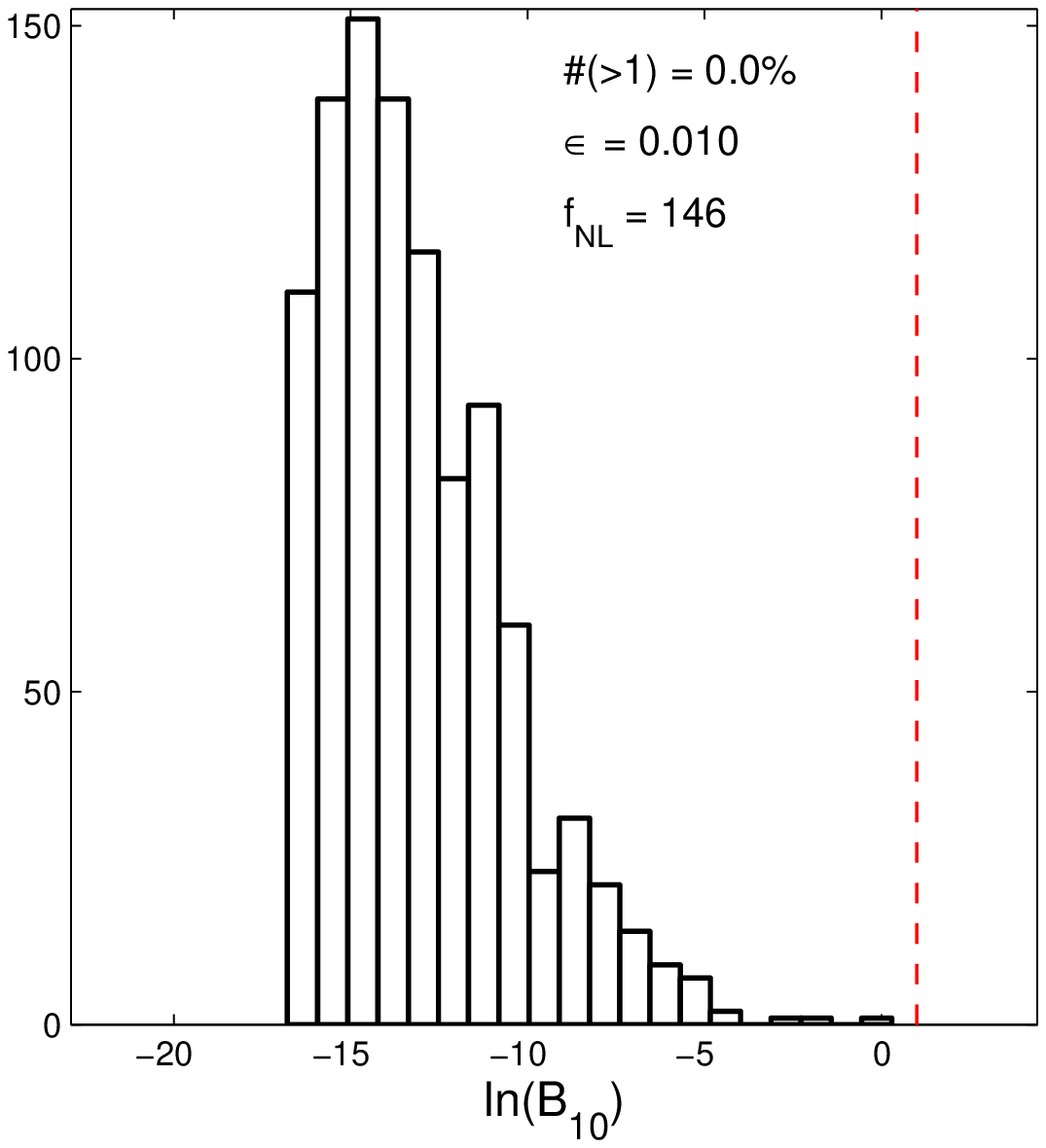}
\includegraphics[width=3.1cm,keepaspectratio]{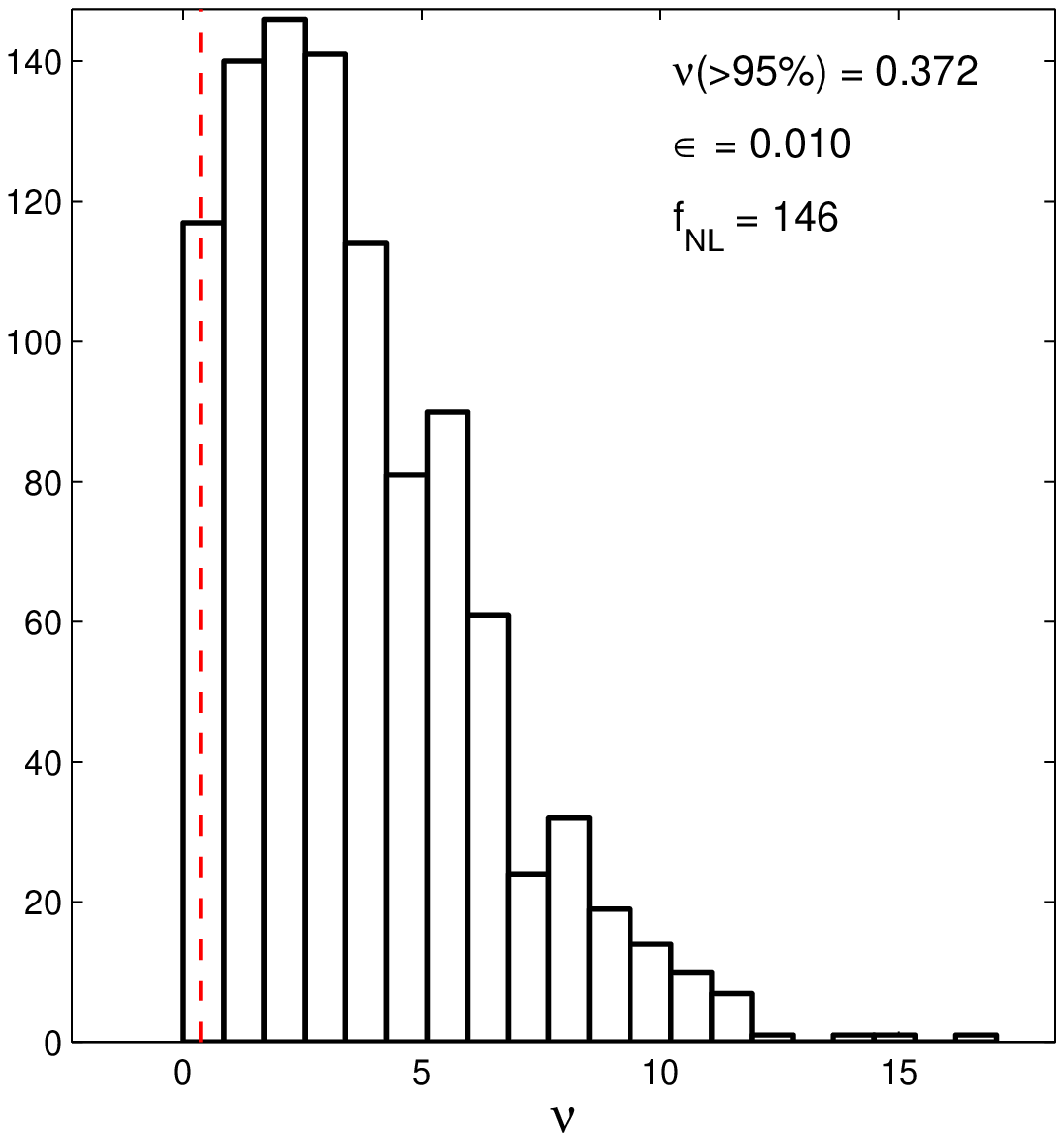}
\includegraphics[width=3.1cm,keepaspectratio]{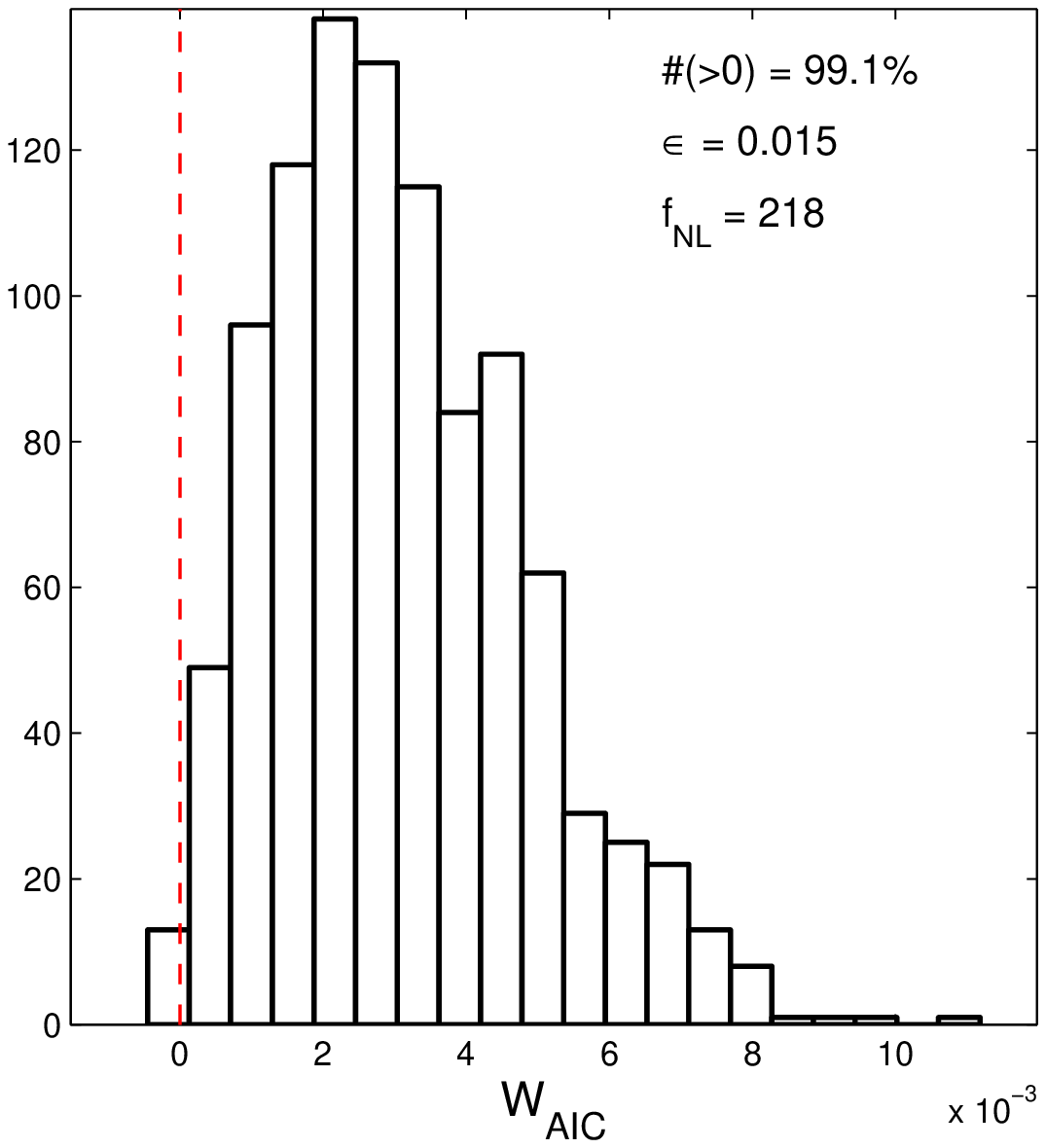}
\includegraphics[width=3.1cm,keepaspectratio]{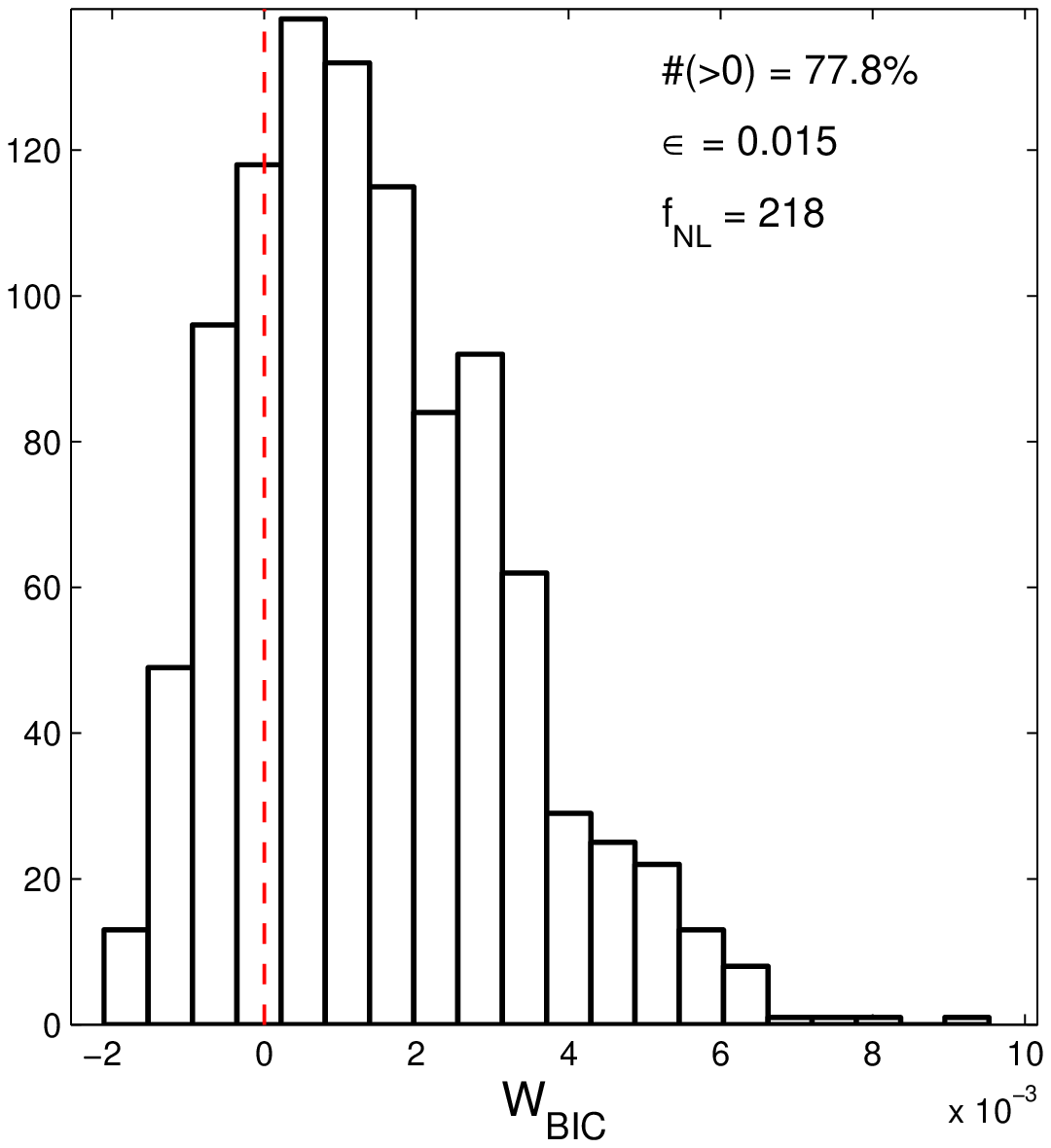}
\includegraphics[width=3.1cm,keepaspectratio]{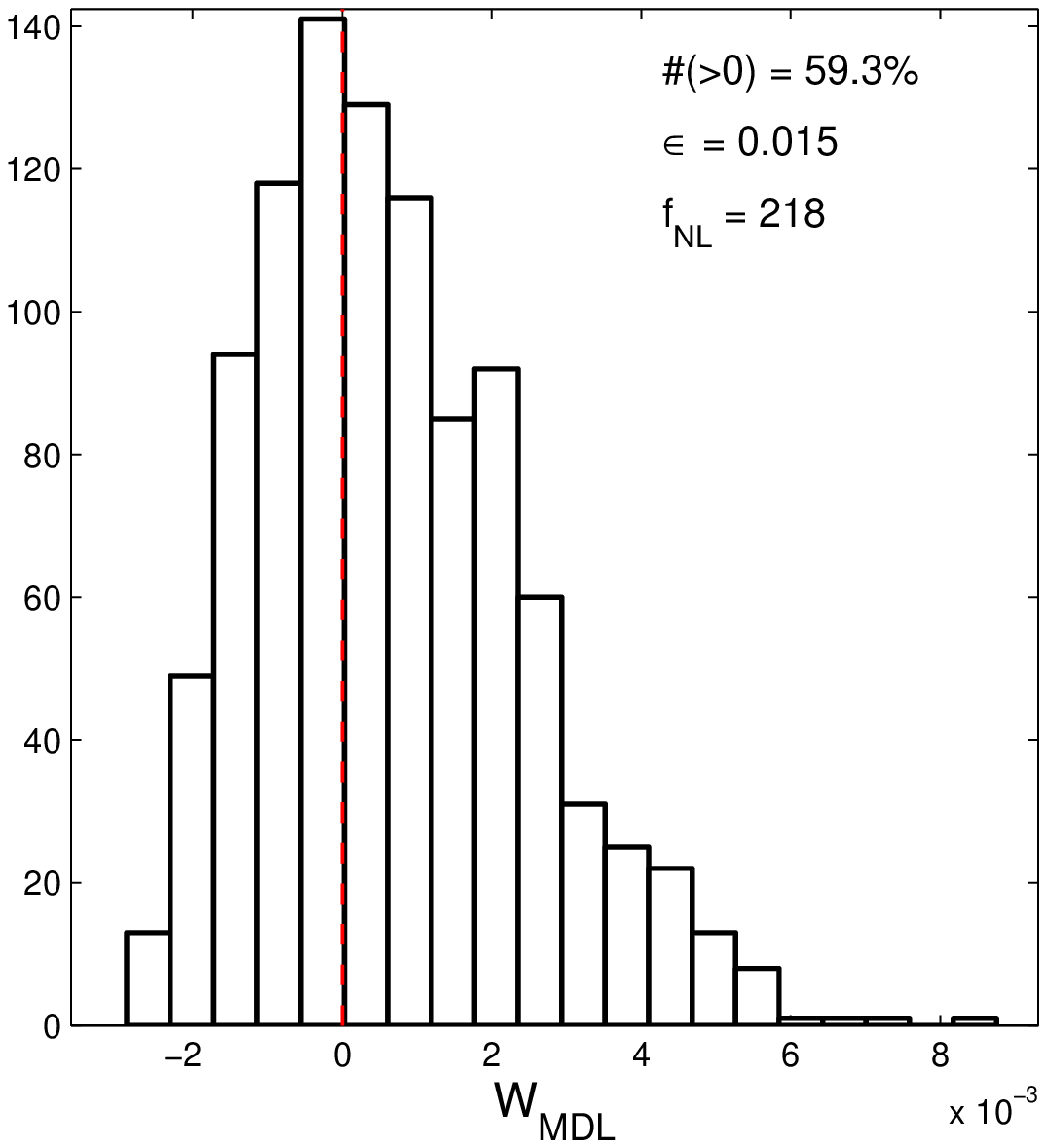}
\includegraphics[width=3.1cm,keepaspectratio]{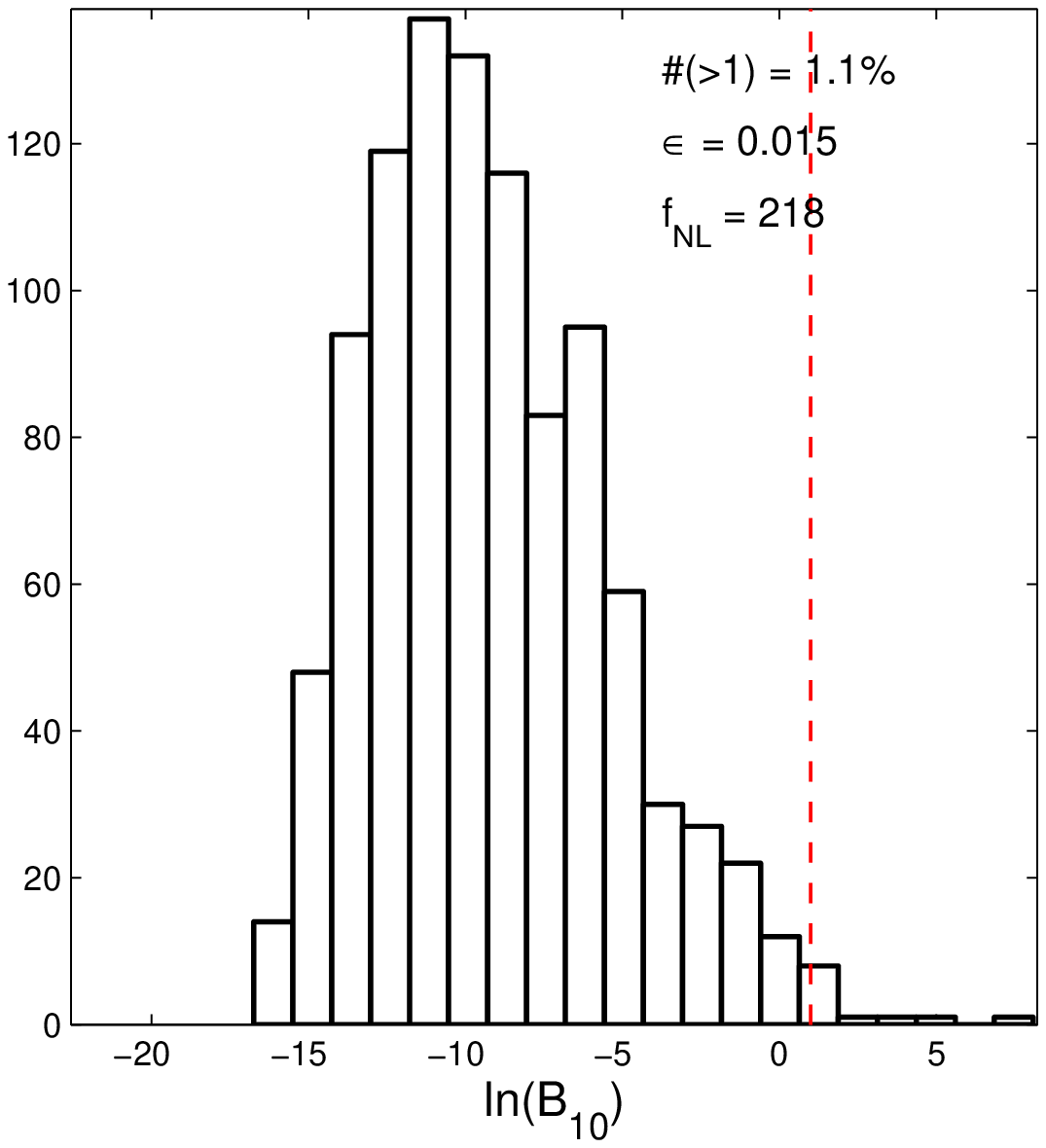}
\includegraphics[width=3.1cm,keepaspectratio]{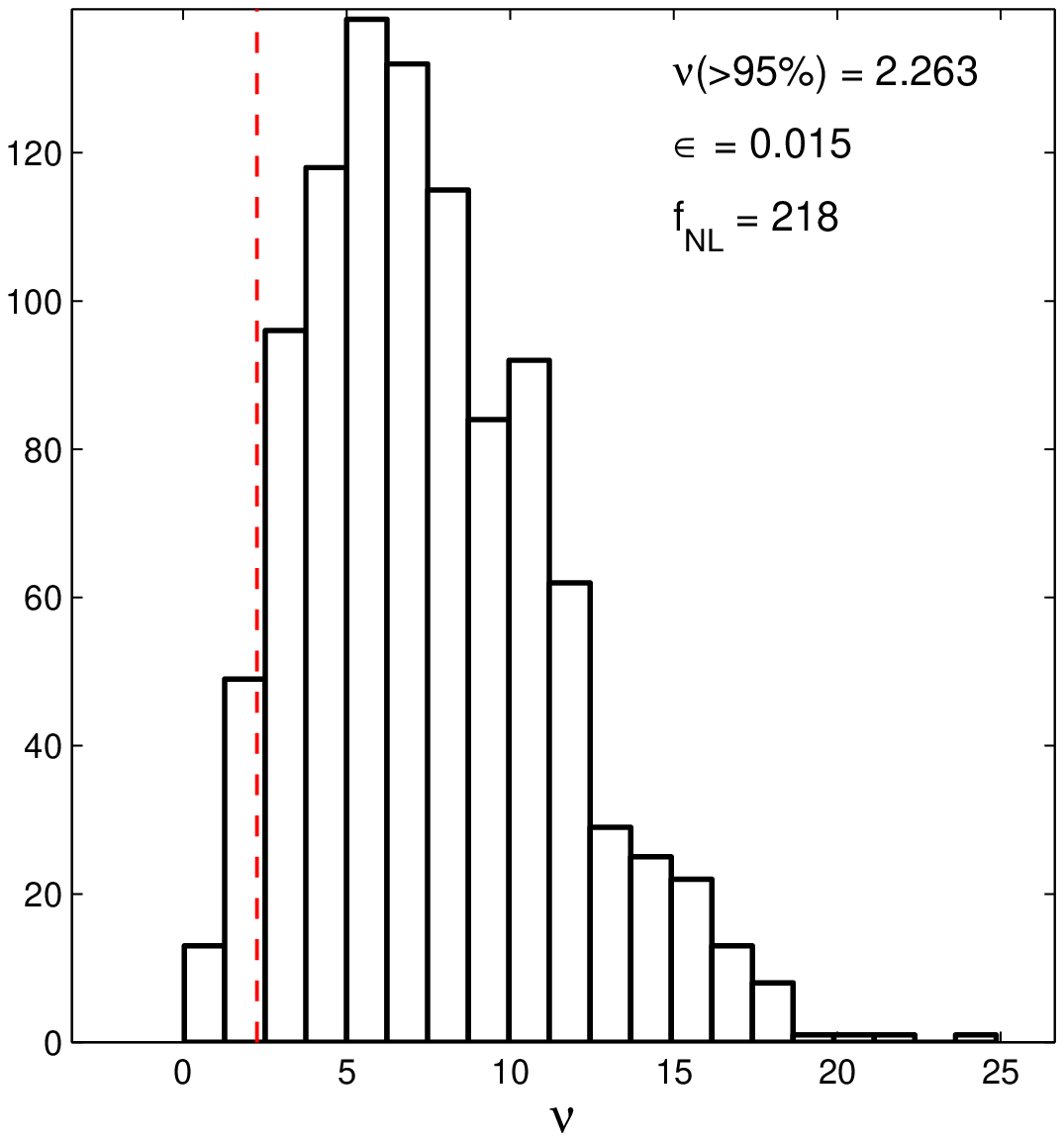}
\includegraphics[width=3.1cm,keepaspectratio]{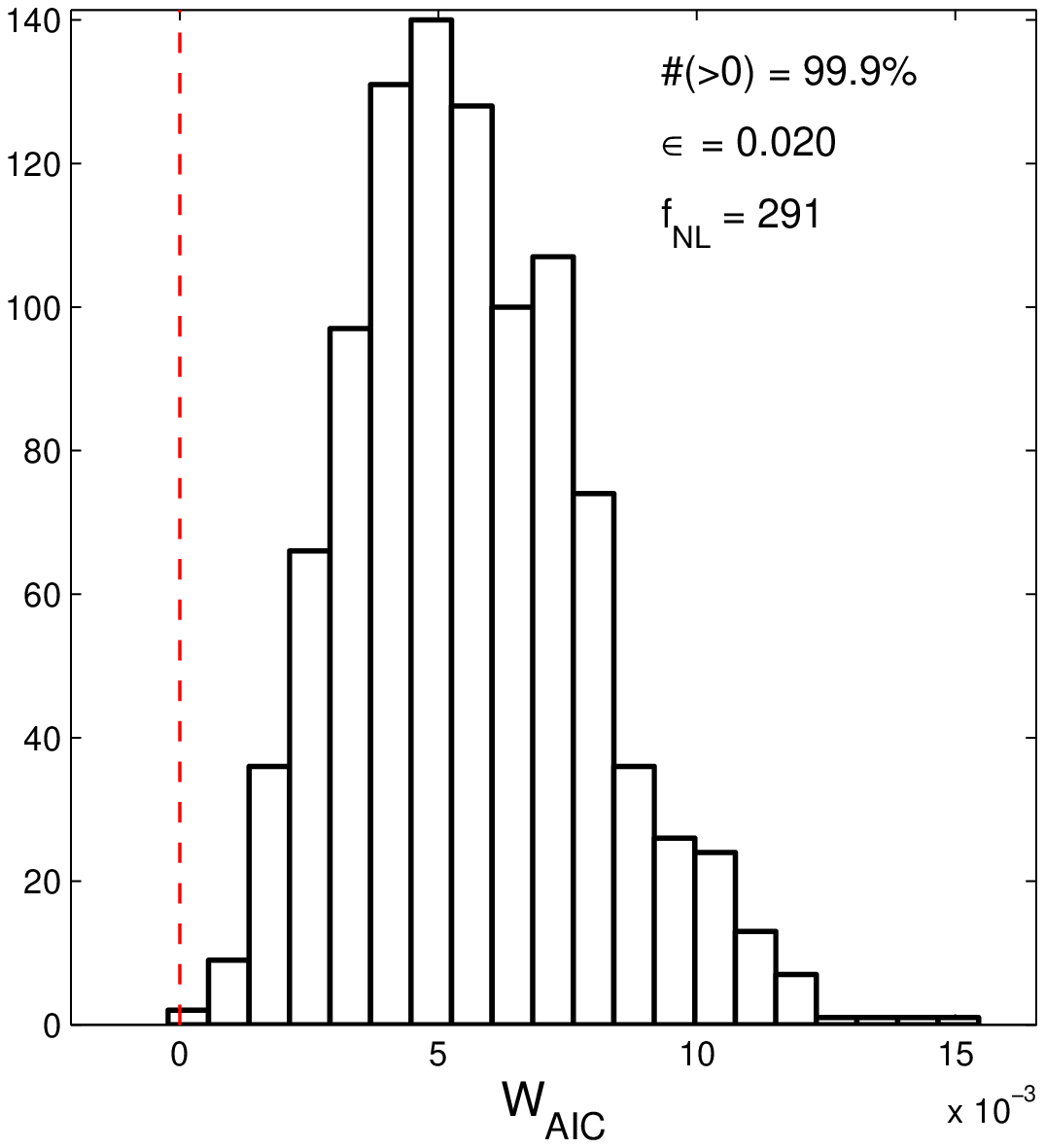}
\includegraphics[width=3.1cm,keepaspectratio]{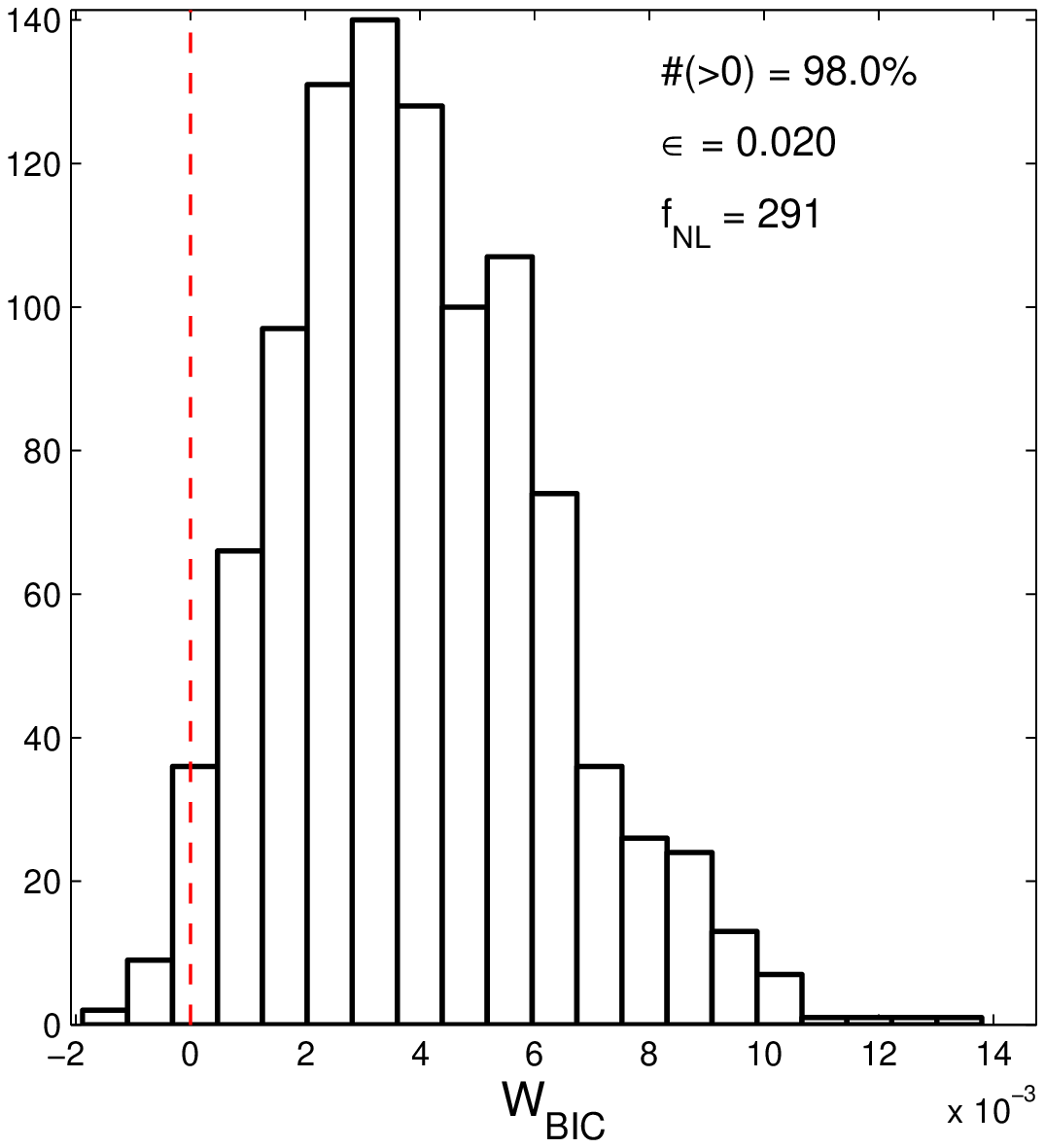}
\includegraphics[width=3.1cm,keepaspectratio]{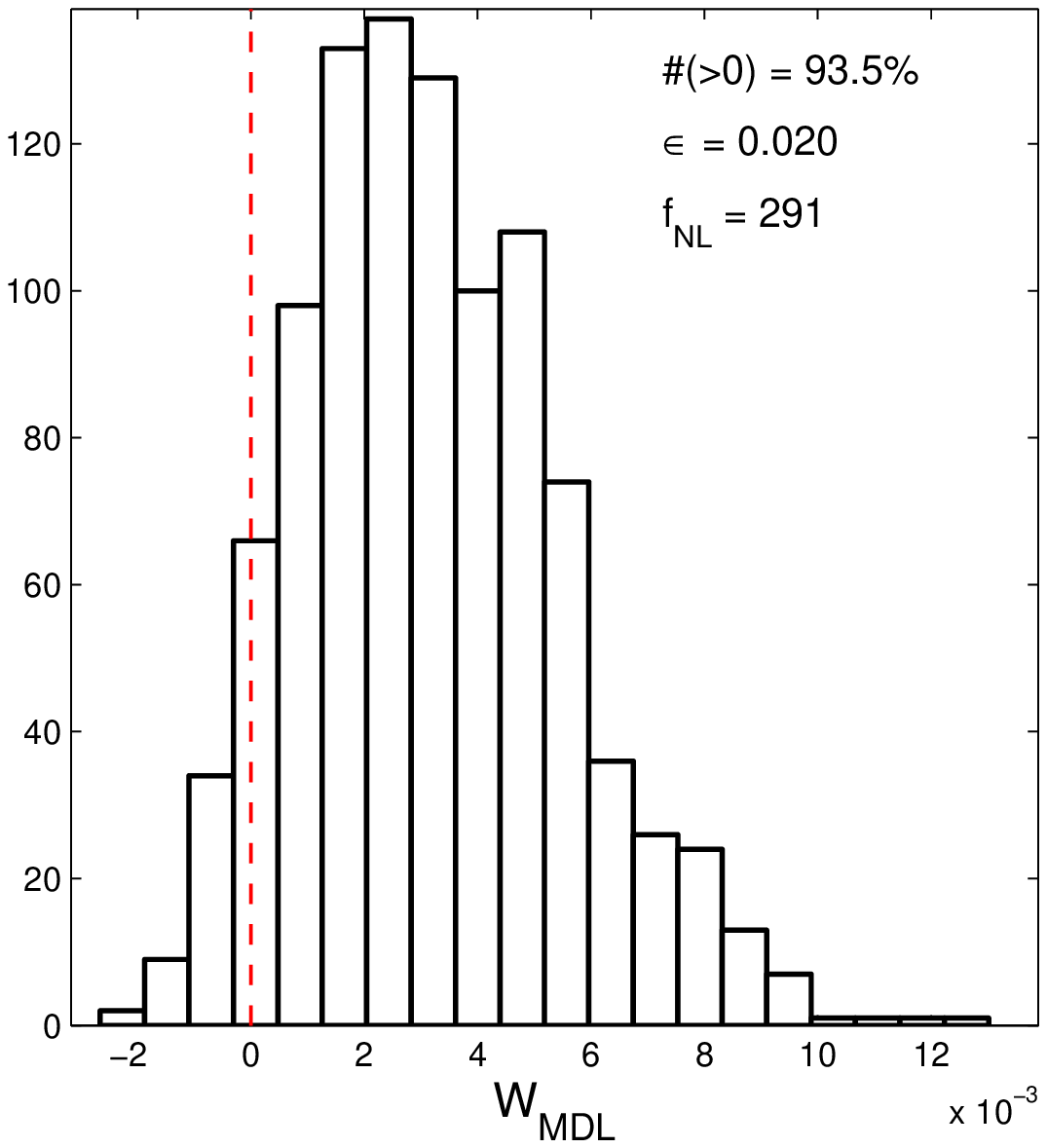}
\includegraphics[width=3.1cm,keepaspectratio]{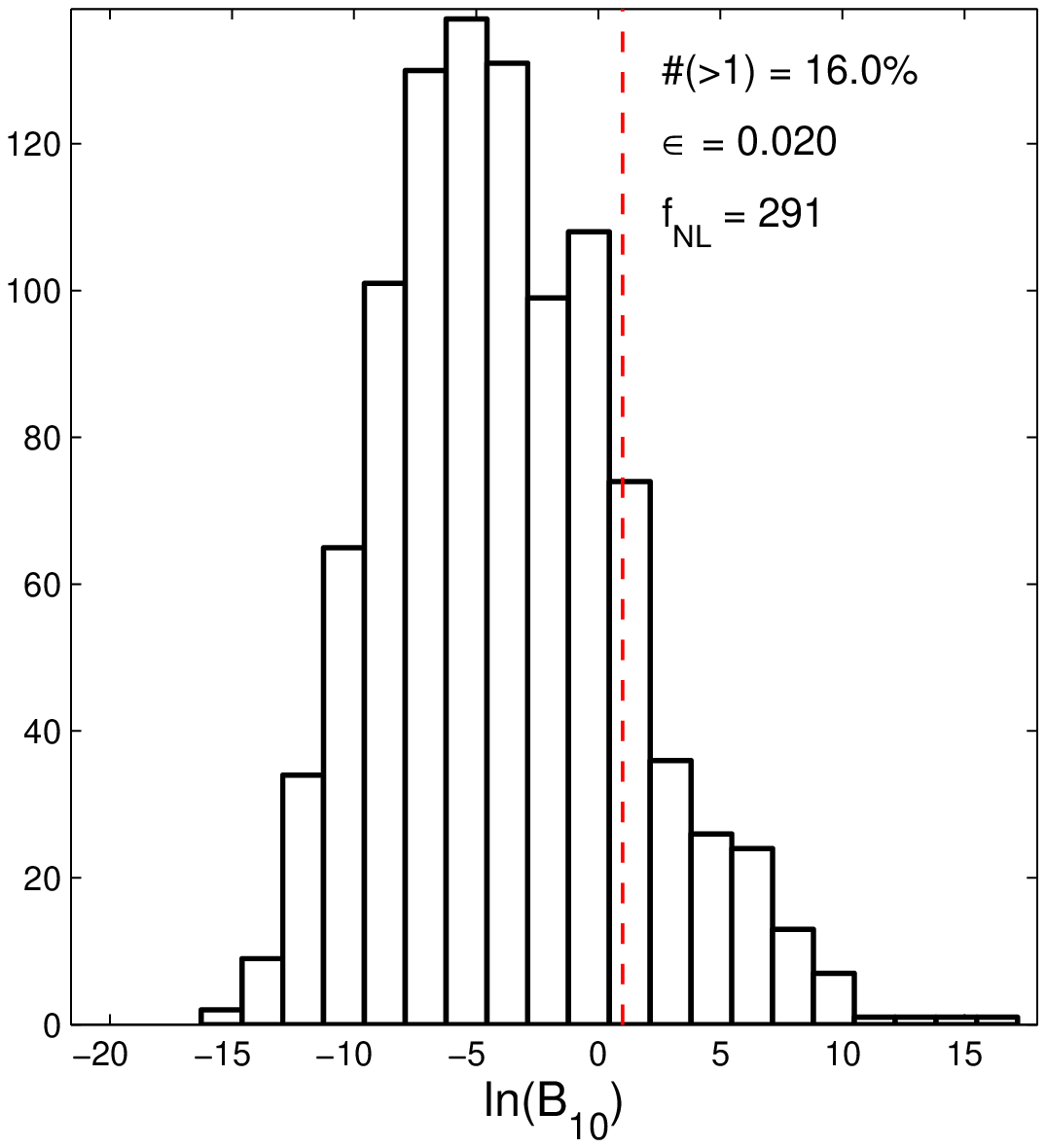}
\includegraphics[width=3.1cm,keepaspectratio]{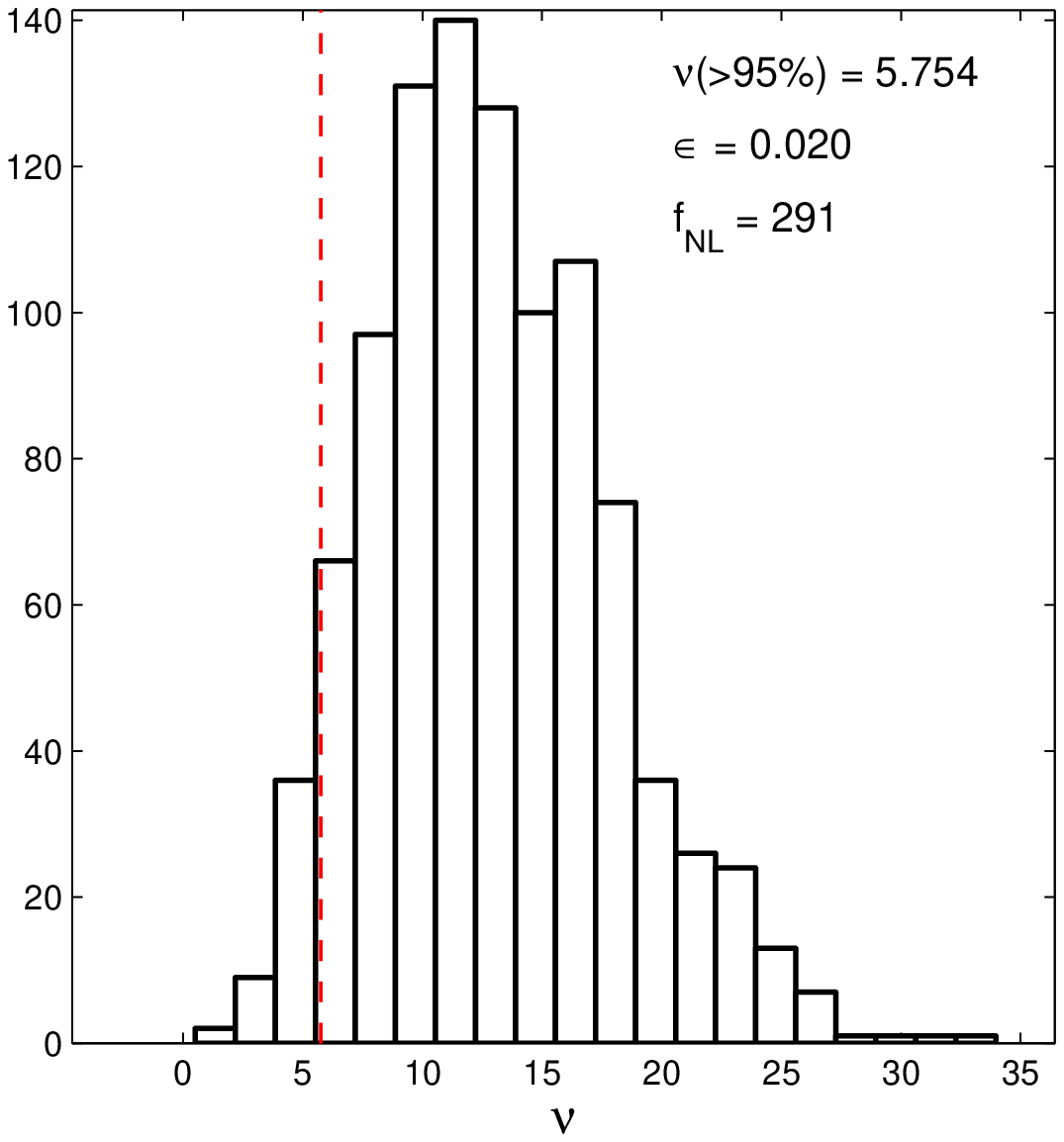}
\includegraphics[width=3.1cm,keepaspectratio]{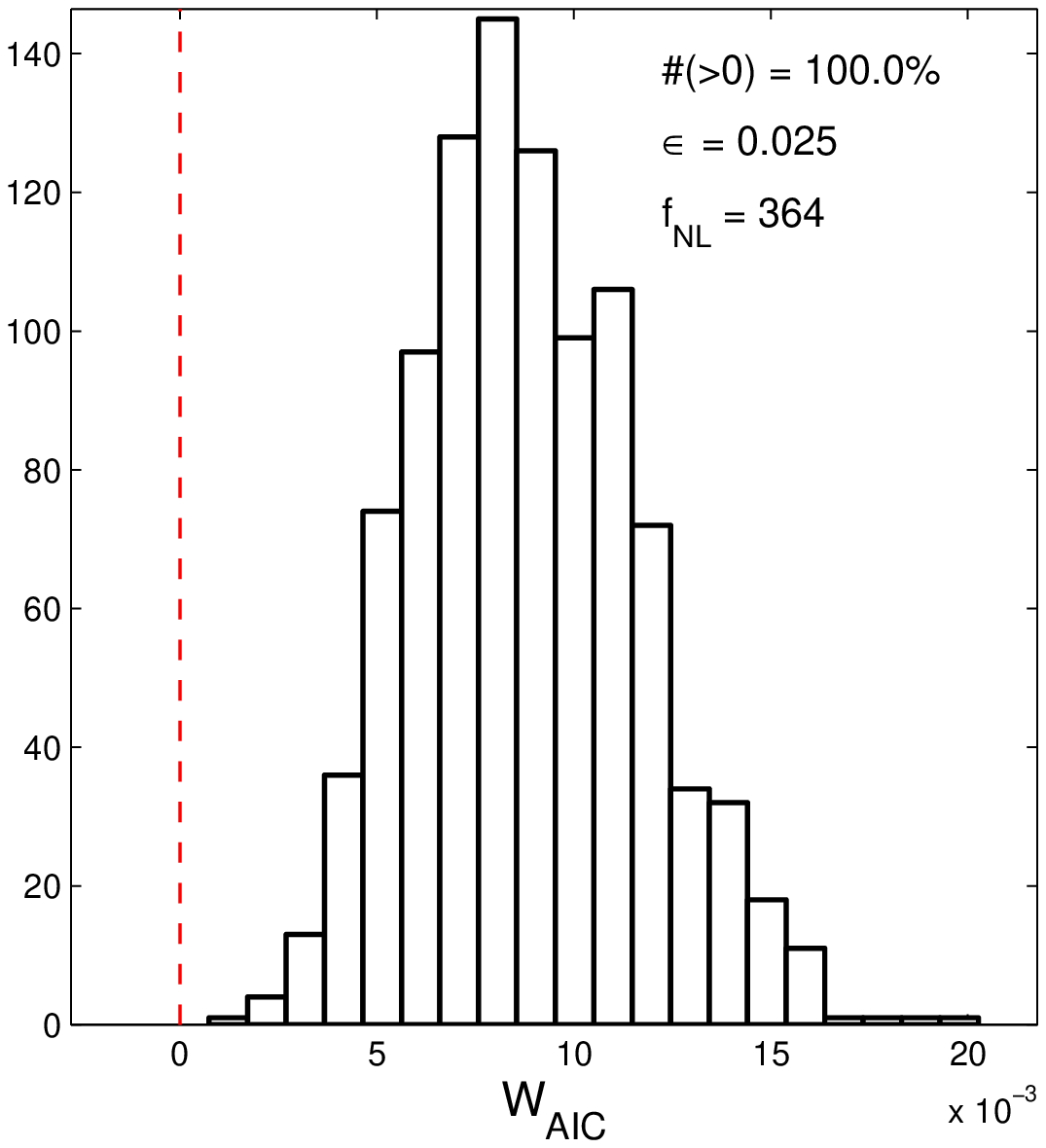}
\includegraphics[width=3.1cm,keepaspectratio]{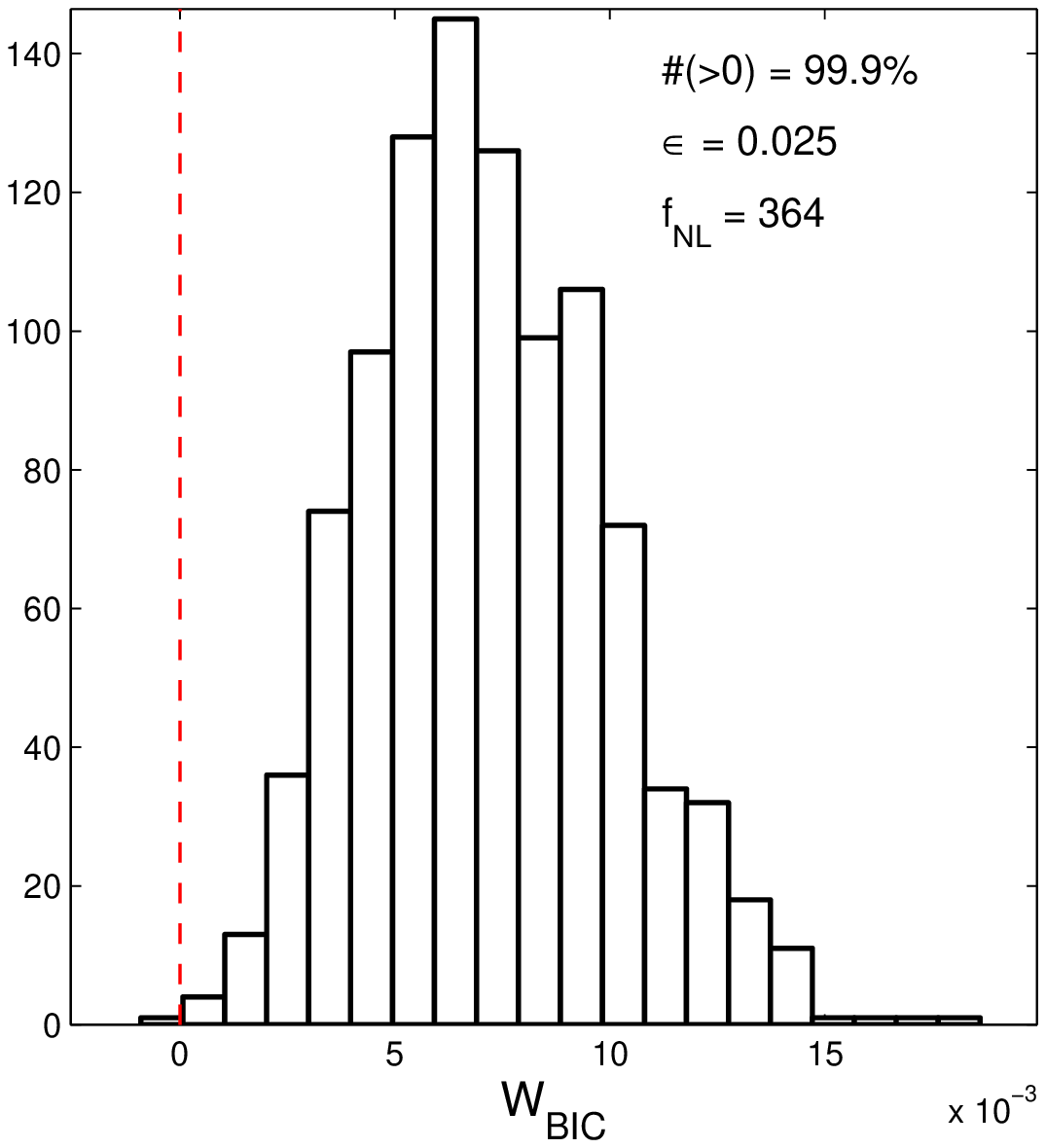}
\includegraphics[width=3.1cm,keepaspectratio]{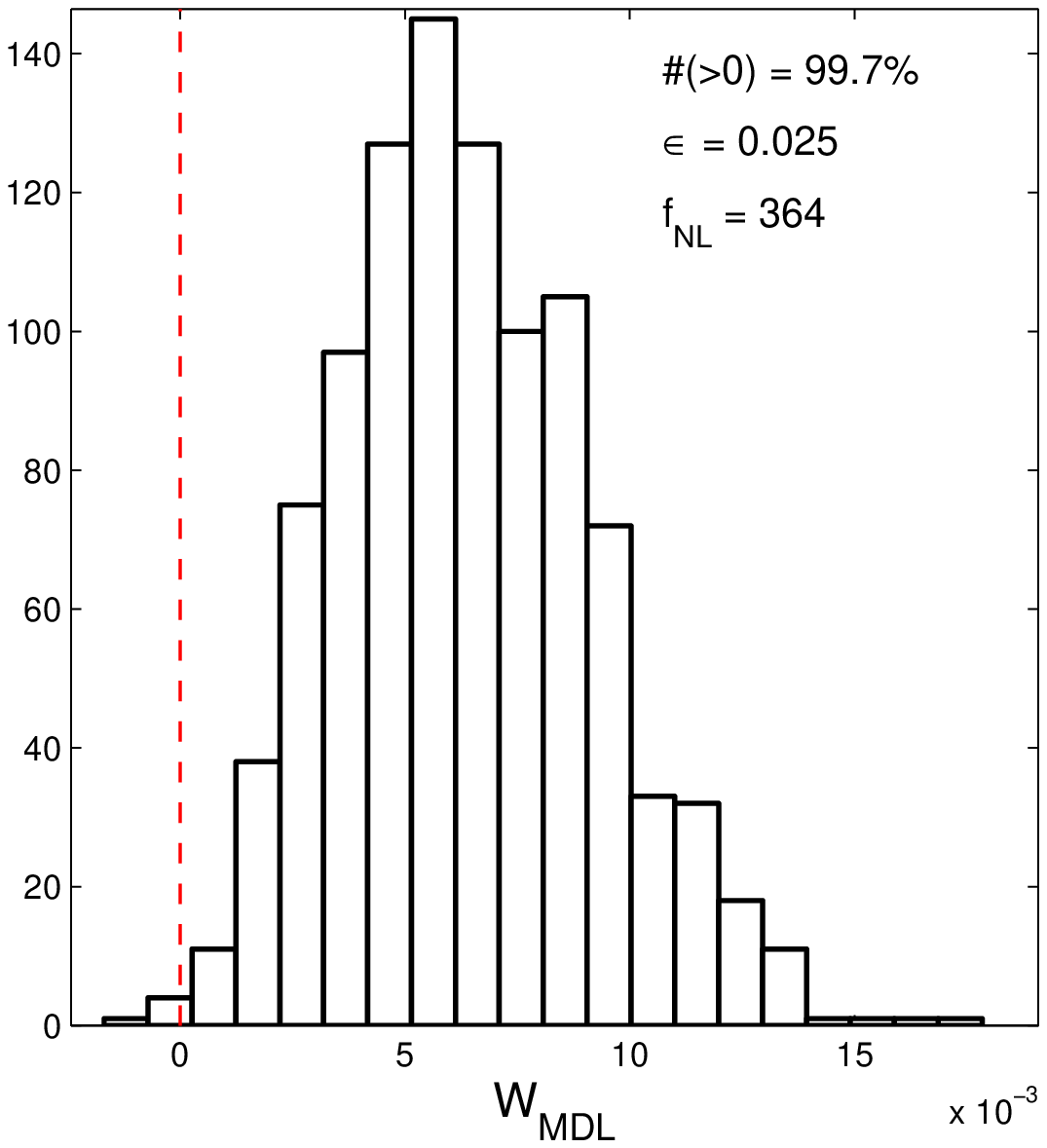}
\includegraphics[width=3.1cm,keepaspectratio]{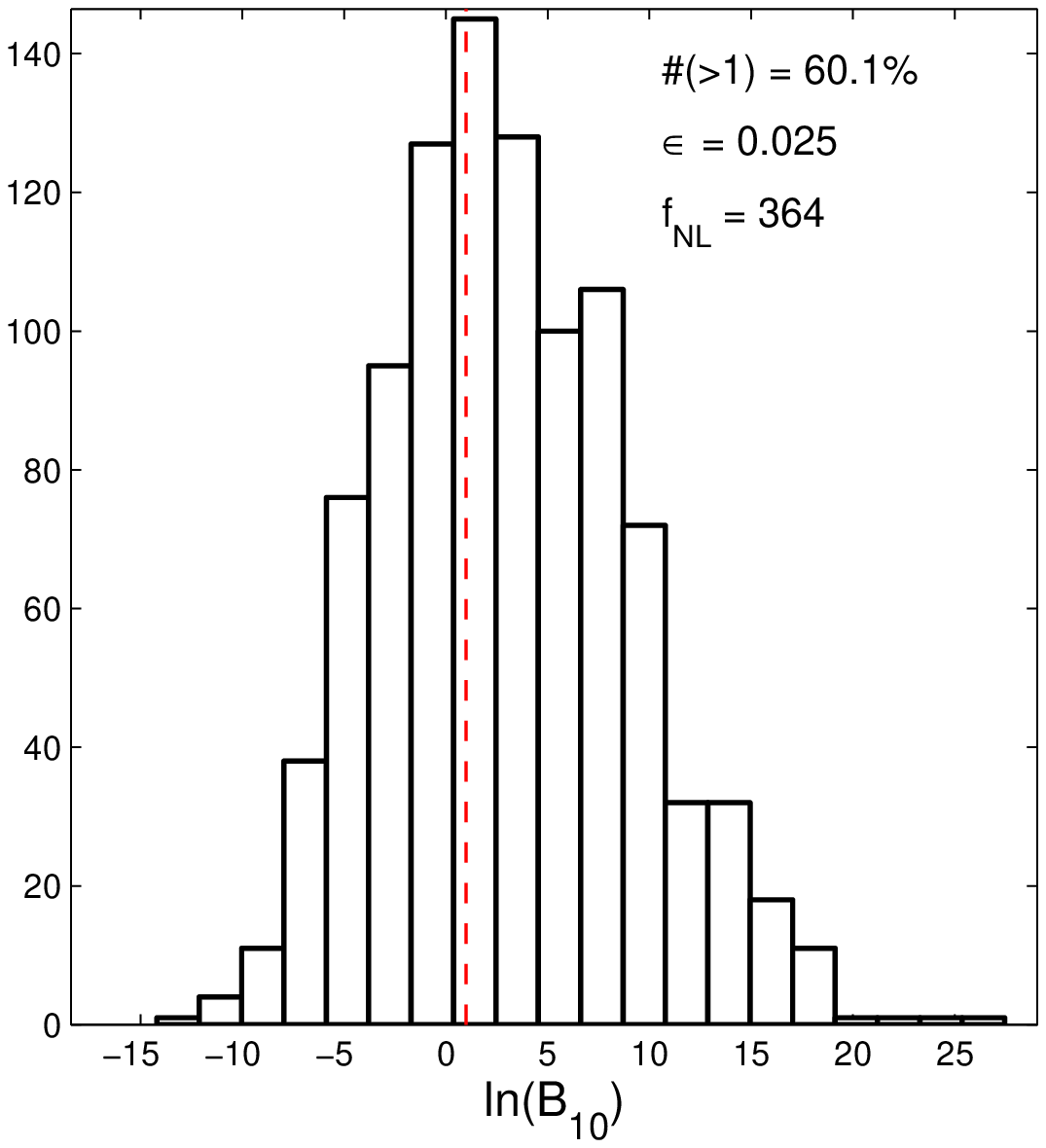}
\includegraphics[width=3.1cm,keepaspectratio]{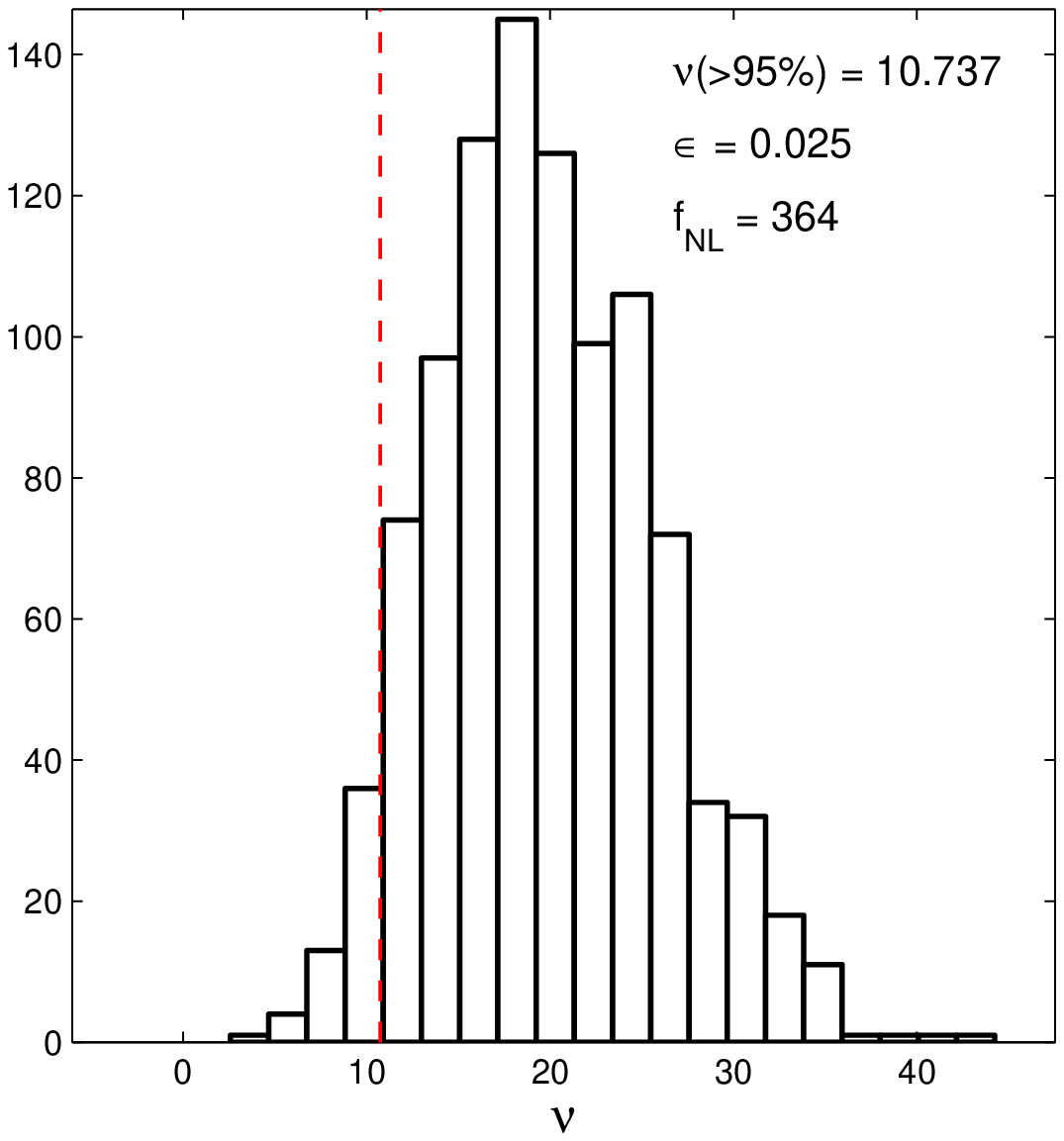}
\caption{\label{fig:selection_sims}
From left ro right, columns show the distribution for several statistics
referred to different model selection criteria: AIC, BIC, MDL, BE and GLRT.
From top to bottom, results for local non-Gaussian models for different
non-linear parameters are give: $\epsilon$ = 0, 0.005, 0.010, 0.015, 0.020 and 0.025.
Panels in columns 1, 2, 3 and 4 present a vertical line, separating the region
where $H_0$ and $H_1$ are preferred (to the left and to the right of the vertical
line, respectively). The vertical line on the last column represents the value
for the $\nu$ parameter for which $95\%$ of the non-Gaussian simulations
are more likely described by $H_1$ rather than $H_0$.}
\end{figure*}

We present the results, graphically, in figure~\ref{fig:selection_sims}.
This plot consist in 6 rows and 5 columns. Each row corresponds to the results obtained for
a given value of the $\epsilon$ parameter (namely, from top to bottom:
$\epsilon$ = 0, 0.005, 0.010, 0.015, 0.020 and 0.025). Each column refers
to a model selection criterion (from left ro right: AIC, BIC, MDL, BE and GLRT).
For the first column (i.e., the AIC case) we plot the distributions (obtained after analysing 1,000 simulations)
of a statistical variable defined as: ${\rm W}_{\rm AIC} \equiv R^2/Q - 4/N$,
notice (from equation~\ref{eq:aic_rule}) that a positve value of ${\rm W}_{\rm AIC}$ implies to
accept $H_1$ against $H_0$. The second column accounts for the distributions of the variable
${\rm W}_{\rm BIC} \equiv R^2/Q - 2\ln{N}/N$, which (from equation~\ref{eq:bic_rule}) also satisfies to
be positive when favoring $H_1$. Equivalently, third column shows the distributions of
${\rm W}_{\rm MDL} \equiv R^2/Q - 4/N\ln{N\Omega\sqrt{Q/\pi}}$, that according to equation~\ref{eq:mdl_rule} is
also positive when $H_1$ is more likely than $H_0$. Fourth column provides $\ln{B_{10}}$, obtained
from equation~\ref{eq:b10_uniform}. Finally, in the fifth column we present the
distributions obtained for $\nu = \left(R^2/N\right)/\left(4/N\right)$ which, according to the GLRT decision
rule (equation~\ref{eq:glrt_rule}) provides a measurement of the strength in accepting
$H_1$ against $H_0$.

In addition to the distributions, we also plot, as an
indication, a vertical line in each panel. For ${\rm W}_{\rm AIC}$, ${\rm W}_{\rm BIC}$
and ${\rm W}_{\rm MDL}$, this vertical line separate the region where $H_0$ is favored (left side)
from the one where $H_1$ is more likely (right side). The percentage of the
1,000 simulations that fall on the $H_1$ region is also reported at each panel.
The vertical line for the ${\rm W}_{\rm BE}$ statistic represents the limit for
which the logarithmic Bayes' factor ($\ln{B_{10}}$) is greater than 1, which, in terms
of the Jeffreys' rule, indicates that $H_1$ is, at least, mildly favored against $H_0$. We also provide the
percentage of the simulations satisfying this condition. Finally, the vertical line for the ${\rm W}_{\rm GLRT}$
statistic indicates the value of $\nu$ for which $95\%$ of the 1,000 simulations favor
$H_1$ instead of $H_0$. This value of $\nu$ is also written in the panels. Notice that, among the asymptotic
model selection criteria (AIC, BIC and MDL),  AIC offers a less restrictive criterion than BIC,
whereas BIC behaves similarly with respect to MDL,
The figure also shows that BE provides a more conservative
criterion than the asymptotic methods.

\section{Application to WMAP 5-year data}
\label{sec:wmap}

We have applied the statistical approaches described in
Section~\ref{sec:statistics} to WMAP 5-year data. In particular, we
have analyzed a co-added CMB map generated from the global noise-weighted
linear combination of the reduced foreground maps for the Q1, Q2,
V1, V2, W1, W2, W3 and W4 difference assemblies~\citep[see][for
details]{gold08}. Weights are normalized to unity and, for each map,
they are proportional to the inverse average noise variance across
the sky. This operation is made at \nside=512 HEALPix resolution,
being degraded afterwards down to \nside=32.

\begin{figure}
\includegraphics[angle=0,width=8cm,keepaspectratio]{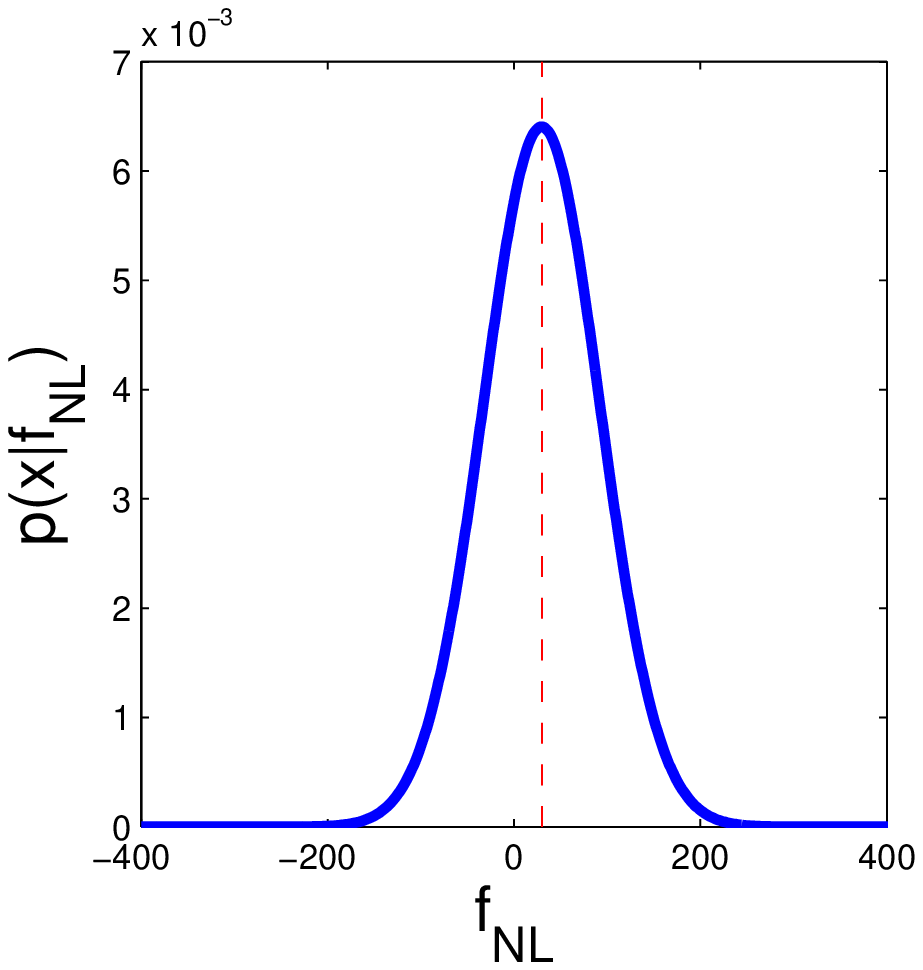}
\caption{\label{fig:data_posterior}This curve represent the
probability given in equation~\ref{eq:pdfx}, i.e. the full pdf of
the WMAP 5-year data $\bmath{x}$ given the non-linear parameter
$\epsilon$ (or, equivalently, f$_\nl)$. Vertical dotted line marks the maximum-likelihood
estimation
$\hat{{\rm f}}_\nl = 30$.}
\end{figure}

Hence, we are in the same conditions as for the analysis on
simulations described in the previous Section and, therefore, the
CMB cross-correlation in WMAP data is given by the $\bmath{\xi}$
correlation matrix already defined in Section~\ref{sec:simulations}.
The estimated full N-pdf of the WMAP 5-year data given
the non-linear parameter $\epsilon$ is showed in
figure~\ref{fig:data_posterior} (indeed, it is given in terms of the
most common f$_\nl$ parameter for allowing a better comparison with
previous works). Maximum-likelihood estimation
(equation~\ref{eq:est_loglike}) provides $\hat{{\rm f}}_\nl = 30$
with an error for the parameter (equation~\ref{eq:error_epsilon}) of
$\sigma_{\hat{{\rm f}}_\nl} = 62$ (compatible with the values
obtained from simulations). Hence our WMAP 5-year data
analysis reports: $\hat{{\rm f}}_\nl = 30 \pm 124$ at 95\%
CL (or, equivalently, $\hat{\epsilon} = 0.019 \pm 0.078$ at 95\%).
This result is compatible with similar works in the literature
reporting WMAP compatibility with Gaussian hypothesis. However, let
us remark that this estimation is more efficient than previous ones
at similar angular resolution,
since it provides a smaller error bar. For instance, \cite{curto07}
performed a Gaussianity test on the WMAP data, at the same HEALPix
resolution, although in a smaller region of the sky (16\% instead of
the 69\% considered in this work). Their f$_\nl$ estimator was based
on the three Minkowski functionals, providing an error bar of
$\approx 200$. The expected $\sigma_{\hat{{\rm f}}_\nl}$ provided
by our maximum-likelihood estimator for a similar observed region would
be $\sigma_{\hat{{\rm f}}_\nl} \approx 124$, i.e., $\approx 40\%$ smaller than the one obtained by the
Minkowski functionals. This result was expected since, as it was
already discussed, the maximum-likelihood estimation of the
non-linear parameter is \emph{optimal}, given the local non-Gaussian model
in equation~\ref{eq:physical_model}.

As it has been mentioned above, this result shows WMAP data
compatibility with the Gaussian hypothesis, since f$_\nl \equiv 0$
can not be rejected at any significant confidence level. Of course,
same conclusions are obtained from the model selection criteria
described in Subsection~\ref{subsec:selection}. Neither AIC, BIC, MDL nor BE criteria
select $H_1$ against $H_0$, whereas GLRT would favor $H_1$ under the very weak
condition for $\nu$ in equation~\ref{eq:glrt_rule} of $\nu \approx 0.1$ (which
implies a likelihood ratio of $\approx 1.1$)

Finally, let us remark that, as it also happened for the simulations analyzed in Section~\ref{sec:simulations},
the contribution of the $S$ term to the value of $Q$ (see equation~\ref{eq:Q}) is not negligible, in
particular, $\vert S/ (-2+J) \vert = 0.74$.
This indicates that,
as it was already mentioned, the cubic term in the model given by equation~\ref{eq:model} is not
\emph{naturally} negligible as compared to the quadratic term and, therefore, $\alpha$ should be chosen
small enough (like the case $\alpha \equiv 0$ considered in this work).

\section{Conclusions}
\label{sec:final}
We have presented a parametric non-Gaussian model for the CMB temperature
fluctuations.
The non-Gaussian model is a local perturbation (up to third order) of the standard CMB Gaussian field
which recovers (for the case of $b \equiv 0$ in equation~\ref{eq:physical_model})
an approximative form of the weak non-linear coupling inflationary model
\citep[e.g.][]{komatsu01,liguori03} at scales larger than the
horizon scale at the recombination time (i.e. above the degree
scale).
For this model, we are able to build the posterior probability of the data given the non-linear
parameter $\epsilon$ (see equation~\ref{eq:model}), from which, in principle, an \emph{optimal estimator}
(i.e., unbiased and with minimum variance) can be derived. Analytical expressions for the maximum-likelihood estimation
of the non-linear parameter ($\hat{\epsilon}$) and its associated error ($\sigma_{\hat{\epsilon}}$) are derived.
In addition, we also discuss an alternative Bayesian
estimation (in terms of the posterior probability of the non-linear parameter), for the hypothetical case
in which we might have some prior information for $\epsilon$. As an example, two cases are addressed: a
non-informative (i.e, uniform) and a Gaussian priors.
We also investigate an issue very much linked to the parameter estimation: the model selection. Indeed, we
discuss several well known techniques to perform hypotheses test, like the Akaike information criterion~\citep[AIC,][]{akaike73},
the Bayesian information criterion~\citep[BIC,][]{schwarz78},
the minimum description length~\citep[MDL,][]{rissanen01}, the generalized likelihood ratio test (GLRT)
and the Bayesian evidence (BE). We derive analytical expressions, for the particular local non-Gaussian model proposed
in this work, for all these model selection techniques.

The performance of both, parameter estimators and model selection criteria, are
investigated by analyzing non-Gaussian simulations, as they could be observed by WMAP. We check that
the maximum-likelihood
estimation provides an unbiased and efficient estimation of the non-linear parameter defining the deviations from
Gaussianity. We find that, for the HEALPix resolution considered in this work (\nside=32), results are
consistent up to a value of $\epsilon = 0.025$, which approximately corresponds (at the Sachs-Wolfe regime)
to a value of the the non-linear coupling parameter f$_{\nl} \approx 350$. This parameter is the one commonly used to described the
weak non-linear coupling inflationary model
\citep[e..g][]{komatsu01}.
We also find that, among the model selection criteria, AIC is the asymptotic method that provides the less
restrictive decision rule, whereas, on the other hand, MDL is the most strict one. We also find that BE, for a uniform
prior given by $\epsilon \in \left[-0.025, 0.025\right]$, is even more restrictive than MDL.

The proposed methodology is applied to WMAP 5-year data. We obtain a value for the non-linear coupling
parameter of $\hat{{\rm f}}_\nl = 30 \pm 124$ at 95\% CL. This result provides a more efficient estimation than
previous works in the literature, at the same angular scales. For instance, comparing with the work by \cite{curto07} using Minkowski functionals, we
can infer that the maximum-likelihood error bar is $\approx 40\%$ smaller than the one obtained with those geometrical estimators.
Application of model selection criteria to WMAP data confirms that standard hypothesis of Gaussianity is favored
against the alternative hypothesis of non-Gaussianity, for the specific local model proposed in this work, and
for the adopted resolution  of $\approx 2^\circ$.

Finally, we would like to comment that, currently, we are extending the technique based on the N-pdf presented in
this work, to deal with a more realistic non-local non-Gaussian model, where higher resolution CMB data are
considered, including as well the effect of anisotropic noise.

\section*{Acknowledgements}
The authors thank Andr{\'e}s Curto, R. Bel\'en Barreiro and Enrique
Mart{\'\i}nez-Gonz{\'a}lez for useful
comments and discussion. We acknowledge partial financial support
from the Spanish Ministerio de Ciencia e Innovaci{\'o}n project
AYA2007-68058-C03-02. PV also acknowledges financial support from
the Ram\'on y Cajal programme. PV thanks to the CNR Istituto de
Scienza e Tecnologie dell'Informazione (ISTI, Pisa) for their warm
hospitality during his research stays in March and June 2008. JLS
acknowledge partial financial support by the Spanish MEC and thanks
the CNR ISTI in Pisa for their hospitality during his sabbatical
leave. The authors acknowledge the computer resources, technical
expertise and assistance provided by the Spanish Supercomputing
Network (RES) node at Universidad de Cantabria. We acknowledge the
use of Legacy Archive for Microwave Background Data Analysis
(LAMBDA). Support for it is provided by the NASA Office of Space
Science. The HEALPix package was used throughout the data analysis
\cite{gorski05}.

\label{lastpage}


\begin{thebibliography}{\protect\citeauthoryear{Mart\'inez-Gonz\'alez \& Sanz}{1990}}

\bibitem[\protect\citeauthoryear{Abramo et al.}{2006}]{abramo06}
Abramo L.R., Bernui A., Ferreira I.S., Villela T., Wuensche C.A.,
2006, Phys. Rev. D, 74, 063506

\bibitem[\protect\citeauthoryear{Ackerman et al.}{2007}]{ackerman07}
Ackerman L., Carroll S.M., Wise M.B. 2007, Phys. Rev. D, 75, 083502

\bibitem[\protect\citeauthoryear{Akaike}{1973}]{akaike73}
Akaike H., 1973, Proceed. of the 2nd International Symposium on
Information Theory (eds. Pertov B.N. \& Czaki F), Akad. Kiado,
Budapest, 267

\bibitem[\protect\citeauthoryear{Babich}{2005}]{babich05}
Babich D., 2005, Phys. Rev. D., 72, 043003

\bibitem[\protect\citeauthoryear{Bartolo et al.}{2004}]{bartolo04}
Bartolo N., Komatsu E., Matarrese S., Riotto A., 2004, Phys. Rep.,
402, 103

\bibitem[\protect\citeauthoryear{Bernui et al.}{2006}]{bernui06}
Bernui A., Villela T., Wuensche C.A., Leonardi R., Ferreira I.,
2006, A\&A, 454, 409

\bibitem[\protect\citeauthoryear{Bernui et al.}{2007}]{bernui07}
Bernui A., Mota B., Rebou\c{c}as M.J., Tavakol R., 2007, A\&A, 464,
479

\bibitem[\protect\citeauthoryear{Bielewicz et al.}{2005}]{bielewicz05}
Bielewicz P., Eriksen H.K., Banday A.J., G\'orski K.M., Lilje P.B.,
2005, ApJ, 635, 750

\bibitem[\protect\citeauthoryear{B\"ohmer \& Mota}{2008}]{bohmer08}
B\"ohmer C.G., Mota D.F., 2008, Phys. Lett. B, 663, 168

\bibitem[\protect\citeauthoryear{Borowiec et al. }{2006}]{borowiec06}
Borowiec A., Godlowski W., Szydlowski M.,  2006, Phys. Rev. D, 74,
043502

\bibitem[\protect\citeauthoryear{Bridges et al.}{2006}]{bridges06}
Bridges M., Lasenby A.N., Hobson M.P., 2006, MNRAS, 369, 1123

\bibitem[\protect\citeauthoryear{Bridges et al.}{2007a}]{bridges07a}
Bridges M., McEwen J.D., Lasenby A.N., Hobson M.P., 2007a, MNRAS,
377, 1473

\bibitem[\protect\citeauthoryear{Bridges et al.}{2007b}]{bridges07b}
Bridges M., Lasenby A.N., Hobson M.P., 2007b, MNRAS, 381, 68

\bibitem[\protect\citeauthoryear{Bridges et al.}{2008}]{bridges08}
Bridges M., McEwen J.D., Cruz M., Lasenby A.N., Hobson M.P., Vielva
P., Mart{\'\i}nez-Gonz\'alez E., 2008, MNRAS, 390, 1372

\bibitem[\protect\citeauthoryear{Cabella et al.}{2005}]{cabella05}
Cabella P., Liguori M., Hansen F.K., Marinucci D., Matarrese S.,
Moscardini L., Vittorio N., 2005, MNRAS, 358, 684

\bibitem[\protect\citeauthoryear{Carvalho et al.}{2008}]{carvalho08}
Carvalho P., Rocha G., Hobson M.P., 2008, MNRAS, submited (preprint
arXiv0802.3916)

\bibitem[\protect\citeauthoryear{Cay\'on et al.}{2003}]{cayon03}
Cay\'{o}n L., Mart{\'\i}nez-Gonz\'alez E., Arg\"ueso F., Banday
A.J., G\'orski K.M., 2003, MNRAS, 339, 1189

\bibitem[\protect\citeauthoryear{Cay\'on et al.}{2005}]{cayon05}
Cay\'{o}n L., Jin J., Treaster A., 2005, MNRAS, 362, 826

\bibitem[\protect\citeauthoryear{Chiang et al.}{2003}]{chiang03}
Chiang L.-Y., Naselsky P.D., Verkhodanov O. V., 2003, ApJ, 590, 65

\bibitem[\protect\citeauthoryear{Chiang \& Naselsky}{2006}]{chiang06}
Chiang L.-Y., Naselsky P. D., 2006, Int. J. Mod. Phys. D, 15, 1283

\bibitem[\protect\citeauthoryear{Coles et al.}{2004}]{coles04}
Coles P., Dineen P., Earl J., Wright D., 2004, MNRAS, 350, 989

\bibitem[\protect\citeauthoryear{Copi et al.}{2004}]{copi04} Copi
C.J., Huterer D., Starkman G.D., 2004, Phys. Rev. D, 70, 043515

\bibitem[\protect\citeauthoryear{Creminelli et al.}{2006}]{creminelli06}
Creminelli P., Nicolis A., Senatore L., Tegmark M., Zaldarriaga M.,
2006, Journal of Cosmology and Astro-Particle Physics, 5, 4

\bibitem[\protect\citeauthoryear{Creminelli et al.}{2007}]{creminelli07}
Creminelli P., Senatore L., Zaldarriaga M.,
2007, Journal of Cosmology and Astro-Particle Physics, 3, 19

\bibitem[\protect\citeauthoryear{Cruz et al.}{2005}]{cruz05} Cruz
M., Mart{\'\i}nez-Gonz\'alez E., Vielva P., Cay\'on L., 2005, MNRAS,
356, 29

\bibitem[\protect\citeauthoryear{Cruz et al.}{2006}]{cruz06} Cruz
M., Tucci M., Mart{\'\i}nez-Gonz\'alez E., Vielva P., 2006, MNRAS,
369, 57

\bibitem[\protect\citeauthoryear{Cruz et al.}{2007a}]{cruz07a} Cruz
M., Cay\'on L., Mart{\'\i}nez-Gonz\'alez E., Vielva P., Jin J.,
2007a, ApJ, 655, 11

\bibitem[\protect\citeauthoryear{Cruz et al.}{2007b}]{cruz07b} Cruz
M., Turok N., Vielva P., Mart{\'\i}nez-Gonz\'alez E., Hobson M.P.,
2007b, Science, 318, 1612

\bibitem[\protect\citeauthoryear{Cruz et al.}{2008}]{cruz08} Cruz
M., Mart{\'\i}nez-Gonz\'alez E., Vielva P. Diego J.M., Hobson M.P.,
Turok N., 2008, MNRAS, 390, 913

\bibitem[\protect\citeauthoryear{Curto et al.}{2007}]{curto07}
Curto A., Aumont J., Mac{\'\i}as-P\'erez, Mart{\'\i}nez-Gonz\'alez
E., Barreiro R.B., Santos D., D\'esert F.-X., Tristram M., 2007,
A\&A, 474, 23

\bibitem[\protect\citeauthoryear{Curto et al.}{2008}]{curto08}
Curto A., Mart{\'\i}nez-Gonz\'alez E., Mukherjee P., Barreiro R.B.,
Hansen F.K., Liguori M., Matarrese S., 2008, MNRAS, in press

\bibitem[\protect\citeauthoryear{Davis et al.}{2007}]{davis07}
Davis T.M. et al., 2007, ApJ, 666, 716

\bibitem[\protect\citeauthoryear{de Oliveira-Costa et al.}{2004}]{deOliveira04}
de Oliveira-Costa A., Tegmark M., Zaldarriaga M., Hamilton A., 2004,
Phys. Rev. D, 69, 063516

\bibitem[\protect\citeauthoryear{Donoghue \& Donoghue}{2005}]{donoghue05}
Donoghue E.P., Donoghue J.F., 2005, Phys. Rev. D, 71, 043002

\bibitem[\protect\citeauthoryear{Eriksen et al.}{2004a}]{eriksen04a}
Eriksen H.K., Hansen F.K., Banday A.J., G\'orski K.M., Lilje P.B.,
2004a, ApJ, 605, 14

\bibitem[\protect\citeauthoryear{Eriksen et al.}{2004b}]{eriksen04b}
Eriksen H.K., Novikov D.I., Lilje P.B., Banday A.J., G\'orski K.M.,
2004b, ApJ, 612, 64

\bibitem[\protect\citeauthoryear{Eriksen et al.}{2005}]{eriksen05}
Eriksen H.K., Banday A.J., G\'orski K.M., Lilje P.B., 2005, ApJ,
622, 58

\bibitem[\protect\citeauthoryear{Eriksen et al.}{2007}]{eriksen07}
Eriksen H.K., Banday A.J., G\'orski K.M., Hansen F.K., Lilje P.B.,
2007, ApJ, 660, L81

\bibitem[\protect\citeauthoryear{Feroz et al.}{2008a}]{feroz08a}
Feroz F., Allanach B.C., Hobson M., Abdus Salam S.S., Trotta R.,
Weber A.M., 2008a, Journ. of High Energy Phys., 10, 64

\bibitem[\protect\citeauthoryear{Feroz et al.}{2008b}]{feroz08b}
Feroz F., Hobson M.P., Zwart J.T.L., Sounders R.D.E., Grainge
K.J.B., 2008b, MNRAS, submited (preprint arXiv0811.1199)

\bibitem[\protect\citeauthoryear{Freeman et al.}{2006}]{freeman06}
Freeman P.E., Genovese C.R., Miller C.J., Nichol R.C., Wasserman L.,
2006, ApJ, 638, 1

\bibitem[\protect\citeauthoryear{Gordon}{2007}]{gordon07}Gordon C.,
2007, ApJ, 656, 636

\bibitem[\protect\citeauthoryear{Gold et al.}{2008}]{gold08}Gold B.,
2008, ApJS, in press

\bibitem[\protect\citeauthoryear{G\'orski et al.}{2005}]{gorski05}
G\'orski K.M., Hivon E., Banday A.J., Wandelt B.D., Hansen F.K.,
Reinecke M., Bartelmann M., 2005, ApJ, 622, 759

\bibitem[\protect\citeauthoryear{Gott et al.}{2007}]{gott07}
Gott J.R., Colley W.N., Park C.-G., Park C., Mugnolo C., 2007,
MNRAS, 377, 1668

\bibitem[\protect\citeauthoryear{Groeneboom \& Eriksen}{2008}]{groeneboom08}
Groeneboom N.E., Eriksen H.K., 2008, ApJ, in press

\bibitem[\protect\citeauthoryear{Hansen et al.}{2004a}]{hansen04a}
Hansen F.K., Cabella P., Marinucci D., Vittorio N., 2004a, ApJ, 607,
L67

\bibitem[\protect\citeauthoryear{Hansen et al.}{2004b}]{hansen04b}
Hansen F.K., Banday A. J., G\'orski K.M., 2004b, MNRAS, 354, 641 0,
063004

\bibitem[\protect\citeauthoryear{Hikage et al.}{2008}]{hikage08}
Hikage C., Matsubara T., Coles P., Liguori M., Hansen F. K.,
Matarrese S., 2008, MNRAS, 389, 1439

\bibitem[\protect\citeauthoryear{Himmetoglu et al.}{2008a}]{himmetoglu08a}
Himmetoglu B., Contaldi C.R., Peloso M., 2008a, (preprint
arXiv0809.2779)

\bibitem[\protect\citeauthoryear{Himmetoglu et al.}{2008b}]{himmetoglu08b}
Himmetoglu B., Contaldi C.R., Peloso M., 2008a, (preprint
arXiv0812.1231)

\bibitem[\protect\citeauthoryear{Hinshaw et al.}{2008}]{hinshaw08}
Hinshaw G. et al., 2008, ApJS, in press

\bibitem[\protect\citeauthoryear{Jaffe et al.}{2006a}]{jaffe06a}
Jaffe T.R., Banday A.J., Eriksen H.K., G\'orski K.M., Hansen F.K.,
2006a, ApJ, 629, L1

\bibitem[\protect\citeauthoryear{Jaffe et al.}{2006b}]{jaffe06b}
Jaffe T.R., Hervik S., Banday A.J., G\'orski K.M., 2006b, ApJ, 644,
701

\bibitem[\protect\citeauthoryear{Jaffe et al.}{2006c}]{jaffe06c}
Jaffe T.R., Banday A.J., Eriksen H.K., G\'orski K.M., Hansen F.K.,
2006c, preprint (arXiv:astro-ph/0606046v1)

\bibitem[\protect\citeauthoryear{Jeffreys}{1961}]{jeffreys61} Jeffreys H.,
1961, Theory of Probability (3rd edition) Oxford university Press

\bibitem[\protect\citeauthoryear{Katz \& Weeks}{2004}]{katz04} Katz
G., Weeks J., 2004, Phys. Rev. D, 70, 063527

\bibitem[\protect\citeauthoryear{Kogo \& Komatsu}{2006}]{kogo06}
Kogo N., Komatsu E., 2006, Phys. Rev. D., 73, 083007

\bibitem[\protect\citeauthoryear{Komatsu et al.}{2001}]{komatsu01}
Komatsu E., Spergel D.N., 2001, Phys. Rev. D, 63, 063002

\bibitem[\protect\citeauthoryear{Komatsu et al.}{2003}]{komatsu03}
Komatsu E., et al., 2003, ApJS, 148, 119

\bibitem[\protect\citeauthoryear{Komatsu et al.}{2008}]{komatsu08}
Komatsu E., et al., 2008, ApJS, in press

\bibitem[\protect\citeauthoryear{Land \& Magueijo}{2005a}]{land05a}
Land K., Magueijo J., 2005a, MNRAS, 357, 994

\bibitem[\protect\citeauthoryear{Land \& Magueijo}{2005b}]{land05b}
Land K., Magueijo J., 2005b, Phys. Rev. Lett., 95, 071301

\bibitem[\protect\citeauthoryear{Land \& Magueijo}{2007}]{land07}
Land K., Magueijo J., 2007, MNRAS, 378, 153

\bibitem[\protect\citeauthoryear{Liguori et al.}{2003}]{liguori03}
Liguori M., Matarrese S., Moscardini L., 2003, ApJ, 597, 57

\bibitem[\protect\citeauthoryear{Liddle \& Lyth}{2000}]{liddle00}
Liddle A., Lyth D. H., 2000, \emph{Cosmological inflation and large-scale structure},
Cambridge University Press

\bibitem[\protect\citeauthoryear{Liddle}{2004}]{liddle04}
Liddle A., 2004, MNRAS, 351, 49

\bibitem[\protect\citeauthoryear{Liddle}{2007}]{liddle07}
Liddle A., 2007, MNRAS, 377, 74

\bibitem[\protect\citeauthoryear{Liddle et al.}{2006}]{liddle06}
Liddle A., Mukherjee P., Parkinson D., Wang Y., 2006, Phys. Rev. D,
74, 123506

\bibitem[\protect\citeauthoryear{Mart\'inez-Gonz\'alez et al.}{2006}]
{martinez06}Mart\'inez-Gonz\'alez E., Cruz M., Cay\'on L., Vielva
P., 2006, New Astron. Rev., 50, 875

\bibitem[\protect\citeauthoryear{McEwen et al.}{2005}]{mcewen05}
McEwen J.D., Hobson M.P., Lasenby A.N., Mortlock D.J., 2005, MNRAS,
259, 1583

\bibitem[\protect\citeauthoryear{McEwen et al.}{2006}]{mcewen06}
McEwen J.D., Hobson M.P., Lasenby A.N., Mortlock D.J., 2006, MNRAS,
371, 50

\bibitem[\protect\citeauthoryear{Monteser{\'\i}n et al.}{2008}]{monteserin08}
Monteser{\'\i}n C., Barreiro R.B., Vielva P., Mart\'inez-Gonz\'alez,
Hobson, M.P., Lasenby A.N., 2008, MNRAS, 387, 209

\bibitem[\protect\citeauthoryear{Mukherjee \& Wang}{2004}]{mukherjee04}Mukherjee
P., Wang Y., 2004, ApJ, 613, 51

\bibitem[\protect\citeauthoryear{Mukherjee et al.}{2006}]{mukherjee06}
Mukherjee P., Parkinson D., Liddle A., 2006, ApJL, 638, 51

\bibitem[\protect\citeauthoryear{Mukherjee \& Liddle}{2008}]{mukherjee08}
Mukherjee P., Liddle A., 2008, MNRAS, 389, 231

\bibitem[\protect\citeauthoryear{Okamoto \& Hu}{2002}]{okamoto02}
Okamoto T., Hu W., 2002, Phys. Rev. D., 66, 063008

\bibitem[\protect\citeauthoryear{Park}{2004}]{park04} Park C.-G.,
2004, MNRAS, 349, 313

\bibitem[\protect\citeauthoryear{Pietrobon et al.}{2008}]{pietrobon08}
Pietrobon D., Amblard A., Balbi A., Cabella P., Cooray A., Mirinucci
D., 2008, Phys. Rev. D, 10, 3504

\bibitem[\protect\citeauthoryear{R\"ath et al.}{2007}]{rath07}
R\"ath C., Schuecker P., Banday A.J., 2007, MNRAS, 380, 466

\bibitem[\protect\citeauthoryear{Rissanen}{2001}]{rissanen01}
Reissanen J., 2001, IEEE Transactions on Information Theory, 47, 5

\bibitem[\protect\citeauthoryear{Schwarz}{1978}]{schwarz78}
Schwarz G., 1978, Ann. Stat., 6, 461

\bibitem[\protect\citeauthoryear{Schwarz et al.}{2004}]{schwarz04}
Schwarz D.J., Starkman G.D., Huterer D., Copi C.J., 2004, Phys. Rev.
Lett., 93, 221301

\bibitem[\protect\citeauthoryear{Spergel et al.}{2007}]{spergel07}
Spergel D.N. et al., 2007, ApJS, 170, 377

\bibitem[\protect\citeauthoryear{Szydlowski \& Godlowski}{2006}]{szydlowski06a}
Szydlowski M., Godlowski W., 2006, Phys. Lett. B, 633, 427

\bibitem[\protect\citeauthoryear{Szydlowski et al}{2006}]{szydlowski06b}
Szydlowski M., Kurek A., Krawiec A., 2006, Phys. Lett. B, 642, 171

\bibitem[\protect\citeauthoryear{Szydlowski et al.}{2008}]{szydlowski08}
Szydlowski M., Godlowski W., Stachowiak T., 2008, Phys. Rev. D, 77, 043530

\bibitem[\protect\citeauthoryear{Tojeiro et al.}{2006}]{tojeiro06}
Tojeiro R., Castro P.G., Heavens A.F., Gupta S., 2006, MNRAS, 365,
265

\bibitem[\protect\citeauthoryear{Vielva et al.}{2004}]{vielva04}
Vielva P., Mart\'inez-Gonz\'alez E., Barreiro R.B., Sanz J.L.,
Cay\'on L., 2004, ApJ, 609, 22

\bibitem[\protect\citeauthoryear{Vielva et al.}{2006}]{vielva06}
Vielva P., Wiaux Y., Mart\'inez-Gonz\'alez E., Vandergheynst P.,
2006, New Astron. Rev., 50, 880

\bibitem[\protect\citeauthoryear{Vielva et al.}{2007}]{vielva07}
Vielva P., Wiaux Y., Mart\'inez-Gonz\'alez E., Vandergheynst P.,
2007, MNRAS, 381, 932

\bibitem[\protect\citeauthoryear{Wiaux et al.}{2006}]{wiaux06}
Wiaux Y., Vielva P., Mart\'inez-Gonz\'alez E., Vandergheynst P.,
2006, Phys. Rev. Lett., 96, 151303

\bibitem[\protect\citeauthoryear{Wiaux et al.}{2008}]{wiaux08}
Wiaux Y., Vielva P., Barreiro R.B., Mart{\'\i}nez-Gonz\'alez,
Vandergheynst P., 2008, MNRAS, 385, 939

\bibitem[\protect\citeauthoryear{Yadav \& Wandelt}{2008}]{yadav08}
Yadav A.P.S., Wandelt B.D., 2008, Phys. Rev. Lett., 100, 181301

\end{thebibliography}
\end{document}